\documentclass[onecolumn,sort&compress,numbers]{els-mrw} 

\usepackage{amsmath,amssymb,amsfonts,amsthm,makeidx,graphicx}
\usepackage{txfonts}
\usepackage{helvet}
\usepackage{soul}

\renewcommand{\Re}{\text{Re }}
\renewcommand{\Im}{\text{Im }}

\begin{document}


\chapter{Regge theory in hadron physics}\label{chap1}

\author[1]{Daniel Winney}%
\author[2,3,4]{Adam P. Szczepaniak}%

\address[1]{Helmholtz Institut f\"ur Strahlen- und Kernphysik (Theorie) and Bethe Center for Theoretical Physics, Universit\"at Bonn, D-53115 Bonn, Germany}
\address[2]{Department of Physics,
Indiana  University,
Bloomington,  IN  47405,  USA}
\address[3]{Center for  Exploration  of  Energy  and  Matter,
Indiana  University,
Bloomington,  IN  47403,  USA}
\address[4]{Theory Center,
Thomas  Jefferson  National  Accelerator  Facility,
Newport  News,  VA  23606,  USA}

\maketitle

\begin{abstract}[Abstract]
We provide a pedagogical introduction to Regge theory as it pertains to the study of hadrons and their interactions. We clarify the fundamental concepts of analyticity in the complex angular momentum plane and their implications for scattering amplitudes. We highlight historical developments that significantly shaped our understanding of scattering theory and the strong interaction, both before and following the discovery of QCD. We end with a review of more recent applications of Regge theory in QCD phenomenology, including describing exchange processes, constraining low-energy amplitudes, and analyzing resonances in the complex angular momentum plane.
\end{abstract}

\begin{keywords}
    \sep Regge theory \sep complex angular momentum \sep hadron exchange \sep Regge trajectory \sep pomeron \sep reggeon 
\end{keywords}

\begin{figure}[h]
        \centering
        \includegraphics[width=0.65\linewidth]{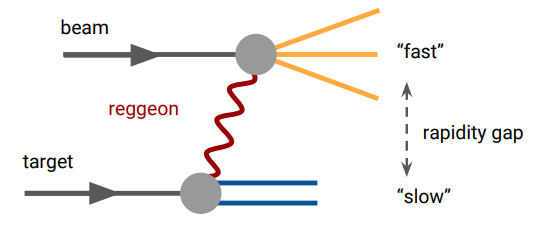}
        \caption{Diagrammatic representation of a diffractive production at high energies from Ref.~\cite{Nys:2018vck}. At small momentum transfers, i.e., $t\sim 0$, the beam particle dissociates and a rapidity gap emerges between the produced particles and the recoiling target.} 
        \label{fig:fast_slow}
\end{figure}


\section*{Objectives}
\begin{itemize}
    \item Explain the core concepts and definitions of Regge theory in both nonrelativistic and relativistic scattering theory.
    \item Describe historical developments in which provided insight to our understanding of strong interactions.
    \item Highlight aspects of QCD which intersect with and emerge from Regge phenomena.
    \item Survey modern techniques and application which use Regge concepts to complement QCD phenomenology.
\end{itemize}
\section{Introduction} \label{sec:intro}

Regge theory\footnote{Pronounced as ``REH-JEH" and not to be confused with the Jamaican music genre, Reggae, which developed independently around the same time.} primarily addresses the fundamental properties of scattering amplitudes as a function of angular momentum. This concept was first introduced, and later named after, Tullio Regge in the context of nonrelativistic potential scattering~\cite{Regge:1959mz, Regge:1960zc}. Specifically, he examined the scattering of a nonrelativistic, spinless particle by a spherically symmetric potential, $V(r)$. In this scenario, the wave function can be expanded into partial waves:
    \begin{equation}
        \label{eq:wavefunction_pwe}
        \Psi(k,r, z) = \sum_{\ell=0}^\infty \, (2\ell +1) \,  P_\ell(z) \, \psi_\ell(k, r) ~,
    \end{equation}
with respect to the wave number $k$ and angular momentum $\ell$.
Here, $P_\ell(z)$ are the Legendre polynomials that depend on the scattering angle, with $z = \cos\theta$. When dealing with particles with spin, the Legendre polynomials are replaced by the Wigner $d$-functions, but the main concepts remain the same. The partial wave satisfies the radial Schrödinger equation (using $c = \hbar = 1$):

    \begin{equation}
        \label{eq:schrodinger_eq}
        \left[\nabla^2 - \frac{\ell\,(\ell+1)}{r^2} + k^2 - V(r) \right] \, \psi_\ell(k,r) = 0 ~,
    \end{equation}
which, for sufficiently well-behaved potentials, can be solved analytically (see treatments in standard textbooks such as Refs.~\cite{Sakurai:2011zz,Newton:1982qc}). The wave function is proportional to the partial wave amplitude $f_\ell(E)$ (which we denote as a function of energy $E=k^2/2m$) in the asymptotic limit:
    \begin{equation}
        \psi_\ell(k,r \to \infty) = f_\ell(E) \, \frac{e^{ikr}}{r}~.
    \end{equation}
Similar to Eq.~\eqref{eq:wavefunction_pwe}, the scattering amplitude as a function of both energy and scattering angle can be reconstructed by summing the partial waves:
    \begin{equation}
        \label{eq:nonrel_pwe}
        F(E,z) = \sum_{\ell=0}^\infty (2\ell+1) \, P_\ell(z) \, f_\ell(E) ~.
        \end{equation}

Regge's groundbreaking idea was to analyze how $f_\ell(E)$ depends on $\ell$ when treated as a continuous and even complex variable. Although the quantized nature of (orbital) angular momentum implies that only nonnegative integer $\ell$ (such as those in the sum of Eq.~\eqref{eq:nonrel_pwe}) can physically contribute to the scattering amplitude, from the perspective of Eq.~\eqref{eq:schrodinger_eq}, $\ell$ acts as an external parameter that can assume any value. The same holds true for energy (or wave number), which, when allowed to take complex values, imposes significant fundamental constraints on the properties of the scattering amplitude (see e.g., Refs.~\cite{Nussenzveig:1972tcd,Newton:1982qc,Martin:1965jj}).
 Specifically, for fixed physical values of $\ell$, $f_\ell(E)$ is an analytic function in the upper-half energy plane except for a branch cut along the positive real axis. If the interaction potential is sufficiently attractive, poles can emerge but must be confined to the negative real axis or on unphysical Riemann sheets, in the case of bound states and resonances respectively.  Since these poles characterize the core dynamics of the interaction, and the most important defining qualities of bound states are their pole location and spin, it seems natural to ask what these amplitudes look like when $E$ is fixed and $\ell$ is varied instead of the other way around.

In his original paper, Regge identified a powerful link between bound states, their spins, and the analytic properties of the scattering amplitude.
Specifically, he demonstrated that by treating the partial waves as functions of both energy and angular momentum, the full scattering amplitude can be shown to be an analytic function across the entire complex $z$-plane (except for possible cuts along the real axis). This builds on the work of Lehmann \cite{Lehmann:1958ita}, which showed that the expansion in Eq.~\eqref{eq:nonrel_pwe} converges only within an ellipse in the complex $z$-plane centered at the origin with foci at $z = \pm1$. This meant that Eq.~\eqref{eq:nonrel_pwe} alone can only confirm the analyticity of scattering amplitudes within this domain, known as the (small) Lehmann ellipse. Martin~\cite{Martin:1965jj} later extended this domain by adding unitarity constraints to cover a larger, yet still finite, ellipse and eventually to any $z$ by Regge. 

\begin{figure}[t]
        \centering
        \includegraphics[width=0.32\linewidth]{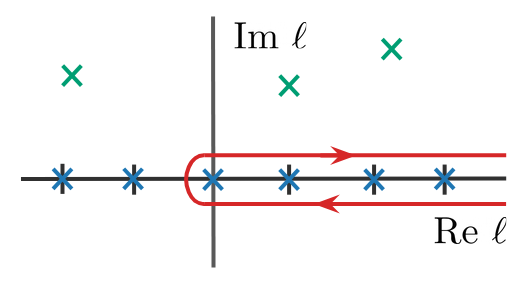}
        \hspace{0.15cm}
        \includegraphics[width=0.32\linewidth]{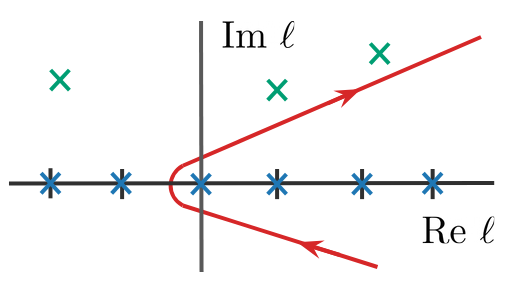}
        \hspace{0.15cm}
        \includegraphics[width=0.32\linewidth]{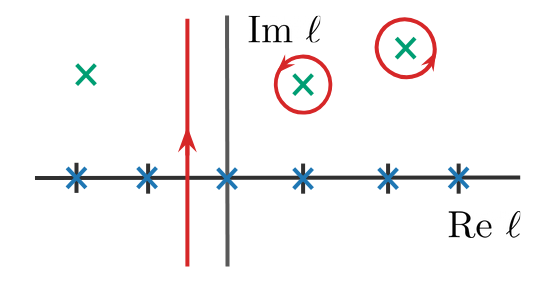}
        \caption{Deformation of the integration contour $C\to C^\prime$ from encircling poles at every integer $\ell \geq 0$ to integrating the imaginary axis with fixed $-1 < \text{Re }\ell < 0$.  We pick up contributions from any possible poles contained in $f(\ell,E)$ in the right-half $\ell$-plane. Poles in the left-half plane with $\Re\ell < -1$ are ``screened" by the background integral.}
        \label{fig:deformed_contour}
    \end{figure}

Using the transform of Watson \cite{Watson:1918} and Sommerfeld \cite{Sommerfeld:1949}, Regge recasts the summation in Eq.~\eqref{eq:nonrel_pwe} as a contour integral.
    \begin{equation}
        \label{eq:sommerfeld_watson}
        F(E,z) = \frac{1}{2i} \int_C d\ell \, \frac{(2\ell+1) \, f(\ell, E) \, P_\ell(-z)}{\sin\pi\ell} ~.
    \end{equation}
The integration path is shown in the left-most panel of Fig.~\ref{fig:deformed_contour}. Here, $f(\ell, E)$ is the analytic continuation of $f_\ell(E)$ to complex values of $\ell$, and in the nonrelativistic framework, can be found by solving Eq.~\eqref{eq:schrodinger_eq} for each $\ell$ along the contour.\footnote{Since Eq.~\eqref{eq:schrodinger_eq} is analytic in $\ell$, and the boundary conditions can be formulated independent of $\ell$, then $\psi_\ell(k,r)$ and $f(\ell,E)$ are analytic in $\ell$ by Poincaré's theorem.} The Legendre polynomials also require analytic continuation, which can be achieved using the Gaussian hypergeometric function,
    \begin{equation}
        \label{eq:Pl_hypergeo}
        P_\ell(-z) = \phantom{}_2F_1\left(-\ell, \ell+1; 1; \frac{1+z}{2}\right) ~,
    \end{equation}
which is entire in $\ell$ and has a branch cut for $z \geq 1$ along the real line.  It is clear that because the factor of $\sin\pi\ell$ is zero at integer values, the integrand will have poles at all integer $\ell$ and, since only those poles at $\ell \geq 0$ lie within the contour, the standard expansion of Eq.~\eqref{eq:nonrel_pwe} follows directly from Eq.~\eqref{eq:sommerfeld_watson} by Cauchy's theorem.  

We observe that for physical $-1 \leq z <1$ (equivalently $0 <\theta \leq \pi$), as $\text{Im }\ell \to \pm\infty$, the ratio $|P_\ell(-z) / \sin\pi\ell| \sim \exp(-\text{Im }\ell \, \theta)$ and the integrand in Eq.~\eqref{eq:sommerfeld_watson} decays exponentially. Therefore, as long as $f(\ell,E)$ is polynomially bounded in the $\ell$-plane, we can deform the contour from $C \to C^\prime$ which involves integrating along $\ell$ with fixed $-1 < \text{Re }\ell <0$ and $\text{Im }\ell \to \pm \infty$ as shown in Fig.~\ref{fig:deformed_contour}. This deformation avoids additional poles from the $\sin\pi\ell$ factor (and the branch cut of the Legendre function), but it may include contributions if $f(\ell,E)$ has poles in the right half of the $\ell$-plane. The resulting expression is a rewrite of Eq.~\eqref{eq:sommerfeld_watson} as:
    \begin{equation}
        \label{eq:transformed_sommerfeld}
        F(E,z) =  \frac{1}{2i} \int_{C^\prime} d\ell \, \frac{(2\ell+1) \, f(\ell,E) \, P_\ell(-z)}{\sin\pi\ell} - \sum_{i} \, \pi \, \beta_i 
        \, \frac{2  \alpha_i +1}{\sin\pi\alpha_i}\, P_{\alpha_i}(-z)  ~,
    \end{equation}
where we generally denote the pole locations and the residues $\alpha_i$ and $\beta_i$, respectively.
In the $E$-plane, the poles of $f_\ell(E)$ correspond to bound states with angular momentum $\ell$, which should remain true even when viewed from the $\ell$-plane. Since Eq.~\eqref{eq:transformed_sommerfeld} is just a different representation of $F(E,z)$, the poles at each $\alpha_i$ should exactly match those in the $E$-plane, and therefore $\alpha$, and by extension $\beta$, must depend on energy. Consequently, the pole position associated with a particle having spin-$\alpha(E)$ traces a path in the $\ell$-plane as $E$ varies. This is known as a moving pole, and the function $\alpha(E)$, called the 
\textit{Regge trajectory}, and $\beta(E)$, the \textit{Regge pole residue}, describe this movement.  Taking the (integer) $\ell$-th projection of Eq.~\eqref{eq:transformed_sommerfeld}, we find:
    \begin{equation}
        \label{eq:PW_of_reggepole}
        f_\ell(E) = \frac{1}{2}\int_{-1}^1 dz \, P_\ell(z) \, F(E,z) =  \left(\frac{2\,\alpha(E) + 1}{\ell + \alpha(E) + 1}\right) \, \frac{\beta(E)}{\alpha(E)-\ell} + (\text{non-pole terms}) ~,
    \end{equation}
Since the same function $\alpha(E)$ appears when projecting onto any $\ell$, all bound states of $F(E,z)$ correspond to the integer values of $\alpha(E)$ which interpolates their pole locations in the $\ell$-plane.

Returning to the integral of Eq.~\eqref{eq:transformed_sommerfeld}, we see that along the new integration path $C^\prime$, we no longer need to limit $z$ to the physical region. In fact, the integral will converge without any additional assumptions on $f(\ell,E)$, for any value of $z$—even those outside the Lehmann ellipse! This way, Eq.~\eqref{eq:transformed_sommerfeld} provides an analytic continuation of the full amplitude to the entire $z$-plane (except along the cut where $z \geq 1$). 

As a final step, we may consider Eq.~\eqref{eq:transformed_sommerfeld} in the limit $z \to \infty$ at a fixed value of $E$, which is typically known as the \textit{Regge limit}. Using $P_\ell(z \to \infty) \propto z^\ell$, the integral is subleading asymptotically because any $\ell^\prime$ along the contour $C^\prime$ has $\Re\ell^\prime < \Re\alpha_i(E)$, and the asymptotic behavior is governed by the pole terms. In this limit, the amplitude behaves as: 

    \begin{equation}
        \label{eq:regge_limit}
        F(E, z \to \infty) \propto \sum_i \, \beta_i(E) \, \frac{2\alpha_i(E)+1}{\sin\pi\alpha_i(E)} \, (-z)^{\alpha_i(E)} ~.
    \end{equation}
This form, especially the power-law behavior $z^{\alpha(E)}$, is widely recognized as a \textit{Regge pole}.
Eq.~\eqref{eq:regge_limit} may seem insignificant at first, since it involves taking an unphysical limit and depends on unphysical values of the angular momentum. However, from Eqs.~\eqref{eq:transformed_sommerfeld} and~\eqref{eq:regge_limit}, we learn that the structure of the full amplitude $F(E,z)$ cannot generally be understood from any finite number of partial waves. Specifically, the analyticity in the $z$-plane typically requires the re-summation of the infinite sum in Eq.~\eqref{eq:nonrel_pwe} outside its radius of convergence through Eq.~\eqref{eq:transformed_sommerfeld}. Moreover, because the partial wave amplitudes are independent of scattering angle, this re-summation is non-trivially connected to the continuation of $f(\ell,E)$ in the complex $\ell$-plane. Finally, in Eq.~\eqref{eq:regge_limit}, it is no longer true that only physical, integer $\ell$ can contribute to the amplitude, as the asymptotic behavior in the Regge limit is dominated by non-integer $\ell \to \alpha(E)$, which correspond to the interpolation between bound state poles. This, in turn, means that one can infer the existence of resonances by examining the asymptotic behavior of scattering amplitudes in this limit.

Before examining the case of relativistic scattering, we briefly look at some specific examples of potentials where the Regge trajectories have been calculated. As mentioned earlier, determining $\alpha(E)$ often only requires knowing the bound state spectrum. A well-known example is Coulomb scattering, $V(r) = - e^2/r$, which allows the scattering amplitude to be solved exactly. In particular, the pole spectrum is known in a closed form, and the trajectory can be found by simply inverting to get $\ell$ as a function of energy~\cite{Singh:1962qgg}:
    \begin{equation}   
        \label{eq:alpha_Coulomb}
        E_{n}(\ell) = \frac{-e^4}{4 \, (n + \ell +1)^2}
        \quad \longrightarrow \quad 
        \alpha_n(E) = - n -1 -\frac{e^2}{2\sqrt{-E}} ~.
    \end{equation}
The trajectory of a
harmonic oscillator potential, $V(r) = \frac{1}{4}\omega r^2$ is similarly calculable~\cite{Aly:1967snr,Kronenfeld:1971rh}:
    \begin{equation}
        \label{eq:alpha_SHO}
        E_{n}(\ell) = \omega \, (2n + \ell + \frac{3}{2}) 
        \quad \longrightarrow \quad \alpha_n(E) = -2n - \frac{3}{2} + \frac{E}{\omega} ~.
    \end{equation}
Here we see that, in both cases, the spectrum is naturally sorted into discrete families lying on parallel trajectories for each nonnegative integer $n$. The two potentials also produce vastly different qualitative behaviors in their trajectories, with Eq.~\eqref{eq:alpha_Coulomb} becoming complex for $E > 0$ and asymptotically approaching a constant, while Eq.~\eqref{eq:alpha_SHO} remains real for $\Im E = 0$ and grows linearly. 
Other potentials have also been considered, e.g. those of square well~\cite{Barut:1962fpy} and Yukawa~\cite{Ahmadzadeh:1963ith,Abbe:1967} type, as well as general algorithms to numerically calculate trajectories from arbitrary potentials~\cite{Burke:1969re}.

\section{$\ell$-analyticity in relativistic scattering} \label{sec:relativistic}

Several years before Regge's seminal paper, Mandelstam, Low, and their contemporaries were already investigating the fundamental structure of relativistic scattering during the early development of what became known as \textit{S-matrix theory}. This interest was fueled by the emergence of a wide variety of relativistic, strongly-interacting particles called hadrons, which could not be described by a field theory similar to quantum electrodynamics. Unlike nonrelativistic scattering, there is no complete ``bottom-up" approach (nor has there been one) that allows for direct calculation of relativistic scattering amplitudes, such as using a relativistic equivalent of the Schr\"odinger equation. The $S$-matrix program, built on ideas introduced by Wheeler~\cite{Wheeler:1937zz} and Heisenberg~\cite{Heisenberg:1943}, aimed instead for a ``top-down" deduction of hadron dynamics by relying as much as possible on experimental observations to constrain general amplitudes that adhere to the fundamental principles of scattering theory.

One early step in this direction was the development of Low's scattering equation in 1955~\cite{Low:1955}, which relates the non-perturbative $S$-matrix to quantities only connected to the asymptotic scatterers and not any internal details of the interaction. The Low equation closely resembles the formulas of Dyson~\cite{Dyson:1949ha} and later by Lehmann, Symanzik, and Zimmermann~\cite{Lehmann:1954rq}, which also aimed to describe scattering consistently using physical (i.e., renormalized) quantities. Unfortunately, it was quickly recognized that Low's equation admits infinitely many solutions in all but the simplest approximations, making it insufficient to uniquely determine the $S$-matrix of a general system~\cite{Castillejo:1955ed, Fukutome:1957}. The problem stems from the possibility of adding any number of poles (known as \textit{CDD poles} after Castillejo, Dalitz, and Dyson, who first identified this ambiguity), which can be incorporated into partial wave amplitudes and correspond to intermediate resonances~\cite{Dyson:1957rgq, Norton:1958vut}. Since these particles decay, they are not asymptotic in the $S$-matrix and therefore remain unconstrained by Low's scattering equation, which only encompasses the initial and final state spectra. This implies that some external, dynamical knowledge about the ``forces" governing the scattering must still be used as input to determine the true $S$-matrix from the set of all possible solutions.

A potentially powerful tool to address this emerged when an additional property, inherent to relativistic scattering, was observed by 
 Gell-Mann and Goldberger~\cite{Gell-Mann:1954wra}: invariance concerning particle-antiparticle transformations, now known as crossing symmetry.
 This symmetry was already implicitly observed in field theories where perturbative amplitudes naturally satisfy the appropriate crossing relations through Lorentz covariance. In the context of the $S$-matrix theory, however, crossing would enable one to use experimental data from one reaction to constrain the unknown dynamics of another.

Consider, for example, the case of $2\to2$ scattering with spinless, distinguishable particles:
    \begin{equation}
        \label{eq:s_channel_reaction}
        a \, (p_a) + b \, (p_b) \longrightarrow c \, (p_c) + d \, (p_d) ~,
    \end{equation}
where $p_i$ denotes their 4-momentum satisfying $p_i^2 = m_i^2$. The kinematics of this reaction is such that it can be described by two Lorentz invariant variables, $s$ and $t$, corresponding to the invariant mass and momentum transfer of the reaction respectively (introduced by Mandelstam in Ref.~\cite{Mandelstam:1958xc}):
    \begin{equation}
        s = (p_a + p_b)^2 = (p_c + p_d)^2 \quad \text{ and } \quad t = (p_a - p_c)^2 = (p_b - p_d)^2 ~.
    \end{equation}
One can define a third variable, $u$, which is not independent because it must satisfy $s + t +u = \sum_i m_i^2$. In the physical region of the reaction, Eq.~\eqref{eq:s_channel_reaction}, we have $s \geq s_\text{th} = (m_a + m_b)^2$ and $t \leq 0$, and this is called the \textit{s-channel reaction}. In relativistic kinematics, we can also now consider the reaction:
    \begin{equation}
        \label{eq:t_channel_reaction}
        a \, (p_a) + \bar{c} \, (-p_c) \longrightarrow \bar{b} \, (-p_b) + d \, (p_d) ~,
    \end{equation}
where the particles $b$ and $c$ are crossed to their antiparticles (denoted by a bar) in the final and initial states, respectively. The reaction becomes physical when $t \geq t_\text{th} = (m_a +m_c)^2$ and $s 
\leq 0$, and is called the 	\textit{t-channel reaction}. Similarly, swapping particles $b \leftrightarrow d$ in Eq.~\eqref{eq:s_channel_reaction} yields the $u$-channel reaction and its corresponding physical region. Crossing symmetry requires that all these reactions are described by the same function $F(s,t)$ evaluated in different regions of the Mandelstam $s-t$ plane~\cite{Kibble:1960zz}.

In a system with bound states, this indicates that any particle propagating in one channel must also necessarily appear as an exchange in every crossed channel. This can have significant consequences because, due to the work of Yukawa~\cite{Yukawa:1935xg}, the exchange of particles was understood to serve as the role of the potentials, i.e., forces, in relativistic scattering. This implied, in principle, that knowing the resonant spectrum of the $t$-channel reaction, for example from experimental phase shifts, allows one to determine the forces that govern the $s$-channel. Crossing symmetry as a dynamical tool to solve Low's equation was successfully used by Chew and Low in elastic $\pi N$ scattering~\cite{Chew:1955zz,Chew:1957zz} (see also Ref.~\cite{Salzman:1957opc}) and in meson photoproduction~\cite{Chew:1956zz} reactions within the so-called ``static limit'' of low energies and small momentum transfer. A formalism for arbitrary kinematics was later developed by Mandelstam in Ref.~\cite{Mandelstam:1958xc}, building upon the early dispersion theory of Goldberger~\cite{Goldberger:1955zz}.
There, Mandelstam realized that a crucial element was missing for the general relativistic theory to satisfy crossing symmetry: simultaneous analyticity in both $s$ and $t$ throughout the entire complex plane\footnote{Commenting on the necessary steps to prove such analyticity, Mandelstam writes ``It is unlikely that such a program will be carried through in the immediate future"~\cite{Mandelstam:1958xc}. The general proof was completed within two years after the timely introduction of Regge theory and its quick adoption to the relativistic case.}.

To illustrate why this is, we can write the analogue of Eq.~\eqref{eq:nonrel_pwe} for the $t$-channel reaction in Eqs.~\eqref{eq:t_channel_reaction} (assuming for simplicity fully symmetric, spinless particles $a = b = c =d$ with mass $m$):
    \begin{equation}
        \label{eq:tchan_expansion}
        F(s, t) = \sum_{\ell=0}^\infty (2\ell+1) \, P_\ell(z_t) \,  f_\ell(t) ~, 
        \quad \text{with} \quad
        z_t(s,t) = \frac{s-u}{4 \, p^2_t} = 1 + \frac{2s}{t-4\,m^2} ~.
    \end{equation}
Suppose the partial waves are primarily influenced by a resonance:
    \begin{equation}
        \label{eq:sum_of_poles}
        f_\ell(t) = \frac{ p_t^{2\ell} \, g_{a\bar{c}} \, g_{\bar{b}d}}{M^2 - t - i M \, \Gamma} ~,
    \end{equation}
where the momentum factors $p_t \equiv p_{a\bar{c}}(t) = p_{\bar{b}d}(t)$ are determined by the analyticity of $f_\ell(t)$ (see, for example, Ref.~\cite{Hara:1964zza})
and each resonance is characterized by its coupling to initial and final states, $g_{a\bar{c},\bar{b}d}$, along with its mass and width, $M$ and $\Gamma$, respectively. Examining Eqs.~\eqref{eq:tchan_expansion} and~\eqref{eq:sum_of_poles}, $F(s,t)$ has a series of poles in $t$ with the $s$ (and $u$) dependence entirely contained in the Legendre functions. Since these functions involve only fixed integer powers of $z_t$, there are no poles (or branch cuts) in $s$ at any finite order in $\ell$. Crossing symmetry now allows us to write a similar expansion for the $s$-channel reaction, with respect to the scattering angle $z_s$. In the fully symmetric case, this is simply Eq.~\eqref{eq:tchan_expansion} with the interchange of $s\leftrightarrow t$ where now the opposite is true: the amplitude has poles in $s$ and is polynomial in $t$.

As long as the summation remains infinite, the expansions in both channels are each exact representations of the full amplitude due to spatial rotation invariance. For crossing symmetry to hold, these must be equal, and the purely kinematic dependence of the Legendre functions in one channel must somehow produce the dynamical dependence of the partial wave in the crossed channel when evaluated in the correct physical region. This suggests that it is generally necessary to consider contributions from all partial waves, since only an infinite polynomial can be re-summed to produce poles\footnote{For example, the re-summation of the infinite geometric series $1 +x +x^2 + \dots = (1-x)^{-1}$ manifests a singularity at $x=1$ whereas any truncated geometric series will not.}. Furthermore, the entire $t$-channel physical region corresponds to an unphysical $s$-channel angle $|z_s| > 1$, meaning that attempting to continue from one reaction to the other also involves considering the partial wave expansion outside its radius of convergence, i.e., outside the Lehmann ellipse. 

As discussed in Sec.~\ref{sec:intro}, such a continuation is possible if $f_\ell(s)$ can also be analytically extended into the complex angular momentum plane. The properties identified by Regge therefore become more significant in relativistic theory, as unphysical scattering angles in one channel are now linked to the physical regions of crossed channel reactions.

Closely related are the implications that resonances in the crossed channel have on unitarity. Based on Mandelstam's dispersive extension of the Chew-Low model for $\pi N$ scattering, Froissart~\cite{Froissart:1961ux} (later generalized by Martin in Ref.~\cite{Martin:1962rt}) formulated a bound on the high-energy behavior of elastic scattering amplitudes. Namely, as $s\to \infty$, the total cross section given by the optical theorem must be bounded above by:
    \begin{equation}
        \label{eq:froissart_bound}
        \sigma_\text{tot}(s\to \infty) = 
    \frac{1}{2s} \, \Im F(s \to \infty, t=0) \leq C \, \log^2 s ~,
    \end{equation}
with some constant $C$. If the sum in Eq.~\eqref{eq:tchan_expansion} contained only a finite spectrum with a maximum spin-$\ell_\text{max}$, then as $s\to\infty$, the exchange of the highest spin particle in the $t$-channel will dominate and produce:
    \begin{equation}
        \label{eq:naive_heb}
        \Im \left[\sum_{\ell=0}^{\ell_\text{max}} (2\ell+1) \, P_\ell(z_t) \, f_\ell(t)\right] \to (2{\ell_\text{max}} +1) \, P_{\ell_\text{max}}(z_t) \, \, \Im f_{\ell_\text{max}}(0) \propto z_t^{\ell_\text{max}}\propto s^{\ell_\text{max}} ~.
    \end{equation}
The Froissart-Martin bound in Eq.~\eqref{eq:froissart_bound} thus requires either that no particles have $\ell >1$ or that the full expansion in Eq.~\eqref{eq:tchan_expansion} leads to less-than-linear asymptotic behavior not apparent from any finite number of terms. Empirically, the latter must be true, as, for example, the spin-2 $f$ and $A2$ mesons (now known as the $f_2(1270)$ and $a_2(1320)$ respectively) were already known to exist by the 60's~\cite{Rosenfeld:1965yz}.
From the definition of $z_t$, one notices that the limit considered by Froissart, i.e., $t = 0$ and $s\to\infty$, is precisely the large-angle limit examined by Regge in Eq.~\eqref{eq:regge_limit}! Therefore, again, if the partial wave amplitude $f_\ell(s)$ is also analytic in $\ell$, the complete amplitude will follow a power law in terms of an energy-dependent Regge trajectory (i.e., instead of a fixed integer power) in order for the elastic amplitude to satisfy Eq.~\eqref{eq:froissart_bound}.

These considerations clearly show that $\ell$-analyticity, as studied by Regge, is key to how relativistic amplitudes satisfy analyticity in energy, crossing symmetry, and unitarity. In the non-relativistic case, however, an analytic continuation $f_\ell(s) \to f(\ell,s)$ to any complex $\ell$ can always be proved to exist (for reasonable potentials) thanks to the Schr\"odinger equation. Since no similar relativistic equation exists, it cannot be proved that the contour deformation in Fig.~\ref{fig:deformed_contour} is well-defined and, therefore, that Regge theory's main results apply to the relativistic $S$-matrix. That is, until 1961, when Gribov~\cite{Gribov:1961fr} and Froissart~\cite{Froissart:1961} proved that the partial wave projection can be written as a dispersion integral of the form:
    \begin{equation}
        \label{eq:finally_gribov}
        f^\pm_\ell(t) =  \frac{1}{\pi} \int_{z_0}^\infty dz \, \, Q_\ell(z) \, \left[ \text{Disc }F(s(t, z) ,t) \pm \, \text{Disc }F(s(t,-z), t)  \right] \equiv f^\pm(\ell, t) ~,
    \end{equation}
which is now called the \textit{Gribov--Froissart projection}. Instead of the typical angular polynomials, the $\ell$-projection occurs through the Legendre functions of the second kind, $Q_\ell(z)$, which are also analytic functions of $\ell$, i.e., through a similar expression in Eq.~\eqref{eq:Pl_hypergeo} and decay uniformly at both large $|z|$ and $|\ell|$. The rest of the integrand consists of the discontinuities of the amplitude across the crossed, i.e., $s$ and $u = s(t,-z)$, channel cuts.
Unlike in non-relativistic scattering, no single function can be constructed to continue all partial waves. Instead, odd and even partial waves must be continued separately, resulting in the two expressions in Eq.~\eqref{eq:finally_gribov} with different relative signs. This sign enforces the necessary symmetry with respect to $z_t \leftrightarrow -z_t$ or, equivalently, $s \leftrightarrow u$. As long as the integration over the discontinuities in Eq.~\eqref{eq:finally_gribov} converges\footnote{In practice, subtractions may be needed in the dispersion relations but this does not spoil the $\ell$-continuation.}, it will define an analytic function that coincides with $f_\ell(t)$ at infinitely many integer points. It then follows from Carlson's theorem that the Gribov--Froissart formula defines a unique analytic continuation $f^\pm(\ell,t)$ for arbitrary $\ell$.

    \begin{figure}[t]
        \centering
        \includegraphics[width=0.65\linewidth]{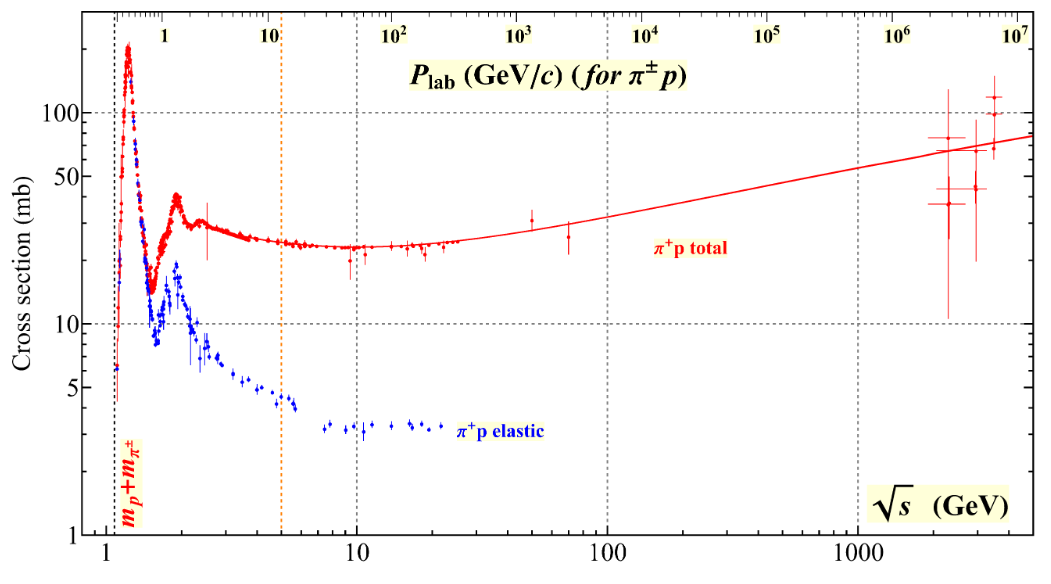}
        \caption{Experimental data for charged pion-proton scattering cross section from Ref.~\cite{ParticleDataGroup:2020ssz}. At small energies the total cross section (in red) is dominated by elastic resonances, while the smooth power-law behavior expected of reggeon exchanges dominates asymptotically. }
        \label{fig:pip_from_pdg}
    \end{figure}

With the existence of a relativistic continuation into the $\ell$-plane, we can revisit and apply the methods of Regge theory in Sec.~\ref{sec:intro} to the relativistic, crossing-symmetric amplitude. Similar to Eqs.~\eqref{eq:transformed_sommerfeld}, the $t$-channel expansion in Eq.~\eqref{eq:tchan_expansion} can now be expressed as:

    \begin{equation}
        F(s,t) = F^+(s,t) + F^-(s,t) ~,
    \end{equation}
with
    \begin{equation}
        \label{eq:tpoles_revealed}
        F^\pm(s, t) = \frac{1}{2i} \int_{C^\prime} d\ell \, (2\ell+1) \, f^\pm(\ell, t) \,\frac{P_\ell(z_t) \pm P_\ell(-z_t)}{2 \, \sin\pi\ell} - \sum_i \, \pi \, \beta_i(t) \, (2\alpha_i(t)+1) \, \left[\frac{P_{\alpha_i(t)}(z_t) \pm P_{\alpha_i(t)}(-z_t)}{2\, \sin\pi\alpha_i(t)}\right] ~. 
    \end{equation}
Just as before, assuming reasonably well-behaved trajectories (now expressed as functions of $t$) the representation $F(s,t)$ extends Eq.~\eqref{eq:tchan_expansion} beyond its radius of convergence for complex $z_t$, and consequently for arbitrary $s$ and $u$, as required for the crossing-symmetric construction of Mandelstam in Ref.~\cite{Mandelstam:1958xc}. In this form, it is clear that the poles in $t$ do not disappear when the amplitude is expanded into $s$-channel partial waves but instead shift onto the complex $\ell$-plane. The poles in $s$ do not appear explicitly in this representation, which only collects poles in the right-half $\ell$-plane and reflects the fact that poles of the crossed channel typically move to the left-half complex plane and are ``screened" by the background integral. Generalizations of this formula, which explicitly show the crossing-symmetric sets of Regge poles by extending the contour to the left were constructed by Mandelstam~\cite{Mandelstam:1959} and further refined by Khuri~\cite{Khuri:1963,Khuri:1963zza}.

Although the motion of the poles will become clearer when we discuss the properties of trajectories more explicitly in Sec.~\ref{sec:regge_poles}, what we find is that in the $t$-channel physical region, $t$ is above threshold, and thus $\alpha(t\geq t_\text{th})$ is complex and near the physical values of $\ell\geq 0$ in the positive right-half plane. As the amplitude is smoothly continued to the $s$-channel physical region, $\alpha(t < t_\text{th})$ becomes real and moves to the left in the $\ell$-plane. The $s$-channel poles, i.e., terms with $\alpha(s)$, do the opposite and move to the right, developing an imaginary part above the respective $s$-channel threshold. The convergence of the expansions with respect to $s$- or $t$-channel partial waves thus reflects which set of moving poles are in the positive, upper quadrant of the $\ell$-plane and are closest to the positive, real $\ell$ axis.

Taking $z_t \to \infty$, as in Eq.~\eqref{eq:regge_limit}, the integral term vanishes asymptotically, and we obtain a remarkable prediction of Regge theory.
    \begin{equation}
        \label{eq:reggeon}
        F(s\to\infty, t< 0) \propto \sum_i \,\pi \, \beta_i(t)  \, \frac{2\alpha_i(t) +1}{\sin\pi\alpha_i(t)} \, \frac{1}{2}\left[1 \pm e^{-i\pi\alpha_i(t)}\right] \, \left(\frac{s}{s_0}\right)^{\alpha_i(t)}  ~,
    \end{equation}
where, we introduce the customary mass scale $s_0 \simeq 2m^2$. Eq.~\eqref{eq:reggeon} is the prototypical Regge pole formula in relativistic scattering and is often referred to as a \textit{reggeon exchange} amplitude\footnote{In phenomenological applications, a factor of $\Gamma(-\alpha(t))$ is often included to cancel all poles at negative integers.}. The most important consequence of this limit is that $\ell$-analyticity requires the $s$-dependence to become asymptotically smooth, i.e., a simple power-law dependence with the exponential dictated by the exchanges in the crossed channel. A classic observation of this phenomenon is the behavior of hadronic cross sections measured at high energies, as seen in Fig.~\ref{fig:pip_from_pdg}.

\begin{wrapfigure}{l}{0.45\textwidth}
        \includegraphics[width=\linewidth]{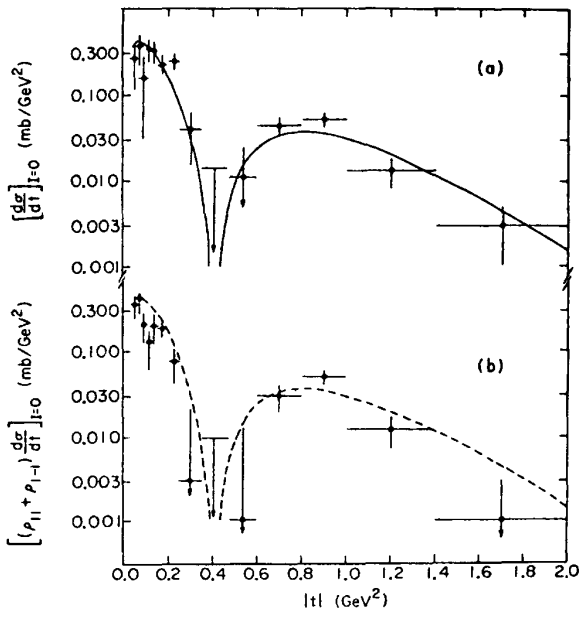}
        \caption{Early evidence for ``Regge dips" in  the total (top) and natural parity component (bottom) of the differential cross section for $\pi^0 p \to \rho^0 p$ in Ref.~\cite{Gordon:1973vc}. The dip corresponds to the first wrong-signature point $\alpha_\rho(t) = 0$.}
        \label{fig:dip}
\end{wrapfigure}

From Eq.~\eqref{eq:reggeon}, it is clear that unitarity does not limit the allowed spins in the physical spectrum but instead simply requires the \textit{Reggeization} of any particles with spin $\ell >1$, that is, these particles all lie on Regge trajectories and are thus re-summed into reggeons asymptotically. Moreover, the Froissart-Martin bound sets a limit on the value of the trajectory at zero momentum transfer:
    \begin{equation}
        \label{eq:intercept_bound}
        \alpha(0) \leq 1 ~,
    \end{equation}
which is often referred to as the \textit{Regge intercept}.
As we will see in the next section, all known Regge trajectories meet this bound, and only a specific trajectory might saturate it.

Comparing Eq.~\eqref{eq:reggeon} to Eq.~\eqref{eq:regge_limit}, crossing symmetry introduces the oscillatory factor:
    \begin{equation}
        \label{eq:signature_factor}
        \frac{1}{2} \left[1 + \tau e^{-i\pi\alpha(t)} \right] ~, 
    \end{equation}
which arises from the need to continue partial waves with either even or odd $\ell$ separately. As a $t$-channel exchange Reggeizes, its ``bare" spin is no longer a useful quantum number. Instead, one typically categorizes reggeons based on their \textit{signature} $\tau$ with trajectories of different signatures generally being independent. This is reflected in Eq.~\eqref{eq:reggeon}, which only features poles of either even or odd $\ell$ (i.e., for $\tau = +1$ and $-1$, respectively), as Eq.~\eqref{eq:signature_factor} induces a zero that cancels the corresponding poles at ``wrong signature" values of $\alpha(t)$. This cancellation of poles manifests as a series of dips visible in the $t$-distribution of differential cross sections, as shown in Fig.~\ref{fig:dip}. These dips are another fundamental prediction of Regge theory, which have been observed across a wide range of processes (see, e.g., Refs.~\cite{Oehme:1968usz,Roskies:1968kak,Arbab:1966nmh}).

One usually also defines the \textit{naturality} $\eta$ of a Regge trajectory, which generalizes the parity quantum number. For a spin-$\ell$ resonance with parity $P$, its naturality is defined as $\eta = P \,(-1)^\ell = \pm1$. As with signature, only particles of a single naturality will lie on the same Regge trajectory, with different naturalities being independent. Together, $\tau$ and $\eta$ allow the identification of individual poles, corresponding to integer values of $\alpha(s)$, back to their intrinsic spin and parity quantum numbers.

Finally, since the function $\beta(t)$ is related to the residue of the $t$-channel pole, it is expected to factorize with respect to the initial state excitation and final state decay~\cite{Arbab:1968urb}. By projecting the partial wave of Eq.~\eqref{eq:reggeon} (analogous to Eq.~\eqref{eq:PW_of_reggepole}), we may expect (see, the $t$-channel reaction of Eq.~\eqref{eq:t_channel_reaction})
    \begin{equation}
        \label{eq:factorization}
        \beta(t) = \beta_{a\bar{c}}(t) \, \beta_{\bar{b}d}(t) ~.
    \end{equation}
When the  the $t$-dependent (i.e, off-shell) residue is evaluated on one of the poles lying on the trajectory, this residue must recover the residue of Eq.~\eqref{eq:sum_of_poles} and thus we have:
    \begin{equation}
        \label{eq:on_shell_residue}
        \beta_{a\bar{c}}(t = M^2) = p^\ell_{a\bar{c}}(M^2) \, g_{a\bar{c}}~,
        \quad
        \beta_{\bar{b}d}(t = M^2) = p^\ell_{\bar{b}d}(M^2) \, g_{\bar{b}d} ~.
    \end{equation}
In systems with spin, this factorization enables constructing residues that depend only on the helicities of initial or final state particles separately~\cite{Frampton:1968rw} and has also been confirmed by high-energy experiments~\cite{Nys:2018vck}.

\section{Hadronic Regge trajectories} \label{sec:regge_poles}

As we have discussed in the previous sections, the central concept in Regge theory is the existence of moving poles in the complex $\ell$-plane\footnote{More complicated models can feature continuous branch cuts in the $\ell$-plane induced by multi-reggeon exchange. These phenomena are expected to be subleading in many processes of interest and will not be discussed here. We instead direct intrepid readers to the relevant chapters in Refs.~\cite{Collins:1977jy,Donnachie:2002en,Gribov:2003nw,Gribov:2009zz}.} which correlate information of the bound state spectrum across energies and spins. It is important to note that, in deriving the primary results of the previous section, we have made no dynamical assumptions of the interaction involved in the scattering. In fact, the structure of amplitudes in complex angular momentum are direct consequences of \textit{any} relativistic theory with crossing symmetry, analyticity, and unitarity.

A curiosity underlying this discussion, is whether or not one can make a distinction of ``elementarity" solely based on the crossing-symmetric $S$-matrix. Specifically, since all bound states can appear as exchanges, then they are all, equally, the ``force carriers" of the theory. The distinction between a composite particle and one thought to be elementary, such as the $\rho$ and $\pi$ mesons respectively, is thus entirely arbitrary. This reasoning was championed by Chew who proposed that all hadrons should be treated as equally composite under the $S$-matrix, a concept later dubbed \textit{nuclear democracy}~\cite{Chew:1961ev,Chew:1962mpd}. While this may no longer seem radical, after all, hadrons are known to indeed be composite quark-gluon objects, the rejection of elementary particles in describing the strong interaction was a wholly new approach to particle physics in the 60's.
The primary basis for this idea was a postulated extension of the maximal analyticity principle\footnote{The principle of maximal analyticity (of the first kind) postulates that the only singularities allowed by nature are those required by unitarity, i.e. resonance poles and threshold cuts (see e.g. Ref.~\cite{Collins:1977jy}).} to the complex $\ell$-plane, later known as maximal analyticity of the second kind~\cite{Chew:1963zze}, requiring that, in light of unitarity and crossing, all poles of the $S$-matrix are Regge poles and thus all hadrons lie on Regge trajectories~\cite{Gribov:1962fx,Barut:1962zz}. If one accepts this, then a hadron appearing at $\alpha(m^2) = \ell$ cannot be ``more elementary" than another at $\alpha(m^{\prime2}) = \ell^\prime$ since they lie on the same trajectory and therefore arise from the same dynamics.

\begin{figure}
        \centering
        \includegraphics[width=0.49\linewidth]{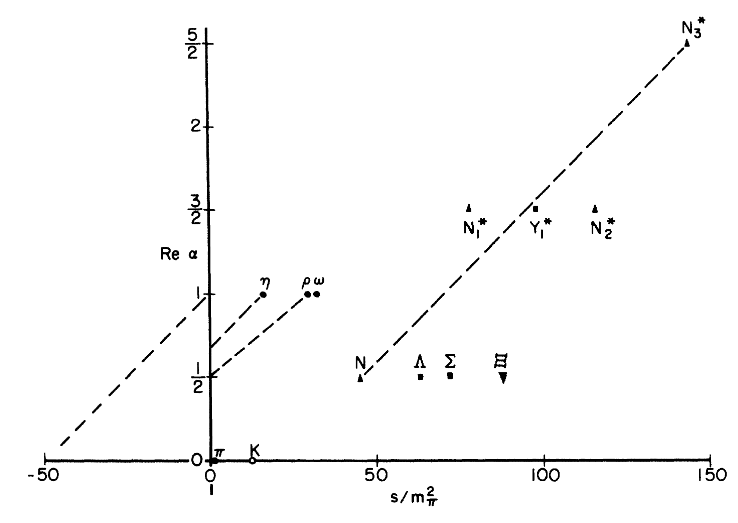}
        \includegraphics[width=0.49\linewidth]{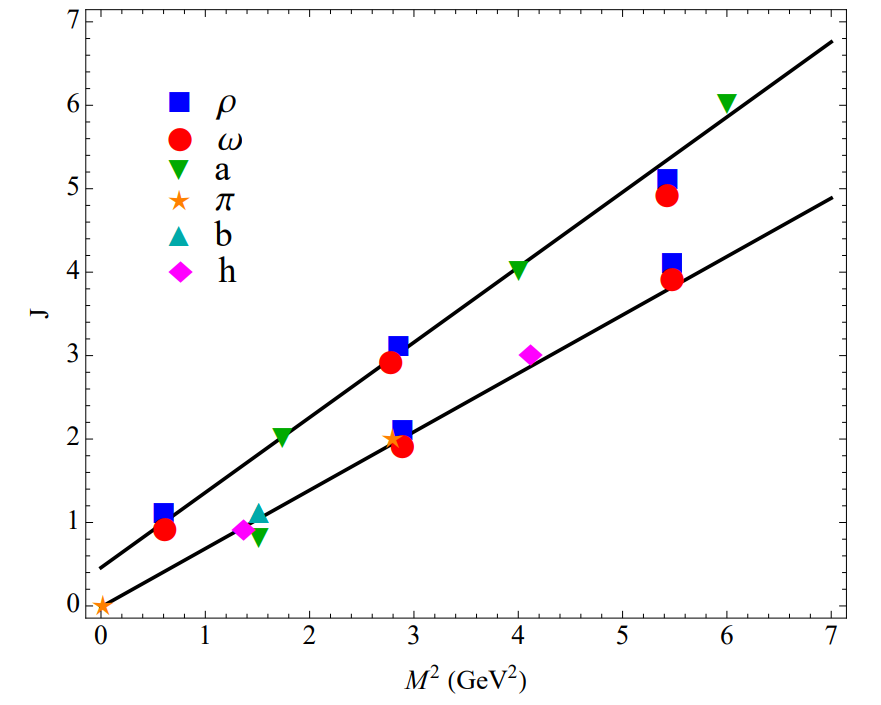}
        \caption{Original 1962 Chew-Frautschi plot from Ref.~\cite{Chew:1962eu} (left) compared to those from the spectrum of light mesons as of 2018 in Ref.~\cite{Mathieu:2018mjw} (right).}
        \label{fig:chew_frautschi}
    \end{figure}

In 1962, Chew and Frautshi made a critical observation in this respect: all hadrons  (known at the time) obey approximately linear relations, when plotting the square of their mass and their spin as in Fig~\ref{fig:chew_frautschi} -- now known as \textit{Chew--Frautschi plots}~\cite{Chew:1962eu}. Further, it was observed that these linear trends featured nearly universal slope such that meson families of opposite signatures seem to lie on the same trajectory, a phenomena called \textit{exchange degeneracy}. The $\rho$ meson trajectory, also generically the reggeon trajectory (as its approximately degenerate with the $\omega$ and $a_2$ trajectories), is seen to be given by: 
    \begin{equation}
        \label{eq:alpha_rho}
        \alpha_\rho(t) \equiv \alpha_\mathbb{R}(t) = 1 + \alpha^\prime_\rho \,(t - m_\rho^2) \approx 0.5 + \alpha^\prime_\mathbb{R} \, t 
        \quad \text{ with } \quad \alpha^\prime_\mathbb{R} = 0.9 \text{ GeV}^{-2} ~,
    \end{equation}
which phenomenologically agreed with data from charge exchange reactions~\cite{Phillips:1965zza}. 
This linear trend pointed to the relativistic Regge trajectories of hadrons rising indefinitely for timelike energies, entirely unlike those given by nonrelativistic exchange potentials. One immediate consequence of a linear $\alpha(t)$ is the prediction that each positive integer value will correspond to a new resonance and thus each trajectory contains an infinite number of particles with increasing mass and spin, known at the time as \textit{Regge recurrences}. 

As pointed out by Van Hove~\cite{vanHove:1967zz} and later Durand~\cite{Durand:1967jrt}, the presence of higher spin recurrences is precisely in line with the Reggeization of Eq.~\eqref{eq:reggeon} and gives a physical intuition as to what reggeons represent. Specifically, the non-diminishing sequence of poles as both spin and energy increase is the reason the partial wave expansion in Eq.~\eqref{eq:tchan_expansion} cannot be truncated. Above a certain energy scale, it is no longer appropriate to describe the scattering as the summation in Eq.~\eqref{eq:tchan_expansion}, since any finite truncation will ignore poles at higher $\ell$ and therefore not reconstruct the full amplitude. Instead, the re-summed reggeon in Eq.~\eqref{eq:reggeon} emerges as the ``propagator" of the simultaneous exchange of all particles which lie on the trajectory in the $t$-channel. Such an interpretation also explains why the exchange of a Reggeized particle acts as a long-range force with an increasing interaction radius unlike a finite-range Yukawa-like potential~\cite{Gribov:1961fm,Gribov:1965hf}.

In their original plot, Chew and Frautschi also included an additional trajectory which contains no hadrons and has $\alpha(0) =1$. As we saw in Eq.~\eqref{eq:intercept_bound}, such a trajectory would saturate the power-law behavior of the unitarity bound and lead to asymptotically constant total cross sections. Experimentally, e.g. in Fig.~\ref{fig:pip_from_pdg}, we actually see very slowly rising cross sections at high energies, a phenomenon also observed in elastic $pp$, $p\bar{p}$, and $K^\pm p$ scattering~\cite{Lindenbaum:1961zz,Baker:1963zzd,Galbraith:1965jk,Barger:1965zz}. From a Regge-theoretic perspective, this necessitates a trajectory with $\alpha(0) = 1$ which then may receive logarithmic corrections (i.e. slower than any power of $s$)\footnote{The literature on high energy diffraction and corrections thereto is vast. As we wish to focus primarily on the trajectories of hadrons, we will not discuss these details here but direct interested readers to Refs.~\cite{Donnachie:2002en,Gribov:2009zz} and references therein.}. In order to avoid resonances with zero mass, this new trajectory must have even signature to cancel the wrong-signature point $\ell = 1$. Since this pole appears in elastic scattering, no additional quantum numbers are exchanged, which further implied this trajectory shares the quantum numbers of the vacuum~\cite{Foldy:1963zz}  and is $s \leftrightarrow u$ (i.e. $z_t \to -z_t$) symmetric. The asymptotic dominance of this pole, then would also result in an equivalence of cross sections for $s$- and $u$-channel reactions:
    \begin{equation}
        \sigma_\text{tot}^{ab}(s\to \infty) = \sigma_\text{tot}^{\bar{a}b}(s\to \infty) ~.
    \end{equation}
This is indeed seen to hold approximately in experiment~\cite{Eden:1971fm,Roy:1972xa} and is known as the \textit{Pomerachuk theorem} after the Soviet physicist who predicted it as a general consequence of crossing symmetry years before~\cite{Pomeranchuck:1958}. The Regge pole responsible for this phenomenon became known as the Pomeranchuck pole, the pomeranchon, or more commonly now, simply the \textit{pomeron}. A classic parameterization of the pomeron trajectory came years later due to the works of Donnachie and Landshoff~\cite{Donnachie:1983hf,Donnachie:1983ff,Donnachie:1992ny} who used
    \begin{equation}
        \label{eq:pomeron_traj}
        \alpha_\mathbb{P}(t) \approx 1.08 + \alpha^\prime_\mathbb{P} \, t 
        \quad \text{ with } \quad
        \alpha^\prime_\mathbb{P} = 0.25 \text{ GeV}^{-2} ~,
    \end{equation}
to describe a wide array of high-energy scattering data. This is a purely phenomenological trajectory as $\epsilon = \alpha_\mathbb{P}(0) -1 = 0.08$ is added to absorb the logarithmic corrections at finite $s$ but will violate unitarity as $s\to \infty$. 

One may notice that, like a typical reggeon, a tower of resonances will emerge if the linear behavior is extrapolated to large, positive arguments. The pomeron is now usually ascribed to gluonic exchanges~\cite{Gell-Mann:1964bha} and thus these resonances would correspond to glueballs, i.e. bound states containing no valence quarks, none of which have thus far been identified unambiguously~\cite{Mathieu:2008me}. An alternative proposal posits the pomeron does not contain any resonances as it is built from entirely non-resonant contributions from many-body inelastic channels~\cite{Freund:1967hw,Harari:1968jw,Harari:1969jcb}, which would require $\alpha_\mathbb{P}(t)$ to not be linear as it must never cross a positive integer. 

Emboldened by the elegant theory emerging from the combination of Regge theory and the relativistic $S$-matrix, Chew conjectured that analyticity, unitarity, and crossing symmetry were constraining enough that only one theory of strong interactions can exist, e.g. precisely the one that gives the spectrum of hadrons in nature~\cite{Chew:1968pj,Chew:1971xwl}. The twofold nature of reggeons as exchange force and bound state, combined with the ``maximal strength" of their interactions (i.e. the saturation of unitarity by the pomeron), suggested that hadrons build themselves through self-interaction in a unique, self-consistent way, i.e. they \textit{bootstrap} themselves. 
Advocates of this hypothesis thus sought self-contained systems of equations resulting from unitarity and involving closed sets of \textit{a priori} unknown parameters (e.g. masses, couplings, etc.). If the bootstrap hypothesis is correct, and sufficient constraints are considered, then these equations can be solved uniquely and allow parameter-free predictions for observable quantities. Early successes came from bootstrap calculations of simple systems with constrained kinematics (so-called ``strip approximations"), such as the $\rho$ resonance parameters~\cite{Zachariasen:1961zz,Collins:1969pg} and, more generally, $\pi\pi$ scattering in $I=0,1$ ~\cite{Collins:1969gq,Webber:1971rm}.

Central to these calculations was determining the trajectory $\alpha(s)$ of various families of resonances, which dictate the dynamics of a reaction. As such, it became necessary to identify the properties of any such trajectory required by $S$-matrix principles. Analyticity of partial waves in energy, for instance, required the same of $\alpha(s)$ which may be written as a dispersion relation~\cite{Collins:1968akw,Degasperis:1970us,Mandelstam:1969dk}:
    \begin{equation}
        \label{eq:alpha_dispersion}
        \alpha(s) = \alpha(0) + \frac{s}{\pi}\int_{s_\text{th}}^\infty ds^\prime \, \frac{\Im \alpha(s^\prime)}{s^\prime \, (s^\prime - s)} ~.
    \end{equation}
Because the trajectory enters amplitudes in the denominator (c.f., Eq.~\eqref{eq:PW_of_reggepole} in analogy to the $N/D$ representation~\cite{Chew:1960iv}), it may only have a right-hand cut\footnote{Models involving coupled channels in which two bare trajectories ``collide" and effectively acquire left-hand cuts are considered in Refs.~\cite{Ball:1969cw,Kaus:1970nc,Ball:1970vy}.}. We write Eq.~\eqref{eq:alpha_dispersion} with a single subtraction such that only the intercept is undetermined by the dispersion relation (but still subject to unitarity constraint of Eq.~\eqref{eq:intercept_bound}). One may expect it  be more appropriate to use two subtractions to compare directly with the linear behavior of Eq.~\eqref{eq:alpha_rho}. Adding a second subtraction, however, fixes the slope and thus the spectrum of resonances external to the hadron dynamics -- and possibly counter to the bootstrap philosophy~\cite{Collins:1968akw}. Requiring maximally one subtraction becomes more concrete when considering the asymptotic bounds imposed on $\alpha(s)$ to avoid essential singularities at infinity in amplitudes of the form in Eq.~\eqref{eq:reggeon}~\cite{Childers:1968vnm,Botke:1972xn,Trushevsky:1975yf}:
    \begin{equation}
        \label{eq:alpha_bounds}
        \Re\alpha(s\to \infty) \, \log s \leq \pi \, \Im \alpha(s\to\infty)
        \quad \text{ with } \quad 
        |\alpha(s\to\infty)| \leq \sqrt{s} \, \log s ~.
    \end{equation}
Evidently, although the Chew-Frautshi plot suggests $\Re\alpha(s)$ is linear, a realistic $\alpha(s)$ is essentially non-linear and complex, requiring one to reconcile the linearity with Eq.~\eqref{eq:alpha_dispersion} via an appropriately chosen $\Im \alpha(s)$. Dynamical calculations imposing unitarity to constrain trajectories were pioneered by Cheng and Sharp~\cite{Cheng:1963} (see also Refs.~\cite{Frautschi:1964,Chu:1968ctr,Epstein:1968vaa}) and used the representation of Eq.~\eqref{eq:PW_of_reggepole} to impose approximate unitarity on a discrete set of poles.

Ultimately, however, bootstrap calculations of more complicated systems very quickly become untractable by requiring many inelastic channels or more trajectories to be considered simultaneously. 
The CDD ambiguities, which plagued Low's original scattering equation, also did not disappear even with the added structure of crossing symmetry and Regge theory~\cite{Atkinson:1969vec} diminishing the possibility of uniqueness hoped for by the bootstrap program. Combined with the rising popularity and success of the parton model~\cite{Feynman:1969wa}, bootstrap calculations saw to a drastic decline in enthusiasm by the end of the 60's. 

\section{Dual resonance models} \label{sec:duality}

Before concluding the discussion of developments in Regge theory prior to the formulation of QCD, in this section, we examine the
 \textit{Dolen--Horn--Schmid (DHS) approach}~\cite{Dolen:1967jr}, which posits a duality between resonances and reggeons.
 This concept originated from the pursuit of self-consistency between ``particles" and ``forces" and led to a new class of amplitudes, the study of which not only outlived the bootstrap program but also played a key role in the shift to viewing hadrons as quark objects.

The discussion of duality began with a new set of self-consistency relations derived in Refs.~\cite{Logunov:1967dy,Igi:1967zza}.  Using the analytic structure of the amplitude at fixed $t$, one can write a dispersion relation, usually in the dimensionless crossing-symmetric variable $\nu = 2 p_t^2 \, z_t /s_0 = (s-u)/2s_0$, with a circular contour of radius $\Lambda$.  Taking $\Lambda \to \infty$, one recovers the usual dispersive representation that reconstructs the amplitude based on the discontinuities across the cuts in $\nu$.
If instead, at some finite energy scale $\Lambda$, $F(\nu,t)$ is well approximated by a sum of Regge poles, then for $\nu \geq \Lambda$ we may expect the difference between $F$ and the approximation to vanish and write (defining $\hat{\beta}(t) =  \tau \pi\, (2\alpha(t)+1) \, \beta(t)$ for convenience):
    \begin{equation}
        \label{eq:FESR}
       \frac{1}{\Lambda} \int_{\nu_\text{th}}^{\Lambda} \Im F(\nu^\prime, t) = \sum_i \, \frac{\hat{\beta}_i(t)}{\alpha_i(t) + 1} \, \Lambda^{\alpha_i(t)}~, 
    \end{equation}
which is the (zeroth moment) \textit{finite energy sum rule} or FESR. Generalizations to higher moments, i.e. sum rules for the derivatives of the amplitude via matching with respect to $\nu^{\prime n} \, \Im F(\nu^\prime,t)$, can also be derived~\cite{Dolen:1967jr,Mathieu:2015gxa}. Examining Eq.~\eqref{eq:FESR}, analyticity requires that the Regge terms on the right-hand side must be related to an average over the low energy amplitude and thus provides a way to constrain the high- or low-energy behavior given the other.

Using the FESRs, Dolen, Horn, and Schmid analyze a type of amplitudes known as interference models, which are now more commonly called \textit{isobar models}. In these models, a complete amplitude is expressed as a sum of poles (either resonant or Reggeized) in each of the $s$, $t$, and $u$ variables~\cite{Vickson:1969ksa}—similar to constructions in theories that use Feynman diagrams.  They argue that such models fail to satisfy the FESRs and, therefore, analyticity by ``double counting" the Regge contributions on both sides of Eq.~\eqref{eq:FESR}.  This suggests that the pole structure of hadronic amplitudes is more restrictive, because, to prevent double counting, the sum over $s$-channel resonances and $t$-channel reggeons must independently satisfy the FESRs and recreate the full amplitude~\cite{chew:1968zz,Schmid:1968zz}. $F(s,t)$ thus needs to be \textit{dual} and cannot have poles in both $s$ and $t$ simultaneously, meaning they cannot be added together. Instead, only one or the other is relevant, depending on the kinematic region where it is evaluated.

Although this requirement may seem excessively restrictive, in a now-famous paper, Veneziano developed the first \textit{dual resonance model} using the Euler Beta function~\cite{Veneziano:1968yb}:
    \begin{equation}
        \label{eq:veneziano}
        F(s,t) = \frac{\Gamma(1-\alpha(s)) \, \Gamma(1-\alpha(t))}{\Gamma(1-\alpha(s) - \alpha(t))}~,
    \end{equation}
and $\alpha(s) = \alpha_0 + \alpha^\prime \, t$ is a real and linear Regge trajectory\footnote{This is known as the narrow resonance approximation and assumes all particles are stable. It is equivalent to treating resonances as Born terms with unitarity as a small perturbation~\cite{Collins:1970as}.}.
 To observe the duality of Eq.~\eqref{eq:veneziano} in action, we first examine the $s$-channel physical region such that $\alpha(s \sim m_\ell^2) \sim \ell \geq 1$ is near a pole.
    \begin{equation}
        \label{eq:venez_pole}
        F(s\sim m_\ell^2, t) = \frac{(-1)^\ell}{\ell!} \frac{1}{\ell - \alpha(s)}\frac{\Gamma(1-\alpha(t))}{\Gamma(1-\ell-\alpha(t)}
        = \frac{(-1)^\ell}{\alpha^\prime \, \ell!} \frac{(-\alpha_0 -\alpha^\prime t)_\ell}{m_\ell^2 - s} ~,
    \end{equation}
where $(x)_\ell = x \, (x-1)\,(x-2) \dots (x-\ell)$ denotes the Pochhammer symbol. Clearly, Eq.~\eqref{eq:venez_pole} describes an isolated pole without any additional ``background term" associated with the $t$-channel. The trajectory in the $t$ channel, $\alpha(t) \propto t$, instead generates the angular polynomial of order $\ell$ in the residue. Note that the resulting residue is not orthogonal to Legendre polynomials and therefore will project onto all spins $\leq \ell$. These are a series of resonances with decreasing spin, all aligning at the same mass, called \textit{daughter poles}. Requiring a real, linear trajectory prevents the angular polynomial from projecting onto spins $> \ell$, resulting in unphysical poles called \textit{ancestors} or at negative masses called \textit{ghosts}. The possibility of a double pole in unphysical kinematics, where $\alpha(s) = \alpha(t) = \ell$, is also avoided by the denominator of Eq.~\eqref{eq:veneziano}.

Looking at the Veneziano amplitude now in the Regge limit, one sees
    \begin{equation}
        F(s\to \infty, t)  = \frac{\pi \, e^{-i\pi\alpha(t)} \, \alpha(s)^{\alpha(t)}}{\Gamma(\alpha(t)) \, \sin\pi\alpha(t)}  = \frac{\pi \, e^{-i\pi\alpha(t)}}{\Gamma(\alpha(t)) \, \sin\pi\alpha(t)} \, \left(\frac{s}{1/\alpha^{\prime}}\right)^{\alpha(t)} ~,
    \end{equation}
which is precisely the expected Regge behavior. This time, $\alpha(s) \propto s$ provides the leading energy dependence, once again, without any additional crossed channel terms. Since Eq.~\eqref{eq:veneziano} is symmetric under $s\leftrightarrow t$, the same limits also apply in the physical region of the $t$ channel. The signature factor from Eq.~\eqref{eq:signature_factor} can be obtained by imposing the necessary $s\leftrightarrow u$ symmetry, i.e. $F(s,t) \pm F(u,t)$.

The Veneziano model marked an important turning point for the bootstrap program, as duality greatly limited the functional space of the amplitudes proposed to be possible within the $S$-matrix. The form of Eq.~\eqref{eq:veneziano} was very appealing because of its simplicity, as it clearly displays crossing symmetry and duality in a closed form, with linear trajectories as seen in nature. Consequently, extensive effort was developed to study Veneziano-like amplitudes, as they were believed to hold the key to understanding the strong interaction (see~\cite{Sivers:1971ig} for a review).
Most notably, the adaptation of Eq.~\eqref{eq:veneziano} to $\pi\pi$ scattering by Shapiro~\cite{Shapiro:1969km} and Lovelace~\cite{Lovelace:1968kjy} shows that the property $\alpha_\rho(0) = 1/2$ (cf. Eq.~\eqref{eq:alpha_rho}) is enough to fulfill the constraints of chiral symmetry explored by Adler~\cite{Adler:1964um}.
Later developments also included a prescription to construct the analogous $N$-particle Veneziano amplitude~\cite{Chan:1969xg}
and the generalization to a larger class of amplitudes called \textit{dual amplitudes with Mandelstam analyticity} (DAMA)~\cite{Friedman:1970wj,Cohen-Tannoudji:1971mlv,Bugrij:1973ph}:
    \begin{equation}
        \label{eq:DAMA}
        F(s,t) = \int_0^1 dx \, g(s,t,x) \, x^{-\alpha(s) - 1} \, (1-x)^{-\alpha(t)-1}  ~,
    \end{equation}
for $g(s,t,x) = g(t,s, 1-x)$ is an arbitrary regular function\footnote{The Veneziano amplitude follows from $g(s,t,x) = 1$ and the integral representation of the Beta function.}. This latter class of models, formed from Eq,~\eqref{eq:DAMA}, enabled more complex trajectories, especially those satisfying Eqs.~\eqref{eq:alpha_dispersion} and \eqref{eq:alpha_bounds}, to be included while still maintaining duality.

\begin{wrapfigure}[27]{l}{0.4\textwidth}
        \includegraphics[width=\linewidth]{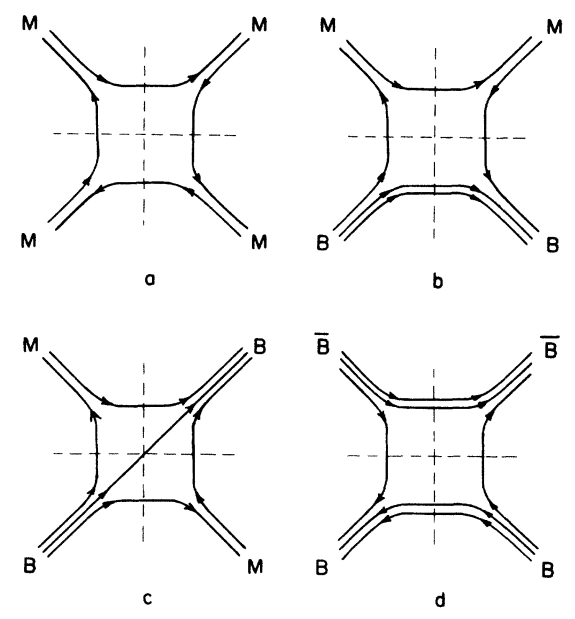}
        \caption{Duality diagrams for meson-meson (a), meson-baryon (b-c) and baryon-antibaryon (d) scattering from Ref.~\cite{Harari:1969oxx}. In (d), the dominance of $t$-channel exchange is explained by the lack of an exotic $qq\bar{q}\bar{q}$ trajectory.}
        \label{fig:duality_diagrams}
\end{wrapfigure}

Despite many subsequent developments following Veneziano's paper, dual resonance models were ultimately too idealized to accurately describe realistic hadronic amplitudes. Specifically, the elegant structure of Eq.~\eqref{eq:veneziano} heavily depends on the zero-width approximation to ignore the dynamical complexities of unitarity. Considerable effort was directed towards ``unitarizing" the Veneziano amplitude (see e.g. Refs.~\cite{Martin:1969tc,Huang:1969ef,Bali:1969fm,Tryon:1971pt,Paciello:1969ry}) to account for finite-width spectra, but with limited success.  
Ultimately, no consistent unitarization scheme was found for the Veneziano amplitude, which greatly limits its applications beyond the Born level. Generalized DAMA models, which relax some restrictions on trajectories, also achieved limited numerical success because of their more complex structure
\footnote{Eq.~\eqref{eq:DAMA} will generally only converge in the unphysical region where both $\Re\alpha(s)$ and $\Re\alpha(t) < -1$. Investigation of resonances requires analytic continuation to the physical regions of the $s$- or $t$-channel~\cite{Bugrij:1973ph}.}.

Still, the idealized Veneziano amplitudes that emerged from the duality program proved to be a pivotal moment for understanding hadrons as composite objects during the early days of QCD. In 1969, Susskind provided a physical interpretation of the Veneziano model by comparing it to a series of harmonic oscillators~\cite{Susskind:1969ha,Susskind:1970qz}, a comparison already evident in Fig.~\ref{fig:chew_frautschi} with Eq.~\eqref{eq:alpha_SHO}. As a result, the structure of hadrons started to resemble ``quarks and antiquarks connected by the continuum limit of a chain of springs." The tower of Regge recurrences and the spectrum of daughters are then naturally interpreted as orbital and radial excitations of those springs\footnote{The harmonic oscillator model, along with later work by Ramond~\cite{Ramond:1971gb}, Neveu, and Schwarz~\cite{Neveu:1971fz}, became the foundation for string theory, see e.g. Refs.~\cite{Becker:2006dvp}.}.

Around the same time, Harari~\cite{Harari:1969oxx} and Rosner~\cite{Rosner:1969bhr} introduced the \textit{duality diagrams} shown in Fig.~\ref{fig:duality_diagrams}. These diagrams schematically demonstrated that the effects of dual exchanges can be attributed to the recombination of components based on the SU(3) assignments of mesons and baryons, namely as $q\bar{q}$ and $qqq$ objects respectively, as proposed by Gell-Mann a few years earlier~\cite{Gell-Mann:1962yej,Gell-Mann:1964ewy}. In a series of papers~\cite{Mandelstam:1969pkr,Mandelstam:1970fd,Mandelstam:1970zy}, Mandelstam showed that predictions from relativistic quark models correspond well with the leading Regge trajectories in both the meson and baryon sectors. Later work by Veneziano~\cite{Veneziano:1976wm} explored the connection between the large-$N_c$ approximation to QCD developed by t'Hooft~\cite{tHooft:1973alw,tHooft:1974pnl} and the zero-width dual models. Specifically, the planar loop diagrams of the latter formalize the duality diagrams of the former in terms of confined quarks. Together with Rossi, they further developed the large-$N_c$ QCD framework into the baryon and multi-quark state sectors~\cite{Rossi:1977cy}.

\begin{figure}[t]
    \centering
    \includegraphics[width=0.6\linewidth]{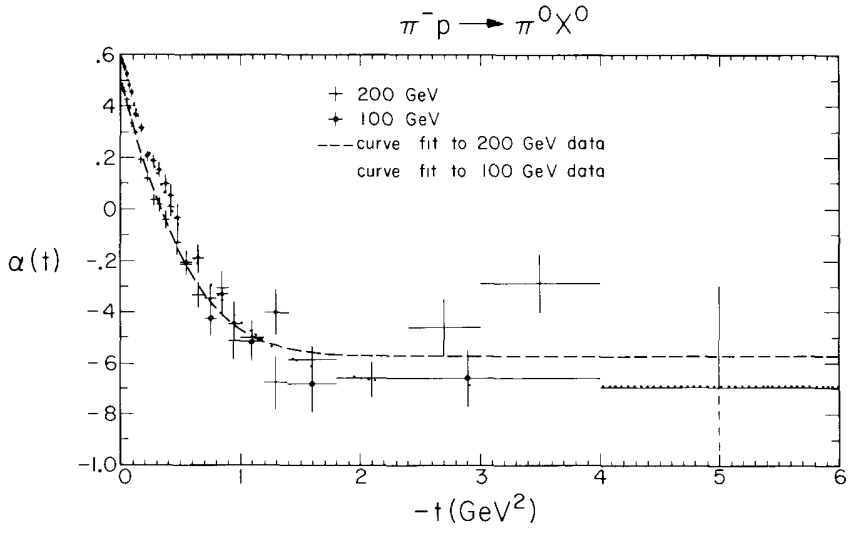}
        \caption{Nonlinear, effective $\rho$ meson trajectory extracted from inclusive neutral pion production at large momentum transfers from Ref.~\cite{Kennett:1986wf}.}
        \label{fig:effective_rho}
\end{figure}

A connection with perturbative parton models was established in Refs.~\cite{Gunion:1972gy,Blankenbecler:1973kt}, which examined the emergence of Regge behavior at small momentum transfer through the re-summation of dressed parton exchange diagrams, the latter of which can be calculated perturbatively at large momentum transfer. 
In particular, constituent counting rules developed in high-energy, fixed-angle scattering~\cite{Brodsky:1973kr,Matveev:1973ra} predicted that Regge trajectories asymptote at large negative arguments to a value depending on the number of constituents exchanged, e.g., for $\alpha_\rho(t \to - \infty) = -1$ ~\cite{Collins:1983fg,Brodsky:1993fv}. A hint of this behavior would be observed in exclusive~\cite{Barnes:1976ek} charged exchange $\pi$ production and later in inclusive~\cite{Kennett:1986wf} final states, the latter extracting the effective $\rho$ trajectory out to $-t \sim 6$ GeV$^2$ as shown in Fig.~\ref{fig:effective_rho}.

Ref.~\cite{Schmidt:1973ew} also investigates the Regge behavior of partons and suggests that DAMA models, such as in Eq.~\eqref{eq:DAMA}, are actually Reggeized versions of the constituent exchange model, with the integration variable representing the momentum fraction of the exchanged partons. This was similarly demonstrated numerically in Ref.~\cite{Coon:1974wh} by analyzing $\pi p$ and $pp$ scattering data in both forward and backward Regge regions and at fixed angles, using either a constituent parton exchange model or a dual resonance model.

\begin{figure}[b]
        \centering
        \includegraphics[width=0.48\linewidth]{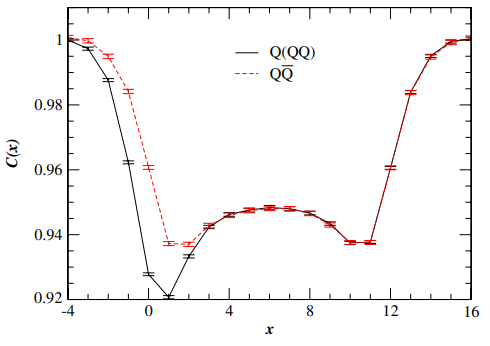}
        \includegraphics[width=0.48\linewidth]{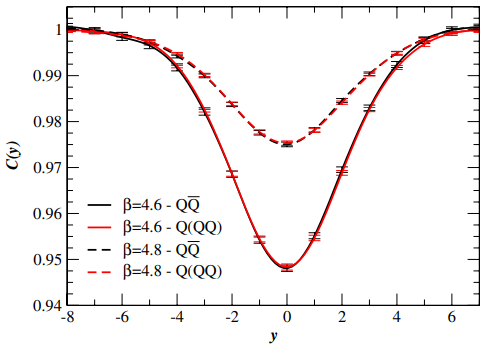}
        \caption{Longitudinal (left) and transverse (right) energy profiles for the gluonic flux tubes of quark-antiquark mesons and quark-diquark baryons calculated on the lattice in Ref.~\cite{Bissey:2009gw}.}
        \label{fig:fluxtubes}
    \end{figure}

By this time, Gross, Wilczek~\cite{Gross:1973id}, and Politzer~\cite{Politzer:1973fx} had published their discoveries of asymptotic freedom and opened the floodgates for perturbative QCD, which quickly overtook the already diminishing $S$-matrix and bootstrap approaches in favor of fundamental quark and gluon descriptions.
Since then, considerable progress has demonstrated an intimate relationship between the quasi-linear Regge trajectories of the Chew-Frautschi plots and the confinement of quarks in QCD, e.g., in Refs.~\cite{McGuigan:1992bi,Kruczenski:2004me,RuizdeElvira:2010cs,Nedelko:2016gdk}. Studies with QCD-based approaches to excited hadron spectra, e.g., from Dyson-Schwinger equations~\cite{Fischer:2014xha}, holographic QCD~\cite{Sonnenschein:2018fph}, and lattice QCD~\cite{Meyer:2004jc}, continue to reinforce and deepen our understanding of hadronic Regge poles.
Moreover, numerical simulations on the lattice~\cite{Greensite:2014bua,Dawid:2019vhl,Dawid:2024uqz,Greensite:2001nx,Bissey:2006bz,Bissey:2009gw} now enable more realistic studies of the ``stringy" flux tube dynamics of gluons (shown in Fig.~\ref{fig:fluxtubes}), which underlie the harmonic oscillator-like Regge trajectories first observed by Chew and Frautschi in 1962.

\section{Modern applications} \label{sec:applications}

As discussed, Regge theory achieved its peak popularity and most of its formal development in the 1960's as an effort to create a self-consistent, dynamical theory for hadrons. 
Considerable evidence over the past half-century, however, points instead to the SU(3) gauge theory of QCD as the true theory of the strong interaction. 
Does this mean that Regge theory and the rest of the $S$-matrix formalism are merely historical relics from a time before QCD? No, in fact, the nonperturbative nature of QCD still prevents first-principles calculations of key soft processes.
Many of the tools developed during the original\footnote{A spiritual successor of the $S$-matrix bootstrap has recently emerged in the realm of mathematical physics and axiomatic QFTs, see e.g. Refs.~\cite{Mizera:2023tfe}.} bootstrap program are universal because they originate from the fundamentals of scattering theory and thus consistently find applications in QCD phenomenology. 

Modern applications of Regge theory aim to connect experimental data with physical quantities, which can then be compared to predictions from lattice simulations or models~\cite{JPAC:2021rxu}.
We highlight several of these applications in the following subsections.

\subsection{Exchange phenomenology} \label{sec:exchange_pheno}

One of the most common uses of Regge theory is in the phenomenology of high-energy exchange reactions. This is not surprising since the most important predictions occur in the Regge limit where $s\to \infty$ at fixed $t$. Scattering reactions in this regime are usually dominated by soft peripheral processes (such as hadron exchange) that cannot be predicted from perturbative QCD. The reggeon exchange in Eq.~\eqref{eq:reggeon} is, however, entirely general because it comes from first principles of relativistic scattering theory\footnote{Regge theory, however, does not predict the overall size or $t$-dependence of the couplings, so phenomenological assumptions are often necessary.}. Dedicated hadron spectroscopy programs at modern experiments such as GlueX~\cite{GlueX:2020idb}, CLAS12~\cite{CLAS:2003umf}, and COMPASS~\cite{COMPASS:2007rjf} have thus led to a resurgence of Regge approaches to analyze high-statistics data of peripheral production. 
These production processes using pion or photon beams, e.g. of the form $\gamma p \to Xp$, enable the flexible creation of meson systems of interest and serve as ideal laboratories for hadron spectroscopy. 
 The dynamics of reggeon exchanges implies that the target particle, usually nucleons, can be kinematically separated by the presence of a \textit{rapidity gap}.
 This is illustrated in Fig.~\ref{fig:fast_slow} and helps limit the influence of potential rescattering between particles in the final state, which can greatly complicate the study of similar systems produced in heavy-hadron decays. Furthermore, production with real photon beams at GlueX allows the creation of arbitrary quantum numbers, including those that seem forbidden by quark models, as well as access to polarized observables that offer additional insights into the spin structure of the interaction. One important set of polarized observables is the \textit{spin density matrix elements} (SDMEs), which parameterize the angular distribution of the lab intensity and provide key insights into the interference between different exchanges~\cite{Schilling:1969um}. 

    \begin{figure}[t]
        \centering
        \includegraphics[width=\linewidth]{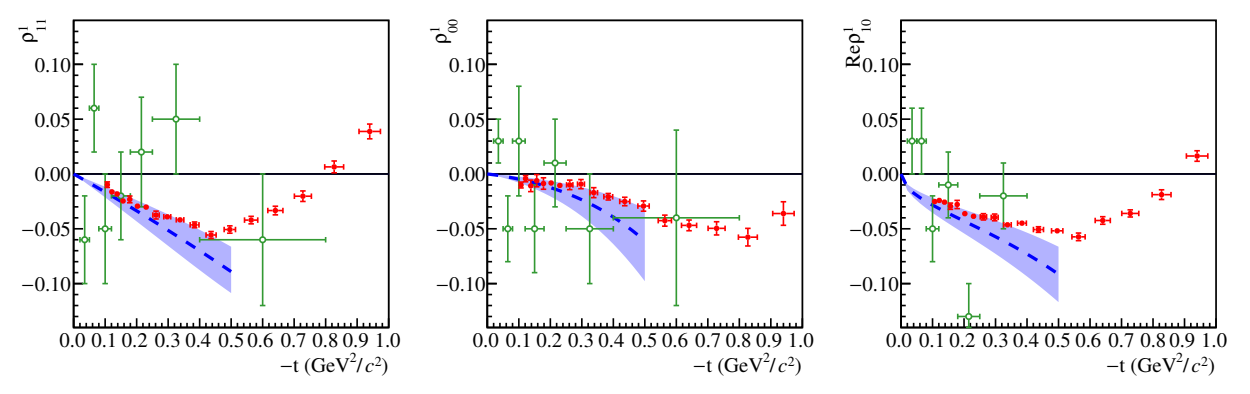}
        \caption{Polarized SDMEs for $\rho^0$ photoproduction. Plots and red data points from Ref.~\cite{GlueX:2023fcq}. Blue curve is predictions from the Regge pole model in Ref.~\cite{Mathieu:2018mjw}.}
        \label{fig:rho_sdmes}
    \end{figure}

The formalism of Sec.~\ref{sec:relativistic} can be extended to include spins so that a general exchange amplitude is given by~\cite{Irving:1977ea}:
    \begin{equation}
        \label{eq:with_spin}
        \langle\lambda_X,\lambda_{N^\prime}|T_\alpha(s,t)|\lambda_\gamma, \lambda_{N}\rangle = \xi_{\lambda\lambda^\prime}(s,t) 
        \, \beta_{\lambda_\gamma\lambda_X}(t) \, \beta_{\lambda_N \lambda_{N^\prime}}(t) \,\frac{1+\tau\,e^{-i\pi\alpha(t)}}{\sin\pi\alpha(t)}
        \, \left(\frac{s}{s_0}\right)^{\alpha(t)}~,
    \end{equation}
in terms of helicities $\{\lambda_\gamma, \lambda_N, \lambda_X, \lambda_{N^\prime}\}$ for the beam (taken to be a photon), target, produced meson, and recoil nucleon respectively. The additional factors 
    \begin{equation}
        \xi_{\lambda\lambda^\prime}(s,t) = \left(\frac{1+z_s(s,t)}{2}\right)^{|\lambda + \lambda^\prime|/2} \, \left(\frac{1-z_s(s,t)}{2}\right)^{|\lambda - \lambda^\prime|/2} \, \left(\frac{-t}{s}\right)^{|\lambda - \lambda^\prime|/2} ~,
    \end{equation}
called the \textit{half-angle factors}, depend only on the net helicities of the initial and final state: $\lambda = \lambda_\gamma - \lambda_N$ and $\lambda^\prime = \lambda_X - \lambda_{N^\prime}$ and contain kinematic singularities associated with the boundary of physical region in the presence of particles with spin (see e.g. Refs.~\cite{Cohen-Tannoudji:1968lnm,Hara:1971kj}). As in Eq.~\eqref{eq:factorization}, the rest of the residue is required to factorize and depend only on the helicities of particles at the ``top" and ``bottom" vertices, i.e. involving the photon and nucleons respectively. A minimal overall dependence of these couplings on $t$ is required by analyticity:
    \begin{equation}
        \beta_{\lambda_1,\lambda_2}(t) \propto \sqrt{-t}^{|\lambda_1-\lambda_2|}~, 
    \end{equation}
but arbitrary additional polynomials of $t$ may also enter and are otherwise completely unconstrained from Regge theory. 

Looking at Eq.~\eqref{eq:with_spin}, the reaction amplitude mainly depends on the trajectory and the two residues. The trajectories are, in principle, process-independent and can usually be determined from quantum number considerations, with the leading trajectories typically being the $\rho$ and pomeron from Eqs.~\eqref{eq:alpha_rho} and \eqref{eq:pomeron_traj}, as well as the $\pi$ trajectory:
    \begin{equation}
        \alpha_\pi(t) \approx \alpha^\prime_\pi \, (t - m_\pi^2) 
        \quad \text{ with } \quad 
        \alpha^\prime_\pi = 0.7 \text{ GeV}^{-2} ~.
    \end{equation}
Although the properties of $\alpha(t)$ discussed in Ref.~\ref{sec:regge_poles} suggest a more complex form, in phenomenological applications, the dominance of the forward direction $-t \approx 0$ at high energies means that linear approximations are usually more than adequate.
Most of the modeling involved in a Regge analysis is therefore in determining an appropriate form for the residues. Fortunately, thanks to the factorization in Eq.~\eqref{eq:factorization}, the top and bottom couplings can be modeled independently and compared across processes with common vertices. A typical starting point for the vertex function is the coupling to the bare (i.e., not Reggeized) exchange, using an effective Lagrangian and on-shell couplings~\cite{Nys:2018vck}. 

\begin{wrapfigure}{r}{0.45\textwidth}
        \includegraphics[width=\linewidth]{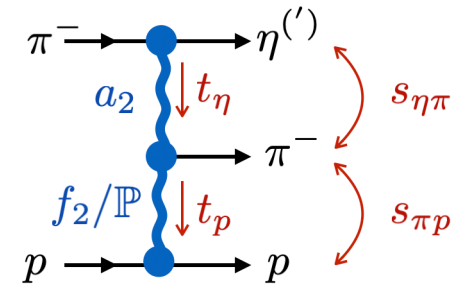}
        \caption{Diagrammatic representation typical double Regge reaction in $\eta^{(\prime)}\pi$ photoproduction from Ref.~\cite{Bibrzycki:2021rwh}. Here both $s_{\eta\pi}$ and $s_{\pi p}$ are large enough that both exchanges Reggeize.} 
        \label{fig:double_regge}
\end{wrapfigure}

Of course, the Regge limit we have explored is formally $s\to \infty$ and therefore $-t/s\to 0$, so it is not certain if the assumptions of Regge pole dominance and factorization apply at finite $s$ of a few tens of GeV$^2$ or $-t$ up to 1 GeV$^2$. In Ref.~\cite{Nys:2018vck}, global charge exchange data (e.g., reactions like $\pi^- p \to \pi^0 \Delta^{++}$, $K^+ n \to K^0 p$, $\pi^+ p \to K^+ \Sigma^+$, etc.) are examined using Reggeized $\rho$, $a_2$, and $K^*$ exchange amplitudes similar to those in Eq.~\eqref{eq:with_spin}.  This analysis demonstrated the importance of Regge pole amplitudes for finite kinematics, with corrections to Regge asymptotics from Regge cuts or daughter exchanges not exceeding $10-20\%$. 

In photon-mediated production, predictions based on Regge models for $\eta^{(\prime)}$~\cite{Mathieu:2017jjs}, vector~\cite{Mathieu:2018xyc}, and tensor~\cite{Mathieu:2020zpm} meson photoproduction have compared favorably with more recent GlueX data in Refs.~\cite{GlueX:2019adl,GlueX:2023fcq,GlueX:2025kma} respectively (the latter also with CLAS12 data in Refs.~\cite{CLAS:2020rdz,CLAS:2020ngl}).
One such comparison of the predicted polarized SDMEs for $\rho^0$ photoproduction is shown in Fig.~\ref{fig:rho_sdmes}. Reactions involving $\pi$ exchange, such as pion photoproduction off a nucleon~\cite{JointPhysicsAnalysisCenter:2024kck} and $\Delta$ baryon~\cite{JointPhysicsAnalysisCenter:2017del} have also garnered significant attention because this is also expected to play a crucial role in searches for light hybrid mesons~\cite{Szczepaniak:2001qz}.  The properties of the amplitude in the $\ell$-plane are critical in $\pi  N$ photoproduction because analytic continuation in $\ell$ is necessary to clearly identify the pion pole and its role in maintaining gauge invariance at high energies~\cite{JointPhysicsAnalysisCenter:2024kck}. 
 The same Regge exchanges are also expected to be the dominant production mechanisms in certain kinematic regions for the photoproduction of heavy exotic mesons, such as the so-called \textit{XYZ}'s~\cite{Brambilla:2019esw,Esposito:2016noz,Guo:2017jvc}, and thus Regge formalism has been applied to order-of-magnitude estimations of their production cross sections for feasibility studies at future facilities~\cite{Albaladejo:2020tzt}.

Although we have mainly focused on exchange diagrams like in Fig.~\ref{fig:fast_slow}, Regge physics can also be applied to more complex reaction topologies such as in central, multi-meson production, i.e., in so-called \textit{double Regge} processes~\cite{Shimada:1978sx} as shown in Fig.~\ref{fig:double_regge}.
Such reactions are particularly interesting for investigating the dynamics of gluons at LHC energies~\cite{Albrow:2010yb}, and more recently in the production of $\eta^{(\prime)}\pi$ in Refs.~\cite{COMPASS:2014vkj,Bibrzycki:2021rwh}. The latter process at high energies can be used to constrain the potential existence of spin-exotic hybrid resonances at low energies through FESRs, as will be discussed in the following subsection.
Even more complex topologies, such as \textit{triple Regge} processes, emerge in inclusive reactions~\cite{Field:1974fg,Ganguli:1980mp} and have also seen recent applications~\cite{Kim:2022zgz,JointPhysicsAnalysisCenter:2024pat,Winney:2022tky} in the study of soft, semi-inclusive production.

\subsection{Analyticity constraints for low-energy amplitudes} \label{sec:constraints}

\begin{wrapfigure}[20]{r}{0.45\textwidth}
        \vspace{-1cm}
        \includegraphics[width=\linewidth]{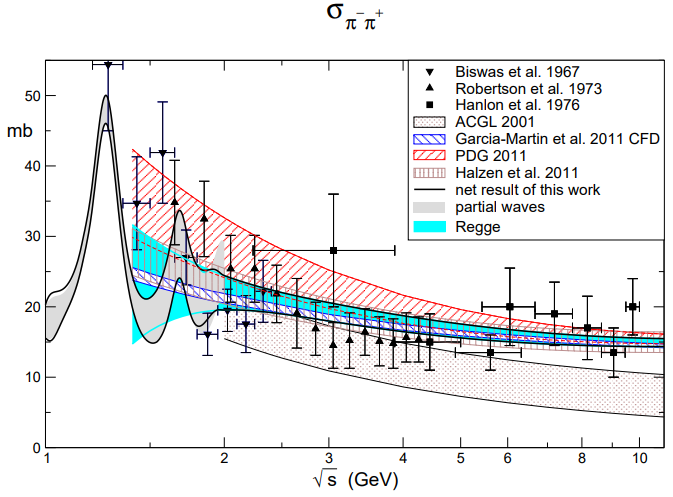}
        \caption{Comparison of parameterizations for $\pi\pi$ total cross section from Ref.~\cite{Caprini:2011ky}. Regge asymptotics are seen to become dominated around $\sqrt{s} \sim 2$ GeV.} 
        \label{fig:pipi}
\end{wrapfigure}

In Sec.~\ref{sec:exchange_pheno}, we discussed how Regge theory offers a clear and efficient framework for parameterizing scattering processes at high energies and forward angles. As explained in Secs.~\ref{sec:regge_poles} and \ref{sec:duality}, the Regge behavior of amplitudes is closely related to analyticity in both energy variables, which allows amplitudes in resonance regions to be constrained through dispersion relations.

The FESRs already introduced in Eq.~\eqref{eq:FESR}, for instance, directly connect the discontinuity amplitude along the unitarity cuts with the leading Regge exchanges.
Although the original purpose of these sum rules as self-contained, dynamic equations ultimately did not succeed, they can still be used to impose the important constraint of analyticity for resonance determination. This is done by requiring that a model for $\Im F(s,t)$ on the left side of Eq.~\eqref{eq:FESR} matches the leading Regge poles known from high-energy experiments. The FESRs have been applied to elastic $\pi N$ scattering in this way in Refs.~\cite{Huang:2008nr,Mathieu:2015gxa}, where the $t$-channel Regge poles are compared and matched with higher-mass baryon resonances in the $\sqrt{s} \gtrsim 2$ GeV region.  A similar methodology is also used in Ref.~\cite{Mathieu:2017jjs,Mathieu:2018mjw} for pion photoproduction, which shows that FESRs can help constrain angular dependence and distinguish between the spin assignments of resonances.  Finally, these constraints are applied in the opposite direction in Ref.~\cite{JPAC:2016lnm}, providing knowledge of the low-energy partial waves to predict the behavior at high energies for $\eta$ and $\eta^\prime$ photoproduction. Since these two reactions are expected to proceed through the same exchanges, the low-energy matching helps constrain the high-energy residues of both, which are otherwise unknown from Regge theory alone.

The FESRs are not the only type of analyticity constraint that requires Regge inputs to estimate amplitudes asymptotically. For example, the FESRs were derived by formulating a general one-variable dispersion relation with a finite contour, placing them within a broader class of dispersive representations for the amplitude at fixed-$t$. A particularly important representation was introduced by Roy~\cite{Roy:1971tc} to calculate the $\pi\pi$ scattering amplitude based solely on knowledge of the amplitude above threshold. These are most commonly expressed as linear integral equations for the partial wave amplitudes and are known as the system of \textit{Roy equations}\footnote{The \textit{GKPY equations} of Ref.~\cite{Garcia-Martin:2011iqs} are closely related and also follow the form in Eq.~\eqref{eq:roy}, with differences primarily in the number of subtractions considered in $k_\ell^I(s)$.} (using the notation of Ref.~\cite{Ananthanarayan:2000ht}):
    \begin{align}
        \label{eq:roy}
        \Re t_\ell^I(s) &= k_\ell^I(s) + \sum_{I^\prime} \sum_{\ell^\prime=0}^{\infty} \textrm{ PV}\!\int_{s_\text{th}}^{\infty} ds^\prime \, K^{II^\prime}_{\ell\ell^\prime}(s,s^\prime) \, \Im t_{\ell^\prime}^{I^\prime}(s^\prime) 
        = k_\ell^I(s) + \sum_{I^\prime} \sum_{\ell^\prime=0}^{\ell_\text{max}} \textrm{ PV}\!\int_{s_\text{th}}^{s_\text{max}} ds^\prime \, K^{II^\prime}_{\ell\ell^\prime}(s,s^\prime) \, \Im t_{\ell^\prime}^{I^\prime}(s^\prime) + d_\ell^I(s) ~,
    \end{align}
where $t^I_\ell(s)$ denotes the $\pi\pi$ partial wave with spin-$\ell$ and isospin-$I$.  In the first equality, for a given number of subtractions contained in polynomial $k_\ell^I(s)$, the partial wave is accurately reconstructed by dispersing over all partial waves (summed over both $\ell$ and $I$) each with a known kernel function $K_{\ell\ell^\prime}^{II^\prime}(s,s^\prime)$.  In practice, considering the entire sum is impractical and is truncated to a finite set of partial waves, as shown in the second equality.  The remaining piece, $d_\ell^I(s)$, also called the \textit{driving term}, absorbs contributions from higher $\ell^\prime > \ell_\text{max}$ and at energies $s > s_\text{max}$. It is usually estimated, at least partly, with Regge parameterizations, such as fits to the $\pi\pi$ total cross section shown in Fig.~\ref{fig:pipi} in Refs.~\cite{Pelaez:2003ky,Caprini:2011ky}. With a suitably chosen driving term, Eq.~\eqref{eq:roy} can be solved~\cite{Ananthanarayan:2000ht} or used as a fitting 
constraint~\cite{Garcia-Martin:2011iqs,Pelaez:2019eqa,Pelaez:2024uav}, and has enabled the determination of low-energy $\pi\pi$ amplitudes with unprecedented precision. More recently, these formalisms have also allowed investigations into the evolution of low-lying $\pi\pi$ resonances at unphysical quark masses~\cite{Cao:2023ntr,Rodas:2023nec} while still satisfying unitarity, crossing, and analyticity constraints.

Analogous equations for non-symmetric systems, based on the works of Steiner~\cite{Steiner:1971ms} and Hite~\cite{Hite:1973pm}, have also been developed and applied to $\pi N$~\cite{Ditsche:2012fv,Becher:2001hv,Hoferichter:2015hva}
and $\pi K$~\cite{Buettiker:2003pp,Pelaez:2020gnd} scattering (the latter also generalized to unphysical quark masses in Refs.~\cite{Cao:2024zuy,Cao:2025hqm}). Likewise, the contributions from the high-energy tails of dispersion integrals are generally estimated using Regge pole analysis. Similar to the $\pi\pi$ case, these serve as powerful analyticity constraints, which have enabled precise determination of $\pi N$ scattering amplitudes and low-lying strange resonance parameters.

\subsection{Hadron spectroscopy in the $\ell$-plane} \label{sec:hadron_spec_ell}
Regge theory is not limited to high energy regions and can also be used to directly parameterize resonances. As shown in Secs.~\ref{sec:relativistic} and~\ref{sec:regge_poles}, Regge trajectories must connect the complex pole positions of resonances that lie on them, providing a way to study resonances through their properties in the complex $\ell$-plane. 
The construction of non-linear Regge trajectories, incorporating unitarity and analyticity constraints, has thus re-emerged in recent decades for studying hadron spectroscopy.

The effort to develop a realistic model for complex $\alpha(s)$ that can meet all the requirements discussed in Ref.~\ref{sec:regge_poles} while also fitting resonance spectra was revived by Ref.~\cite{Fiore:2000fp}. 
In particular, certain DAMA models (see Sec.~\ref{sec:duality}) impose a stricter constraint on the asymptotic growth of $\alpha(s)$ than Eq.~\eqref{eq:alpha_bounds} and require $\Re\alpha(s)$ to level off to a constant.
If this is true, then the infinite tower of higher spin resonances actually ends above a certain energy scale. This phenomenon is still consistent with the apparently linearly rising trajectories below the termination energy (and all other properties discussed in Sec.~\ref{sec:regge_poles}) if  
\begin{equation}
        \Im\alpha(s) \approx \sum_i g_i \, \sqrt{s - s_i} \, \left(\frac{s-s_i}{s}\right)^{\Re\alpha(s_i)} \, \theta(s-s_i) ~,
    \end{equation}
for a series of inelastic thresholds $s_i$ and couplings $g_i$.

The exponential power involves $\Re\alpha(s)$ to ensure the correct behavior required by analyticity as $s \to s_i$. The energy range in which the trajectory is approximately linear is thus created by the opening of more and more inelastic thresholds. Above a saturation scale, i.e., the last relevant threshold, the $\Im\alpha(s)$ continues to grow as required by unitarity, but $\Re\alpha(s)$ quickly asymptotes. This saturation phenomenon is argued to be produced by the deconfinement of the quarks at high temperatures~\cite{Szanyi:2023ano} or quark pair-creation in so-called ``string breaking" processes~\cite{Brisudova:1998wq,Kholodkov:1991hx,Bali:2005fu}, but has not been proven to be required of physical hadron trajectories.

The complex angular momentum plane also helps identify resonances with exotic features through unconventional Regge behavior. Trajectories of mesons like the $\rho$ are observed to follow the typical linear pattern shown in Fig.~\ref{fig:chew_frautschi}, reflecting their string-like structure as a $q\bar{q}$ state. If a trajectory shows significantly different behavior in the 
$\ell$-plane, it may suggest a non-standard composition, such as a multi-quark state or a hadronic molecule.

\begin{wrapfigure}[22]{r}{0.45\textwidth}
        \includegraphics[width=\linewidth]{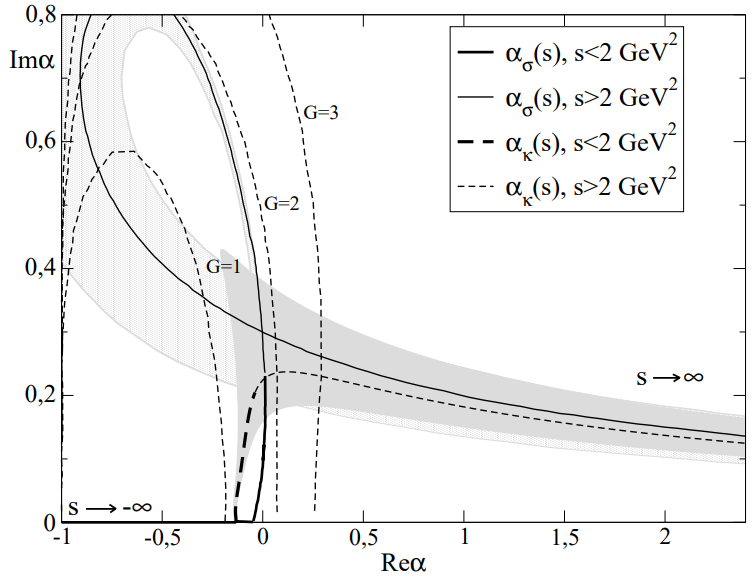}
        \caption{Behavior of the Regge trajectories for the $\sigma$ and $\kappa$ mesons. At low energies, both of these resemble those of Yukawa potentials instead of typical linear trajectories.} 
        \label{fig:alpha_yukawa}
\end{wrapfigure}

A fairly effective way to extract Regge trajectories for this comparison is to parameterize partial waves as $\ell$-poles constrained by unitarity. If a partial wave is dominated by a single resonance, for example in the case of low-lying resonances near threshold, then one can write~\cite{Cheng:1963}.
    \begin{equation}
        \label{eq:CRP}
        f_\ell(s) \approx \frac{\beta(s)}{\ell - \alpha(s)} 
        \quad \text{ with } \quad 
        \Im \alpha(s) = \rho(s) \, \beta(s) \, \theta(s-s_\text{th}) ~.
    \end{equation}
where $\Im\alpha(s)$ is determined by requiring elastic unitarity of the form $\Im f_\ell(s) = \rho(s) \, |f_\ell(s)|^2$. Fixing the imaginary part, in turn, determines $\alpha(s)$ through a dispersion relation as shown in Eq.~\eqref{eq:alpha_dispersion}. For a suitable model of $\beta(s)$, one can then derive a system of coupled equations for $f_\ell(s)$ and therefore $\alpha(s)$, with any free parameters fixed by requiring the pole to reproduce the resonance parameters. The \textit{constrained Regge pole} model of Eq.~\eqref{eq:CRP} has been employed to calculate the trajectories of the $\rho$ and $\sigma$ mesons~\cite{Londergan:2013dza}, $f_2^{(\prime)}$ and $K^*$~\cite{Carrasco:2015fva}, and $\kappa$ mesons~\cite{Pelaez:2017sit} to show that most of these mesons follow a similar pattern of roughly linear and mainly real Regge trajectories. The trajectories of the $\sigma$ and $\kappa$ are very shallow, with a rapidly increasing imaginary part, resembling those of a Yukawa potential as shown in Fig.~\ref{fig:alpha_yukawa}, at least near the pole. The different behaviors in the $\ell$-plane have been argued to suggest that both of these mesons are more extended objects, likely originating from meson-meson interactions rather than interactions between a $q\bar{q}$ pair~\cite{Pelaez:2017sit}.

    \begin{figure}[b]
        \centering
        \includegraphics[width=\linewidth]{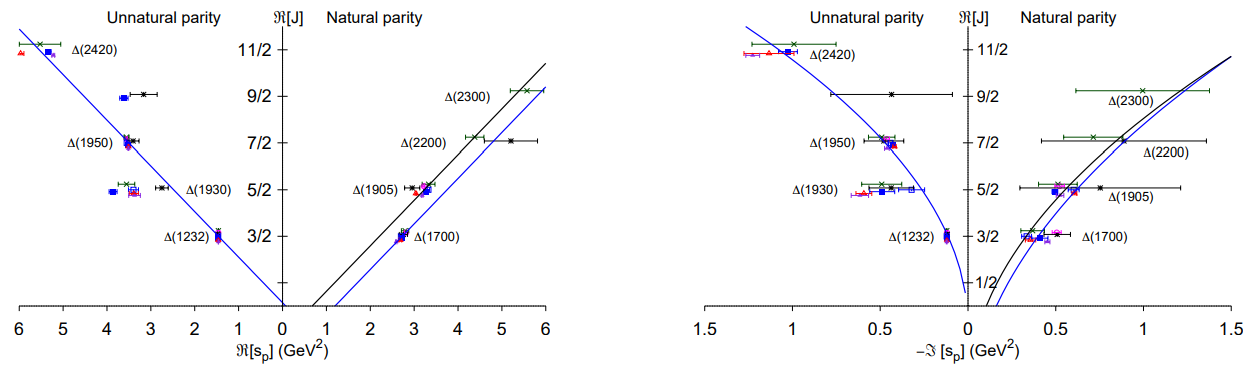}
        \caption{Real part of the $\Delta$ Regge trajectories for both parities evaluated in the complex plane as a function of $\Re s$ (left) and $\Im s$ (right) from Ref.~\cite{JPAC:2018zjz} The data points are correspond to the pole parameters extracted by various partial wave analyses. }
        \label{fig:deltas}
    \end{figure}

A similar methodology has also been used in the baryon sector for the Regge trajectories of $\Lambda$ and $\Sigma$ baryons in Ref.~\cite{Fernandez-Ramirez:2015fbq} and for $N^*$ and $\Delta$ baryons in Refs.~\cite{Fiore:2004xb,JPAC:2018zjz}. 
 Considering fermion spectra involves analyzing trajectories in parity doublets, as baryons obey additional symmetries that relate states differing only by parity, known as MacDowell symmetry~\cite{MacDowell:1959zza}. Just like in the meson sector, there is a noticeable regularity in most baryon resonances that follow similar trajectories, as shown in Fig~\ref{fig:deltas}, indicating that most originate from the common dynamics of typical $qqq$ baryons. In contrast, the parity doublet of $N(1680)$ and $N(1720)$ is believed to probably not align along the same trajectories in~\cite{JPAC:2018zjz}. The parameters of these states were already noted to be overestimated by quark model predictions in~\cite{Loring:2001kx,Capstick:1986ter}, indicating contributions from non-trivial components beyond $qqq$. The Regge analysis of hyperons similarly identified at least one of the two resonances which form the $\Lambda(1405)$\footnote{For a survey of the phenomenology and structure of the $\Lambda(1405)$, see Refs.~\cite{Meissner:2020khl,Mai:2020ltx} and references therein.} to not lie on any regular trajectories and thus suggests a different dynamical origin as conventional baryons. 
 
 Finally, the construction of more complex amplitudes that satisfy the $\ell$-analyticity by design has also been explored in Ref.~\cite{Stamen:2024gfz} to establish more stringent constraints on the Regge trajectories than the approximation in Eq.~\eqref{eq:CRP}. Specifically, an amplitude formed from crossing symmetric combinations of terms such as:

    \begin{equation}
        \label{eq:hypergeo}
        F(\alpha(t), \nu) = \Gamma(\ell_\text{min} - \alpha(t)) \, \nu^{\ell_\text{min}} \, \phantom{}_2\tilde{F}_1\left[\ell_\text{min}+1, \ell_\text{min}-\alpha(t); \ell_\text{min}+1-\alpha(t), \nu\right] ~,
    \end{equation}
with respect to the dimensionless, variable $\nu$ appearing in Eq.~\eqref{eq:FESR} and the regularized hypergeometric function $\phantom{}_2\tilde{F}_1$, will result in an amplitude which is simultaneously analytic in $\ell$ as well as all Mandelstam variables. A trajectory defined by Eq.~\eqref{eq:alpha_dispersion} with
    \begin{equation}
        \label{eq:imalpha_hypergeo}
        \Im \alpha(s) = \frac{\gamma}{\pi} \log\left[1 + \frac{\pi}{\gamma} \, \rho(s) \,\beta(s) \right] \, \theta(s-s_\text{th}) ~,
     \end{equation}
for constant $\gamma >1$ and an arbitrary real function $\beta(s)$, bounded above by $s^{\Re\alpha(s)}$, is also shown to satisfy all properties identified in Sec.~\ref{sec:regge_poles}.

    \begin{figure}[t]
        \centering
        \includegraphics[width=0.4\linewidth]{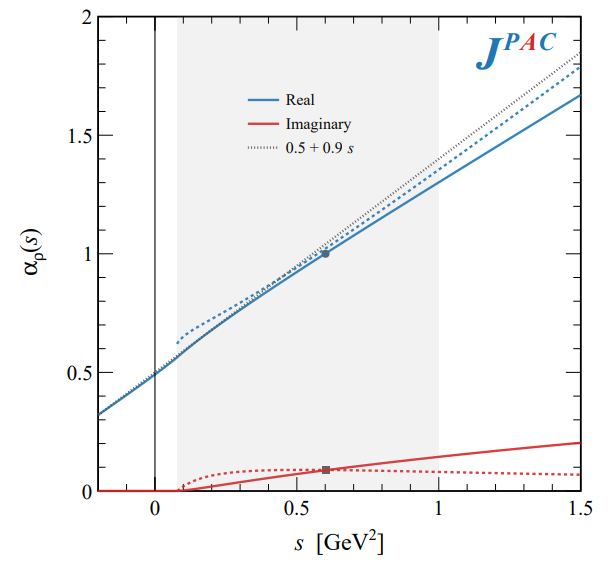}
        \hspace{1cm}
        \includegraphics[width=0.4\linewidth]{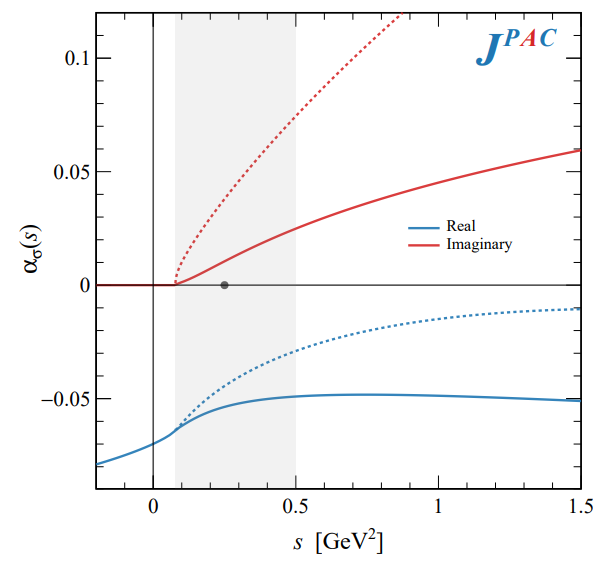}
        
        \caption{Trajectories of the $\rho$ (right) and $\sigma$ (left) mesons as in Ref.~\cite{Stamen:2024gfz}. The solid lines are calculated using Eq.~\eqref{eq:imalpha_hypergeo} while is the CRP model Eq.~\eqref{eq:CRP} calculated in Ref.~\cite{Londergan:2013dza}.}
        \label{fig:rho_sigma}
    \end{figure}

The amplitude constructed from Eq.~\eqref{eq:hypergeo} and this trajectory therefore demonstrates well-defined Regge and fixed-angle limits, as well as complex resonance poles for all $\ell \geq \ell_\text{min}$ without ancestors, fully utilizing the all-energies constraint of analyticity\footnote{Interestingly, the model constructed in Ref.~\cite{Stamen:2024gfz} satisfies all FESRs but does not feature duality as discussed in Sec.~\ref{sec:duality}.}. The construction is applied to $\pi\pi$ scattering to extract the $\rho$ and $\sigma$ trajectories shown in Fig.~\ref{fig:rho_sigma}. Just as in Ref.~\cite{Londergan:2013dza}, the $\rho$ follows a rising trajectory which is approximately linear at low energies. The $\sigma$ trajectory, however, does not rise indefinitely, as the real part logarithmically turns over toward negative infinity at high energies. The turnover of the trajectory indicates that, unlike typical mesons, the $\sigma$ does not have an associated tower of Regge recurrences linked to orbital excitations, further suggesting an exotic composition.

\section{Conclusions} \label{sec:conclusions}

The theory of quarks and gluons is now established as the foundation of the strong interaction. However, many key open questions related to the generation of resonances and the nature of soft interactions at high energies remain elusive to perturbative solutions. Without a rigorous non-perturbative solution to QCD, hadron phenomenology depends on the use of multiple frameworks to deepen our understanding of hadron dynamics. As discussed in this article, Regge physics is one such tool in the phenomenologist's toolkit, arising from fundamental scattering theory and providing powerful constraints on scattering amplitudes that are closely linked to analyticity, crossing symmetry, and unitarity through angular momentum in the complex plane.

The last few decades have experienced a renaissance for Regge theory since its peak during the bootstrap program and its subsequent decline at the start of QCD. With new, high-statistics production data from current and next-generation experiments, Regge poles continue to be valuable tools for describing the dynamics of peripheral scattering. Regge concepts are also increasingly used both directly and indirectly to constrain, extract, and analyze the properties of resonances. Particularly, the potential existence of exotic bound states is an exciting and rapidly expanding field that has also incorporated Regge theory to study their production at high energies and their properties in complex angular momentum.

\begin{ack}[Acknowledgments]%
We would like to thank all of our collaborators in the Joint Physics Analysis Center (JPAC). This material is based upon work supported by the U.S. Department of Energy, Office of Science, Office of Nuclear Physics under contracts DE-AC05-06OR23177,  DE-FG02-87ER40365. 
\end{ack}


\bibliographystyle{elsarticle-num} 
\bibliography{references}

@article{Regge:1959mz,
    author = "Regge, T.",
    title = "{Introduction to complex orbital momenta}",
    doi = "10.1007/BF02728177",
    journal = "Nuovo Cim.",
    volume = "14",
    pages = "951",
    year = "1959"
}

@article{Regge:1960zc,
    author = "Regge, T.",
    title = "{Bound states, shadow states and Mandelstam representation}",
    doi = "10.1007/BF02733035",
    journal = "Nuovo Cim.",
    volume = "18",
    pages = "947--956",
    year = "1960"
}

@book{Newton:1982qc,
    author = "Newton, R. G.",
    title = "{Scattering theory of waves and particles}",
    publisher = "Springer-Verlag",
    address = "New York, NY",
    year = "1982"
}

@book{Sakurai:2011zz,
    author = "Sakurai, Jun John and Napolitano, Jim",
    title = "{Modern Quantum Mechanics}",
    doi = "10.1017/9781108587280",
    isbn = "978-0-8053-8291-4, 978-1-108-52742-2, 978-1-108-58728-0",
    publisher = "Cambridge University Press",
    series = "Quantum physics, quantum information and quantum computation",
    month = "10",
    year = "2020"
}

@article{Lehmann:1958ita,
    author = "Lehmann, H.",
    title = "{Analytic properties of scattering amplitudes as functions of momentum transfer}",
    doi = "10.1007/bf02859794",
    journal = "Nuovo Cim.",
    volume = "10",
    number = "4",
    pages = "579--589",
    year = "1958"
}

@article{Martin:1965jj,
    author = "Martin, Andre",
    title = "{Extension of the axiomatic analyticity domain of scattering amplitudes by unitarity. 1.}",
    doi = "10.1007/BF02720568",
    journal = "Nuovo Cim. A",
    volume = "42",
    pages = "930--953",
    year = "1965"
}

@book{Nussenzveig:1972tcd,
    author = "Nussenzveig, H. M.",
    title = "{Causality and dispersion relations}",
    isbn = "978-0-12-523050-6, 978-0-08-095604-6",
    publisher = "Academic Press",
    address = "New York, London",
    year = "1972"
}

@article{Watson:1918,
    author = "Watson, G. N.",
    title = "{The diffraction of electric waves by the earth}",
    doi = "10.1098/rspa.1918.0050",
    journal = "Proc. R. Soc. Lond. A",
    volume = "95",
    pages = "83–99",
    year = "1918"
}

@book{Sommerfeld:1949,
    author = "Sommerfeld, A.",
    title = "{Partial Differential Equations in Physics}",
    isbn = "978-0-12-654658-3",
    publisher = "Academic Press",
    address = "New York, London",
    year = "1949"
}

@book{Gribov:2003nw,
    author = "Gribov, V. N.",
    title = "{The theory of complex angular momenta: Gribov lectures on theoretical physics}",
    doi = "10.1017/CBO9780511534959",
    isbn = "978-0-521-03703-7, 978-0-521-81834-6, 978-0-511-05504-1",
    publisher = "Cambridge University Press",
    series = "Cambridge Monographs on Mathematical Physics",
    month = "6",
    year = "2007"
}

@book{Collins:1977jy,
    author = "Collins, P. D. B.",
    title = "{An Introduction to Regge Theory and High Energy Physics}",
    doi = "10.1017/9781009403269",
    isbn = "978-1-009-40326-9, 978-1-009-40329-0, 978-1-009-40328-3, 978-0-521-11035-8",
    publisher = "Cambridge University Press",
    year = "1977"
}

@book{Gribov:2009zz,
    author = "Gribov, Vladimir N.",
    title = "{Strong interactions of hadrons at high emnergies: Gribov lectures on 
Theoretical Physics}",
    isbn = "978-1-107-41130-2, 978-0-521-85609-6, 978-0-511-45145-4",
    publisher = "Cambridge University Press",
    month = "10",
    year = "2012"
}

@book{Donnachie:2002en,
    author = "Donnachie, S. and Dosch, Hans Gunter and Nachtmann, O. and Landshoff, P.",
    title = "{Pomeron physics and QCD}",
    isbn = "978-0-511-06050-2, 978-0-521-78039-1, 978-0-521-67570-3",
    publisher = "Cambridge University Press",
    volume = "19",
    month = "12",
    year = "2004"
}

@article{Barut:1962fpy,
    author = "Barut, A. O. and Calogero, F.",
    title = "{Singularities in Angular Momentum of the Scattering Amplitude for a Class of Soluble Potentials}",
    doi = "10.1103/PhysRev.128.1383",
    journal = "Phys. Rev.",
    volume = "128",
    number = "3",
    pages = "1383",
    year = "1962"
}

@article{Aly:1967snr,
    author = {Aly, H. H. and M\"uller, H. J. W. and Schilcher, K.},
    title = "{Regge trajectories for even-power potentials}",
    doi = "10.1016/0550-3213(67)90010-7",
    journal = "Nucl. Phys. B",
    volume = "3",
    pages = "401--406",
    year = "1967"
}

@article{Kronenfeld:1971rh,
    author = "Kronenfeld, R.",
    title = "{Regge poles and bound states of nonrelativistic scattering systems}",
    doi = "10.1119/1.1986369",
    journal = "Am. J. Phys.",
    volume = "39",
    pages = "1056--1068",
    year = "1971"
}

@article{Ahmadzadeh:1963ith,
    author = "Ahmadzadeh, Akbar and Burke, Philip G. and Tate, Cecil",
    title = "{Regge Trajectories for Yukawa Potentials}",
    doi = "10.1103/PhysRev.131.1315",
    journal = "Phys. Rev.",
    volume = "131",
    number = "3",
    pages = "1315",
    year = "1963"
}

@article{Singh:1962qgg,
    author = "Singh, Virendra",
    title = "{Analyticity in the Complex Angular Momentum Plane of the Coulomb Scattering Amplitude}",
    doi = "10.1103/PhysRev.127.632",
    journal = "Phys. Rev.",
    volume = "127",
    number = "2",
    pages = "632",
    year = "1962"
}

@article{Gell-Mann:1954wra,
    author = "Gell-Mann, Murray and Goldberger, M. L.",
    title = "{Scattering of low-energy photons by particles of spin 1/2}",
    doi = "10.1103/PhysRev.96.1433",
    journal = "Phys. Rev.",
    volume = "96",
    pages = "1433--1438",
    year = "1954"
}

@article{Castillejo:1955ed,
    author = "Castillejo, L. and Dalitz, R. H. and Dyson, F. J.",
    title = "{Low's scattering equation for the charged and neutral scalar theories}",
    doi = "10.1103/PhysRev.101.453",
    journal = "Phys. Rev.",
    volume = "101",
    pages = "453--458",
    year = "1956"
}

@article{Low:1955,
    author = "Low, F. E.",
    title = "{Boson-Fermion Scattering in the Heisenberg Representation}",
    doi = "10.1103/PhysRev.97.1392",
    journal = "Phys. Rev.",
    volume = "97",
    pages = "1392",
    year = "1955"
}

@article{Dyson:1957rgq,
    author = "Dyson, F. J.",
    title = "{Meaning of the solutions of Low's scattering equation}",
    doi = "10.1103/physrev.106.157",
    journal = "Phys. Rev.",
    volume = "106",
    number = "1",
    pages = "157--159",
    year = "1957"
}

@article{Atkinson:1969vec,
    author = "Atkinson, D. and Warnock, R. L.",
    title = "{Persistence of the castillejo-dalitz-dyson ambiguity in relativistic crossing-symmetric amplitudes}",
    doi = "10.1103/PhysRev.188.2098",
    journal = "Phys. Rev.",
    volume = "188",
    pages = "2098--2111",
    year = "1969"
}

@article{Abbe:1967,
    author = "Abbe, W. J.",
    title = "{The creation of bound states by a Yukawa potential}",
    doi = "10.1007/BF02827746",
    journal = "Il Nuovo Cimento A",
    volume = "50",
    pages = "364--367",
    year = "1967"
}

@article{Burke:1969re,
    author = "Burke, P. G. and Tate, C.",
    title = "{A program for calculating regge trajectories in potential scattering}",
    doi = "10.1016/0010-4655(69)90003-4",
    journal = "Comput. Phys. Commun.",
    volume = "1",
    pages = "97--105",
    year = "1969"
}

@article{Yukawa:1935xg,
    author = "Yukawa, Hideki",
    title = "{On the Interaction of Elementary Particles I}",
    reportNumber = "PRINT-92-0144",
    doi = "10.1143/PTPS.1.1",
    journal = "Proc. Phys. Math. Soc. Jap.",
    volume = "17",
    pages = "48--57",
    year = "1935"
}

@article{Norton:1958vut,
    author = "Norton, Richard and Klein, Abraham",
    title = "{Significance of the redundant solutions of the Low-Wick equation}",
    doi = "10.1103/physrev.109.991",
    journal = "Phys. Rev.",
    volume = "109",
    number = "3",
    pages = "991--995",
    year = "1958"
}

@article{Chew:1955zz,
    author = "Chew, G. F. and Low, F. E.",
    title = "{Effective range approach to the low-energy p wave pion - nucleon interaction}",
    doi = "10.1103/PhysRev.101.1570",
    journal = "Phys. Rev.",
    volume = "101",
    pages = "1570--1579",
    year = "1956"
}

@article{Mandelstam:1958xc,
    author = "Mandelstam, S.",
    title = "{Determination of the pion - nucleon scattering amplitude from dispersion relations and unitarity. General theory}",
    doi = "10.1103/PhysRev.112.1344",
    journal = "Phys. Rev.",
    volume = "112",
    pages = "1344--1360",
    year = "1958"
}

@article{Heisenberg:1943,
    author = "Heisenberg, W.",
    title = "{Die ,,beobachtbaren Gr\"oßen in der Theorie der Elementarteilchen}",
    doi = "10.1007/BF01329800",
    journal = "Z. Physik",
    volume = "120",
    pages = "513--538",
    year = "1943",
    note = "(in German)"
}

@article{Lehmann:1954rq,
    author = "Lehmann, H. and Symanzik, K. and Zimmermann, W.",
    title = "{On the formulation of quantized field theories}",
    doi = "10.1007/BF02731765",
    journal = "Nuovo Cim.",
    volume = "1",
    pages = "205--225",
    year = "1955"
}

@article{Dyson:1949ha,
    author = "Dyson, F. J.",
    title = "{The S matrix in quantum electrodynamics}",
    doi = "10.1103/PhysRev.75.1736",
    journal = "Phys. Rev.",
    volume = "75",
    pages = "1736--1755",
    year = "1949"
}

@article{Fukutome:1957,
    author = "Fukutome, H.",
    title = "{Low's Scattering Equation and S-matrix}",
    doi = "10.1143/PTP.17.383",
    journal = "Prog. of Theo. Phy.",
    volume = "17",
    number = "3",
    pages = "383--400",
    year = "1957"
}

@article{Salzman:1957opc,
    author = "Salzman, George and Salzman, Freda",
    title = "{Solutions of the Static Theory Integral Equations for Pion-Nucleon Scattering in the One-Meson Approximation}",
    doi = "10.1103/PhysRev.108.1619",
    journal = "Phys. Rev.",
    volume = "108",
    number = "6",
    pages = "1619",
    year = "1957"
}

@article{Chew:1956zz,
    author = "Chew, G. F. and Low, F. E.",
    title = "{Theory of Photomeson Production at Low Energies}",
    doi = "10.1103/PhysRev.101.1579",
    journal = "Phys. Rev.",
    volume = "101",
    pages = "1579--1587",
    year = "1956"
}

@article{Gribov:1961fr,
    author = "Gribov, V. N.",
    title = "{Partial waves with complex orbital angular momenta and the asymptotic behavior of the scattering amplitude}",
    journal = "Zh. Eksp. Teor. Fiz.",
    volume = "41",
    pages = "1962",
    year = "1961"
}

@article{Chew:1962mpd,
    author = "Chew, Geoffrey F.",
    title = "{S-Matrix Theory of Strong Interactions without Elementary Particles}",
    doi = "10.1103/RevModPhys.34.394",
    journal = "Rev. Mod. Phys.",
    volume = "34",
    number = "3",
    pages = "394--401",
    year = "1962"
}

@article{Froissart:1961ux,
    author = "Froissart, Marcel",
    title = "{Asymptotic behavior and subtractions in the Mandelstam representation}",
    doi = "10.1103/PhysRev.123.1053",
    journal = "Phys. Rev.",
    volume = "123",
    pages = "1053--1057",
    year = "1961"
}

@article{Hara:1964zza,
    author = "Hara, Yasuo",
    title = "{Analyticity Properties of Helicity Amplitudes and Construction of Kinematical Singularity-Free Amplitudes for Any Spin}",
    doi = "10.1103/PhysRev.136.B507",
    journal = "Phys. Rev.",
    volume = "136",
    pages = "B507--B514",
    year = "1964"
}

@article{Cohen-Tannoudji:1968lnm,
    author = "Cohen-Tannoudji, G. and Morel, A. and Navelet, H.",
    title = "{Kinematical singularities, crossing matrix and kinematical constraints for two-body helicity amplitudes}",
    doi = "10.1016/0003-4916(68)90243-1",
    journal = "Annals Phys.",
    volume = "46",
    number = "2",
    pages = "239--316",
    year = "1968"
}

@article{Kibble:1960zz,
    author = "Kibble, T. W. B.",
    title = "{Kinematics of General Scattering Processes and the Mandelstam Representation}",
    doi = "10.1103/PhysRev.117.1159",
    journal = "Phys. Rev.",
    volume = "117",
    pages = "1159--1162",
    year = "1960"
}

@article{Rosenfeld:1965yz,
    author = "Rosenfeld, Arthur H. and Barbaro-Galtieri, Angela and Barkas, Walter H. and Bastien, Pierre L. and Kirz, Janos and Roos, Matts",
    title = "{Data on particles and resonant states}",
    doi = "10.1103/RevModPhys.37.633",
    journal = "Rev. Mod. Phys.",
    volume = "37",
    pages = "633--651",
    year = "1965"
}

@article{Martin:1962rt,
    author = "Martin, A.",
    title = "{Unitarity and high-energy behavior of scattering amplitudes}",
    doi = "10.1103/PhysRev.129.1432",
    journal = "Phys. Rev.",
    volume = "129",
    pages = "1432--1436",
    year = "1963"
}

@article{Chew:1961ev,
    author = "Chew, G. F. and Frautschi, Steven C.",
    title = "{Principle of Equivalence for All Strongly Interacting Particles Within the S Matrix Framework}",
    doi = "10.1103/PhysRevLett.7.394",
    journal = "Phys. Rev. Lett.",
    volume = "7",
    pages = "394--397",
    year = "1961"
}

@article{Chew:1962eu,
    author = "Chew, G. F. and Frautschi, Steven C.",
    title = "{Regge Trajectories and the Principle of Maximum Strength for Strong Interactions}",
    doi = "10.1103/PhysRevLett.8.41",
    journal = "Phys. Rev. Lett.",
    volume = "8",
    pages = "41--44",
    year = "1962"
}

@article{Barut:1962zz,
    author = "Barut, A. O. and Zwanziger, D. E.",
    title = "{Complex Angular Momentum in Relativistic S-Matrix Theory}",
    doi = "10.1103/PhysRev.127.974",
    journal = "Phys. Rev.",
    volume = "127",
    pages = "974--977",
    year = "1962"
}

@article{Khuri:1963zza,
    author = "Khuri, N. N.",
    title = "{Regge Poles, Power Series, and a Crossing-Symmetric Watson-Sommerfeld Transformation}",
    doi = "10.1103/PhysRev.132.914",
    journal = "Phys. Rev.",
    volume = "132",
    pages = "914--926",
    year = "1963"
}

@article{Khuri:1963,
    author = "Khuri, N. N.",
    title = "{Crossing-Symmetric Watson-Sommerfeld Transformation}",
    doi = "10.1103/PhysRevLett.10.420",
    journal = "Phys. Rev. Lett.",
    volume = "10",
    number = "9",
    pages = "420--423",
    year = "1963"
}

@unpublished{Froissart:1961,
    author = "Froissart, M.",
    title = "{Report at La Jolla conference on weak and strong interactions}",
    year = "1961",
    note = "(unpublished)"
}

@article{Mandelstam:1959,
    author = "Mandelstam, M.",
    title = "{An extention of the Regge formula}",
    doi = "10.1016/0003-4916(62)90218-X",
    journal = "Ann. Phys.",
    volume = "19",
    number = "2",
    pages = "254--261",
    year = "1959"
}

@article{Arbab:1966nmh,
    author = "Arbab, Farzam and Chiu, Charles B.",
    title = "{Association between the Dip in the \ensuremath{\pi}\ensuremath{-}p\textrightarrow{}\ensuremath{\pi}0n High-Energy Angular Distribution and the Zero of the \ensuremath{\rho} Trajectory}",
    doi = "10.1103/PhysRev.147.1045",
    journal = "Phys. Rev.",
    volume = "147",
    number = "4",
    pages = "1045",
    year = "1966"
}

@article{ParticleDataGroup:2020ssz,
    author = "Zyla, P. A. and others",
    collaboration = "Particle Data Group",
    title = "{Review of Particle Physics}",
    doi = "10.1093/ptep/ptaa104",
    journal = "PTEP",
    volume = "2020",
    number = "8",
    pages = "083C01",
    year = "2020"
}

@article{Harari:1969jcb,
    author = "Harari, H. and Zarmi, Y.",
    title = "{Duality and the pomeranchuk singularity}",
    doi = "10.1103/PhysRev.187.2230",
    journal = "Phys. Rev.",
    volume = "187",
    pages = "2230--2244",
    year = "1969"
}

@article{Harari:1968jw,
    author = "Harari, Haim",
    title = "{Pomeranchuk trajectory and its relation to low-energy scattering amplitudes}",
    doi = "10.1103/PhysRevLett.20.1395",
    journal = "Phys. Rev. Lett.",
    volume = "20",
    pages = "1395--1398",
    year = "1968"
}

@article{vanHove:1967zz,
    author = "van Hove, L.",
    title = "{Regge pole and single particle exchange mechanisms in high energy collisions}",
    doi = "10.1016/0370-2693(67)90488-1",
    journal = "Phys. Lett.",
    volume = "24",
    pages = "183--184",
    year = "1967"
}

@article{Phillips:1965zza,
    author = "Phillips, Roger J. and Rarita, William",
    title = "{Regge-Pole Models for High-Energy pi-N, K-N, and anti-K-N Scattering}",
    doi = "10.1103/PhysRev.139.B1336",
    journal = "Phys. Rev.",
    volume = "139",
    pages = "B1336--B1347",
    year = "1965"
}

@article{Oehme:1968usz,
    author = "Oehme, R.",
    title = "{The problem of dips in Regge pole amplitudes}",
    doi = "10.1016/0370-2693(68)90148-2",
    journal = "Phys. Lett. B",
    volume = "28",
    pages = "122--124",
    year = "1968"
}

@article{Roskies:1968kak,
    author = "Roskies, R.",
    title = "{Puzzle of regge dips}",
    doi = "10.1103/PhysRev.175.1933",
    journal = "Phys. Rev.",
    volume = "175",
    pages = "1933--1941",
    year = "1968"
}

@article{Goldberger:1955zz,
    author = "Goldberger, Marvin L.",
    title = "{Causality Conditions and Dispersion Relations. 1. Boson Fields}",
    doi = "10.1103/PhysRev.99.979",
    journal = "Phys. Rev.",
    volume = "99",
    pages = "979--985",
    year = "1955"
}

@article{Chew:1957zz,
    author = "Chew, G. F. and Goldberger, M. L. and Low, F. E. and Nambu, Y.",
    title = "{Application of Dispersion Relations to Low-Energy Meson-Nucleon Scattering}",
    doi = "10.1103/PhysRev.106.1337",
    journal = "Phys. Rev.",
    volume = "106",
    pages = "1337--1344",
    year = "1957"
}

@unpublished{Chew:1963zze,
    author = "Chew, G. F.",
    title = "{MAXIMUM ANALYTICITY OF THE SECOND DEGREE}",
    year = "1963",
    doi = "10.2172/4010271",
    reportNumber = "UCRL-10786",
    note = "(technical report)"
}

@article{Chew:1968pj,
    author = "Chew, G. F.",
    title = "{*bootstrap* - a scientific idea?}",
    doi = "10.1126/science.161.3843.762",
    journal = "Science",
    volume = "161",
    pages = "762--765",
    year = "1968"
}

@article{Chew:1971xwl,
    author = "Chew, G. F.",
    title = "{Hadron bootstrap hypothesis}",
    doi = "10.1103/PhysRevD.4.2330",
    journal = "Phys. Rev. D",
    volume = "4",
    pages = "2330--2335",
    year = "1971"
}

@article{Mathieu:2018mjw,
    author = "Mathieu, V. and Nys, J. and Fern\'andez-Ram\'\i{}rez, C. and Hiller Blin, A. N. and Jackura, A. and Pilloni, A. and Szczepaniak, A. P. and Fox, G.",
    collaboration = "JPAC",
    title = "{Structure of Pion Photoproduction Amplitudes}",
    eprint = "1806.08414",
    archivePrefix = "arXiv",
    primaryClass = "hep-ph",
    reportNumber = "JLAB-THY-18-2755",
    doi = "10.1103/PhysRevD.98.014041",
    journal = "Phys. Rev. D",
    volume = "98",
    number = "1",
    pages = "014041",
    year = "2018"
}

@article{Durand:1967jrt,
    author = "Durand, Loyal",
    title = "{Connection between Regge-Pole and Single-Particle Exchange Models for High-Energy Reactions}",
    doi = "10.1103/PhysRev.161.1610",
    journal = "Phys. Rev.",
    volume = "161",
    number = "5",
    pages = "1610--1611",
    year = "1967"
}

@article{Gribov:1965hf,
    author = "Gribov, V. N. and Ioffe, B. L. and Pomeranchuk, I. Ya.",
    title = "{What is the range of interactions at high-energies}",
    journal = "Yad. Fiz.",
    volume = "2",
    pages = "768--776",
    year = "1965"
}

@article{Barger:1965zz,
    author = "Barger, V. and Olsson, M.",
    title = "{Analysis of Total Cross-Section Differences at High Energy}",
    doi = "10.1103/PhysRevLett.15.930",
    journal = "Phys. Rev. Lett.",
    volume = "15",
    pages = "930--934",
    year = "1965"
}

@article{Lindenbaum:1961zz,
    author = "Lindenbaum, S. J. and Love, W. A. and Niederer, J. A. and Ozaki, S. and Russell, J. J. and Yuan, L. C. L.",
    title = "{Antiproton-Proton and Proton-Proton Total Cross Sections from 4 to 20 Bev/c}",
    doi = "10.1103/PhysRevLett.7.185",
    journal = "Phys. Rev. Lett.",
    volume = "7",
    pages = "185--188",
    year = "1961"
}

@article{Baker:1963zzd,
    author = "Baker, W. F. and Cool, R. L. and Jenkins, E. W. and Kycia, T. F. and Phillips, R. H. and Read, A. L.",
    title = "{K+-p and K--p Total Cross Sections in the Momentum Range 3-19 BeV/c}",
    doi = "10.1103/PhysRev.129.2285",
    journal = "Phys. Rev.",
    volume = "129",
    pages = "2285--2291",
    year = "1963"
}

@article{Galbraith:1965jk,
    author = "Galbraith, W. and Jenkins, E. W. and Kycia, T. F. and Leontic, B. A. and Phillips, R. H. and Read, Anthony Lincoln and Rubinstein, R.",
    title = "{Total cross-sections of protons, anti-protons, and pi and K mesons on hydrogen and deuterium in the momentum range 6-GeV/c to 22-GeV/c}",
    doi = "10.1103/PhysRev.138.B913",
    journal = "Phys. Rev.",
    volume = "138",
    pages = "B913--B920",
    year = "1965"
}

@article{Pomeranchuck:1958,
    author = "Pomeranchuck, I. Ya.",
    title = "{EQUALITY OF THE NUCELON AND ANTINUCLEON TOTAL INTERACTION CORSS SECTION AT HIGH ENERGIES}",
    journal = "Sov. Phys. JETP",
    volume = "7",
    pages = "499--501",
    year = "1958"
}

@article{Barnes:1976ek,
    author = "Barnes, A. V. and Mellema, D. J. and Tollestrup, A. V. and Walker, R. L. and Dahl, O. I. and Johnson, R. A. and Kenney, R. W. and Pripstein, M.",
    title = "{Pion Charge Exchange Scattering at High-Energies}",
    reportNumber = "FERMILAB-PUB-76-130-E, CALT-68-548, LBL-4830",
    doi = "10.1103/PhysRevLett.37.76",
    journal = "Phys. Rev. Lett.",
    volume = "37",
    pages = "76",
    year = "1976"
}

@article{Gribov:1962fx,
    author = "Gribov, V. N.",
    title = "{Analytic properties of the partial wave amplitudes and the asymptotic behavior of the scattering amplitude}",
    doi = "10.1016/0029-5582(63)90256-6",
    journal = "Sov. Phys. JETP",
    volume = "15",
    pages = "873",
    year = "1962"
}

@article{Gribov:1961fm,
    author = "Gribov, V. N.",
    title = "{Asymptotic behaviour of the scattering amplitude at high energies}",
    doi = "10.1016/0029-5582(61)90457-6",
    journal = "Nucl. Phys.",
    volume = "22",
    pages = "249--261",
    year = "1961"
}

@article{Donnachie:1983hf,
    author = "Donnachie, A. and Landshoff, P. V.",
    title = "{p p and anti-p p Elastic Scattering}",
    reportNumber = "M/C TH-83/13, DAMTP 83/9",
    doi = "10.1016/0550-3213(84)90283-9",
    journal = "Nucl. Phys. B",
    volume = "231",
    pages = "189--204",
    year = "1984"
}

@article{Donnachie:1992ny,
    author = "Donnachie, A. and Landshoff, P. V.",
    title = "{Total cross-sections}",
    eprint = "hep-ph/9209205",
    archivePrefix = "arXiv",
    reportNumber = "CERN-TH-6635-92",
    doi = "10.1016/0370-2693(92)90832-O",
    journal = "Phys. Lett. B",
    volume = "296",
    pages = "227--232",
    year = "1992"
}

@article{Donnachie:1983ff,
    author = "Donnachie, A. and Landshoff, P. V.",
    title = "{Multi - Gluon Exchange in $p p$ Elastic Scattering}",
    reportNumber = "DAMTP 82/33",
    doi = "10.1016/0370-2693(83)91215-7",
    journal = "Phys. Lett. B",
    volume = "123",
    pages = "345--348",
    year = "1983"
}

@article{Gell-Mann:1964bha,
    author = "Gell-Mann, M. and Goldberger, M. and Low, F.",
    title = "{The Vacuum Trajectory in Conventional Field Theory}",
    doi = "10.1103/RevModPhys.36.640",
    journal = "Rev. Mod. Phys.",
    volume = "36",
    number = "2",
    pages = "640--649",
    year = "1964"
}

@article{Mathieu:2008me,
    author = "Mathieu, Vincent and Kochelev, Nikolai and Vento, Vicente",
    title = "{The Physics of Glueballs}",
    eprint = "0810.4453",
    archivePrefix = "arXiv",
    primaryClass = "hep-ph",
    doi = "10.1142/S0218301309012124",
    journal = "Int. J. Mod. Phys. E",
    volume = "18",
    pages = "1--49",
    year = "2009"
}

@article{Freund:1967hw,
    author = "Freund, Peter G. O.",
    title = "{finite energy sum rules and bootstraps}",
    doi = "10.1103/PhysRevLett.20.235",
    journal = "Phys. Rev. Lett.",
    volume = "20",
    pages = "235--237",
    year = "1968"
}

@article{Arbab:1968urb,
    author = "Arbab, F. and Jackson, John David",
    title = "{Factorization, kinematic singularities, and conspiracies}",
    doi = "10.1103/PhysRev.176.1796",
    journal = "Phys. Rev.",
    volume = "176",
    pages = "1796--1808",
    year = "1968"
}

@article{Frampton:1968rw,
    author = "Frampton, P. H.",
    title = "{Factorization, kinematic factors of the Regge residue function and conspiracy}",
    doi = "10.1016/0550-3213(68)90188-0",
    journal = "Nucl. Phys. B",
    volume = "7",
    pages = "507--526",
    year = "1968"
}

@article{Nys:2018vck,
    author = "Nys, J. and Hiller Blin, A. N. and Mathieu, V. and Fern\'andez-Ram\'\i{}rez, C. and Jackura, A. and Pilloni, A. and Ryckebusch, J. and Szczepaniak, A. P. and Fox, G.",
    collaboration = "JPAC",
    title = "{Global analysis of charge exchange meson production at high energies}",
    eprint = "1806.01891",
    archivePrefix = "arXiv",
    primaryClass = "hep-ph",
    reportNumber = "JLAB-THY-18-2736",
    doi = "10.1103/PhysRevD.98.034020",
    journal = "Phys. Rev. D",
    volume = "98",
    number = "3",
    pages = "034020",
    year = "2018"
}

@article{Collins:1969gq,
    author = "Collins, P. D. B. and Johnson, R. C.",
    title = "{Bootstrap of pion pion scattering in the unitarized strip approximation}",
    doi = "10.1103/PhysRev.182.1755",
    journal = "Phys. Rev.",
    volume = "182",
    pages = "1755--1761",
    year = "1969"
}

@article{Collins:1969pg,
    author = "Collins, P. D. B. and Johnson, R. C.",
    title = "{Rho bootstrap in the unitarized strip approximation}",
    doi = "10.1103/PhysRev.177.2472",
    journal = "Phys. Rev.",
    volume = "177",
    pages = "2472--2481",
    year = "1969"
}

@article{Zachariasen:1961zz,
    author = "Zachariasen, Fredrik",
    title = "{Self-Consistent Calculation of the Mass and Width of the J=1, T=1, pipi Resonance}",
    doi = "10.1103/PhysRevLett.7.112",
    journal = "Phys. Rev. Lett.",
    volume = "7",
    pages = "112--113",
    year = "1961"
}

@article{Webber:1971rm,
    author = "Webber, B. R.",
    title = "{Bootstrap calculations of pi pi scattering using the mandelstam iteration}",
    doi = "10.1103/PhysRevD.3.1971",
    journal = "Phys. Rev. D",
    volume = "3",
    pages = "1971--1980",
    year = "1971"
}

@article{Chew:1960iv,
    author = "Chew, Geoffrey F. and Mandelstam, Stanley",
    title = "{Theory of low-energy pion pion interactions}",
    doi = "10.1103/PhysRev.119.467",
    journal = "Phys. Rev.",
    volume = "119",
    pages = "467--477",
    year = "1960"
}

@article{Degasperis:1970us,
    author = "Degasperis, A. and Predazzi, E.",
    title = "{Dynamical calculation of regge trajectories}",
    doi = "10.1007/BF02892141",
    journal = "Nuovo Cim. A",
    volume = "65",
    pages = "764--782",
    year = "1970"
}

@article{Mandelstam:1969dk,
    author = "Mandelstam, Stanley",
    title = "{Rising regge trajectories and dynamical calculations}",
    journal = "Comments Nucl. Part. Phys.",
    volume = "3",
    number = "3",
    pages = "65--72",
    year = "1969"
}

@article{Chu:1968ctr,
    author = "Chu, Shu-Yuan and Epstein, G. and Kaus, P. and Slansky, R. C. and Zachariasen, F.",
    title = "{Crossing-symmetric rising regge trajectories}",
    doi = "10.1103/PhysRev.175.2098",
    journal = "Phys. Rev.",
    volume = "175",
    pages = "2098--2105",
    year = "1968"
}

@article{Epstein:1968vaa,
    author = "Epstein, G",
    title = "{Rising meson trajectories}",
    doi = "10.1103/PhysRev.166.1633",
    journal = "Phys. Rev.",
    volume = "166",
    pages = "1633--1637",
    year = "1968"
}

@article{Kaus:1970nc,
    author = "Kaus, P. and Zachariasen, F.",
    title = "{Colliding regge poles and cuts}",
    doi = "10.1103/PhysRevD.1.2962",
    journal = "Phys. Rev. D",
    volume = "1",
    pages = "2962--2967",
    year = "1970"
}

@article{Ball:1969cw,
    author = "Ball, J. S. and Zachariasen, F.",
    title = "{Left-hand cuts in regge trajectories}",
    doi = "10.1103/PhysRevLett.23.346",
    journal = "Phys. Rev. Lett.",
    volume = "23",
    pages = "346--348",
    year = "1969"
}

@article{Ball:1970vy,
    author = "Ball, J. S. and Marchesini, G. and Zachariasen, F.",
    title = "{Phenomenological implications of colliding regge poles and cuts}",
    doi = "10.1016/0370-2693(70)90701-X",
    journal = "Phys. Lett. B",
    volume = "31",
    pages = "583--588",
    year = "1970"
}

@article{Collins:1968akw,
    author = "Collins, P. D. B. and Johnson, R. C. and Squires, E. J.",
    title = "{Heavy bosons, Regge trajectories and dynamical theories}",
    doi = "10.1016/0370-2693(68)90348-1",
    journal = "Phys. Lett. B",
    volume = "26",
    pages = "223--225",
    year = "1968"
}

@article{Childers:1968vnm,
    author = "Childers, R. W.",
    title = "{Asymptotic Behavior of Infinitely Rising Trajectories}",
    doi = "10.1103/PhysRevLett.21.868",
    journal = "Phys. Rev. Lett.",
    volume = "21",
    number = "12",
    pages = "868",
    year = "1968"
}

@article{Bugrij:1973ph,
    author = "Bugrij, A. I. and Cohen-Tannoudji, G. and Jenkovszky, Laszlo L. and Kobylinsky, N. A.",
    title = "{Dual amplitudes with mandelstam analyticity}",
    doi = "10.1002/prop.19730210902",
    journal = "Fortsch. Phys.",
    volume = "21",
    pages = "427--506",
    year = "1973"
}

@article{Trushevsky:1975yf,
    author = "Trushevsky, A. A.",
    title = "{Asymptotic Behavior of Boson Regge Trajectories}",
    reportNumber = "ITF-75-81E",
    doi = "10.15407/ujpe66.2.97",
    journal = "Ukr. Fiz. Zh. (Ukr. Ed. )",
    volume = "22",
    number = "3",
    pages = "353--362",
    year = "1977"
}

@article{Botke:1972xn,
    author = "Botke, J. C.",
    title = "{Infinitely rising trajectories and the existence of the mandelstam representation}",
    doi = "10.1016/0550-3213(72)90537-8",
    journal = "Nucl. Phys. B",
    volume = "40",
    pages = "141--150",
    year = "1972"
}

@article{Cheng:1963,
    author = "Cheng, H. and Sharp, D.",
    title = "{Formulation and Numerical Solutions of a Set of Dynamical Equations for the Regge Pole Parameters}",
    doi = "10.1103/PhysRev.132.1854",
    journal = "Phys. Rev.",
    volume = "132",
    pages = "1854--1866",
    year = "1963"
}

@article{Frautschi:1964,
    author = "Frautschi, S. C. and Kaus, P. E. and Zachariasen, F.",
    title = "{Method for te Self-Consistent Determination of Regge Pole Parameters}",
    doi = "10.1103/PhysRev.133.B1607",
    journal = "Phys. Rev.",
    volume = "133",
    pages = "B1607--B1615",
    year = "1964"
}

@article{JPAC:2021rxu,
    author = "Albaladejo, Miguel and others",
    collaboration = "JPAC",
    title = "{Novel approaches in hadron spectroscopy}",
    eprint = "2112.13436",
    archivePrefix = "arXiv",
    primaryClass = "hep-ph",
    reportNumber = "LA-UR-21-31664, JLAB-THY-22-3459",
    doi = "10.1016/j.ppnp.2022.103981",
    journal = "Prog. Part. Nucl. Phys.",
    volume = "127",
    pages = "103981",
    year = "2022"
}

@article{Mizera:2023tfe,
    author = "Mizera, Sebastian",
    title = "{Physics of the analytic S-matrix}",
    eprint = "2306.05395",
    archivePrefix = "arXiv",
    primaryClass = "hep-th",
    doi = "10.1016/j.physrep.2023.10.006",
    journal = "Phys. Rept.",
    volume = "1047",
    pages = "1--92",
    year = "2024"
}

@article{Dolen:1967jr,
    author = "Dolen, R. and Horn, D. and Schmid, C.",
    title = "{Finite energy sum rules and their application to pi N charge exchange}",
    doi = "10.1103/PhysRev.166.1768",
    journal = "Phys. Rev.",
    volume = "166",
    pages = "1768--1781",
    year = "1968"
}

@article{Logunov:1967dy,
    author = "Logunov, A. A. and Soloviev, L. D. and Tavkhelidze, A. N.",
    title = "{Dispersion sum rules and high-energy scattering}",
    doi = "10.1016/0370-2693(67)90487-X",
    journal = "Phys. Lett. B",
    volume = "24",
    pages = "181--182",
    year = "1967"
}

@article{Igi:1967zza,
    author = "Igi, Keiji and Matsuda, Satoshi",
    title = "{New Sum Rules and Singularities in the Complex J Plane}",
    doi = "10.1103/PhysRevLett.18.625",
    journal = "Phys. Rev. Lett.",
    volume = "18",
    pages = "625--627",
    year = "1967"
}

@article{Mathieu:2015gxa,
    author = "Mathieu, V. and Danilkin, I. V. and Fern\'andez-Ram\'\i{}rez, C. and Pennington, M. R. and Schott, D. and Szczepaniak, A. P. and Fox, G.",
    title = "{Toward Complete Pion Nucleon Amplitudes}",
    eprint = "1506.01764",
    archivePrefix = "arXiv",
    primaryClass = "hep-ph",
    reportNumber = "JLAB-THY-15-2056",
    doi = "10.1103/PhysRevD.92.074004",
    journal = "Phys. Rev. D",
    volume = "92",
    number = "7",
    pages = "074004",
    year = "2015"
}

@article{Chew:1968zz,
    author = "Chew, G. F. and Pignotti, A.",
    title = "{Dolen-Horn-Schmid Duality and the Deck Effect}",
    doi = "10.1103/PhysRevLett.20.1078",
    journal = "Phys. Rev. Lett.",
    volume = "20",
    pages = "1078--1081",
    year = "1968"
}

@article{Vickson:1969ksa,
    author = "Vickson, R. G.",
    title = "{Generalized interference model}",
    doi = "10.1103/PhysRev.188.2154",
    journal = "Phys. Rev.",
    volume = "188",
    pages = "2154--2158",
    year = "1969"
}

@article{Stamen:2024gfz,
    author = "Stamen, D. and Winney, D. and Rodas, A. and Fern\'andez-Ram\'\i{}rez, C. and Mathieu, V. and Monta\~na, G. and Pilloni, A. and Szczepaniak, A. P.",
    title = "{Toward a unified description of hadron scattering at all energies}",
    eprint = "2409.09172",
    archivePrefix = "arXiv",
    primaryClass = "hep-ph",
    reportNumber = "JLAB-THY-24-4182",
    doi = "10.1103/PhysRevD.110.114023",
    journal = "Phys. Rev. D",
    volume = "110",
    number = "11",
    pages = "114023",
    year = "2024"
}

@article{Schmid:1968zz,
    author = "Schmid, Christoph",
    title = "{Direct-Channel Resonances from Regge-Pole Exchange}",
    doi = "10.1103/PhysRevLett.20.689",
    journal = "Phys. Rev. Lett.",
    volume = "20",
    pages = "689--691",
    year = "1968"
}

@article{Harari:1969oxx,
    author = "Harari, H.",
    title = "{Duality Diagrams}",
    doi = "10.1103/PhysRevLett.22.562",
    journal = "Phys. Rev. Lett.",
    volume = "22",
    pages = "562--565",
    year = "1969"
}

@article{Veneziano:1968yb,
    author = "Veneziano, G.",
    title = "{Construction of a crossing - symmetric, Regge behaved amplitude for linearly rising trajectories}",
    doi = "10.1007/BF02824451",
    journal = "Nuovo Cim. A",
    volume = "57",
    pages = "190--197",
    year = "1968"
}

@article{Lovelace:1968kjy,
    author = "Lovelace, C.",
    title = "{A novel application of regge trajectories}",
    doi = "10.1016/0370-2693(68)90255-4",
    journal = "Phys. Lett. B",
    volume = "28",
    pages = "264--268",
    year = "1968"
}

@article{Shapiro:1969km,
    author = "Shapiro, J. A.",
    title = "{Narrow-resonance model with regge behavior for pi pi scattering}",
    doi = "10.1103/PhysRev.179.1345",
    journal = "Phys. Rev.",
    volume = "179",
    pages = "1345--1353",
    year = "1969"
}

@article{Collins:1970as,
    author = "Collins, P. D. B. and Mir, K. L.",
    title = "{Unitarity in dual models}",
    doi = "10.1016/0550-3213(70)90364-0",
    journal = "Nucl. Phys. B",
    volume = "19",
    pages = "509--524",
    year = "1970"
}

@article{Susskind:1969ha,
    author = "Susskind, Leonard",
    title = "{Harmonic-oscillator analogy for the veneziano model}",
    doi = "10.1103/PhysRevLett.23.545",
    journal = "Phys. Rev. Lett.",
    volume = "23",
    pages = "545--547",
    year = "1969"
}

@article{Susskind:1970qz,
    author = "Susskind, Leonard",
    title = "{Structure of hadrons implied by duality}",
    doi = "10.1103/PhysRevD.1.1182",
    journal = "Phys. Rev. D",
    volume = "1",
    pages = "1182--1186",
    year = "1970"
}

@article{Cohen-Tannoudji:1971mlv,
    author = "Cohen-Tannoudji, G. and Henyey, F. and Kane, Gordon L. and Zakrzewski, W. J.",
    title = "{Dual, crossing-symmetric amplitude with mandelstam analyticity}",
    doi = "10.1103/PhysRevLett.26.112",
    journal = "Phys. Rev. Lett.",
    volume = "26",
    pages = "112--115",
    year = "1971"
}

@article{Friedman:1970wj,
    author = "Friedman, M. H. and Nath, Pran and Srivastava, Y. N.",
    title = "{Amplitudes with mandelstam analyticity and dual structure}",
    doi = "10.1103/PhysRevLett.24.1317",
    journal = "Phys. Rev. Lett.",
    volume = "24",
    pages = "1317--1320",
    year = "1970"
}

@article{Martin:1969tc,
    author = "Martin, Andre",
    title = "{Smoothing veneziano}",
    doi = "10.1016/0370-2693(69)90240-8",
    journal = "Phys. Lett. B",
    volume = "29",
    pages = "431--432",
    year = "1969"
}

@article{Huang:1969ef,
    author = "Huang, Kerson",
    title = "{Unitarization of the veneziano model, pomeranchuk singularity, and the pion mass}",
    doi = "10.1103/PhysRevLett.23.903",
    journal = "Phys. Rev. Lett.",
    volume = "23",
    pages = "903--906",
    year = "1969"
}

@article{Bali:1969fm,
    author = "Bali, N. F. and Coon, Darryl D. and Dash, J. W.",
    title = "{Unstable particles, two-body inelastic unitarity, and veneziano's model}",
    doi = "10.1103/PhysRevLett.23.900",
    journal = "Phys. Rev. Lett.",
    volume = "23",
    pages = "900--903",
    year = "1969"
}

@article{Sivers:1971ig,
    author = "Sivers, D. and Yellin, J.",
    title = "{Review of recent work on narrow resonance models}",
    doi = "10.1103/RevModPhys.43.125",
    journal = "Rev. Mod. Phys.",
    volume = "43",
    pages = "125--188",
    year = "1971"
}

@article{Rosner:1969bhr,
    author = "Rosner, Jonathan L.",
    title = "{Graphical Form of Duality}",
    doi = "10.1103/PhysRevLett.22.689",
    journal = "Phys. Rev. Lett.",
    volume = "22",
    pages = "689--692",
    year = "1969"
}

@article{Tryon:1971pt,
    author = "Tryon, E. P.",
    title = "{Inverse amplitude (K matrix) method for unitarizing Veneziano amplitudes, with application to pi pi scattering}",
    doi = "10.1103/PhysRevD.4.1202",
    journal = "Phys. Rev. D",
    volume = "4",
    pages = "1202--1215",
    year = "1971"
}

@article{Paciello:1969ry,
    author = "Paciello, M. L. and Sertorio, L. and Taglienti, B.",
    title = "{Veneziano amplitude with complex trajectories}",
    doi = "10.1007/BF02754920",
    journal = "Nuovo Cim. A",
    volume = "63",
    pages = "1026--1034",
    year = "1969"
}

@article{Blankenbecler:1973kt,
    author = "Blankenbecler, Richard and Brodsky, Stanley J. and Gunion, J. F. and Savit, R.",
    title = "{THE CONNECTION BETWEEN REGGE BEHAVIOR AND FIXED ANGLE SCATTERING}",
    reportNumber = "SLAC-PUB-1294",
    doi = "10.1103/PhysRevD.8.4117",
    journal = "Phys. Rev. D",
    volume = "8",
    pages = "4117",
    year = "1973"
}

@article{Gunion:1972gy,
    author = "Gunion, J. F. and Brodsky, Stanley J. and Blankenbecler, Richard",
    title = "{Composite Theory of Large Angle Scattering and New Tests of Parton Concepts}",
    reportNumber = "SLAC-PUB-1037",
    doi = "10.1016/0370-2693(72)90022-6",
    journal = "Phys. Lett. B",
    volume = "39",
    pages = "649--653",
    year = "1972"
}

@article{Schmidt:1973ew,
    author = "Schmidt, Michael G.",
    title = "{TOWARDS A DYNAMICAL INTERPRETATION OF DUAL AMPLITUDES WITH MANDELSTAM ANALYTICITY}",
    reportNumber = "SLAC-PUB-1173",
    doi = "10.1016/0370-2693(73)90388-2",
    journal = "Phys. Lett. B",
    volume = "43",
    pages = "417--421",
    year = "1973"
}

@article{Gell-Mann:1964ewy,
    author = "Gell-Mann, Murray",
    title = "{A Schematic Model of Baryons and Mesons}",
    doi = "10.1016/S0031-9163(64)92001-3",
    journal = "Phys. Lett.",
    volume = "8",
    pages = "214--215",
    year = "1964"
}

@article{Gell-Mann:1962yej,
    author = "Gell-Mann, Murray",
    title = "{Symmetries of baryons and mesons}",
    doi = "10.1103/PhysRev.125.1067",
    journal = "Phys. Rev.",
    volume = "125",
    pages = "1067--1084",
    year = "1962"
}

@article{Chan:1969xg,
    author = "Chan, Hong-Mo",
    title = "{A generalised veneziano model for the n-point function}",
    doi = "10.1016/0370-2693(69)90342-6",
    journal = "Phys. Lett. B",
    volume = "28",
    pages = "425--428",
    year = "1969"
}

@article{Ramond:1971gb,
    author = "Ramond, Pierre",
    title = "{Dual Theory for Free Fermions}",
    reportNumber = "FERMILAB-PUB-70-008-T, FERMILAB-PUB-70-008-THY, NAL-THY-8",
    doi = "10.1103/PhysRevD.3.2415",
    journal = "Phys. Rev. D",
    volume = "3",
    pages = "2415--2418",
    year = "1971"
}

@article{Neveu:1971fz,
    author = "Neveu, A. and Schwarz, J. H.",
    title = "{Tachyon-free dual model with a positive-intercept trajectory}",
    doi = "10.1016/0370-2693(71)90669-1",
    journal = "Phys. Lett. B",
    volume = "34",
    pages = "517--518",
    year = "1971"
}

@article{Mandelstam:1970fd,
    author = "Mandelstam, S.",
    title = "{Relativistic quark model based on the veneziano representation. ii. general trajectories}",
    doi = "10.1103/PhysRevD.1.1734",
    journal = "Phys. Rev. D",
    volume = "1",
    pages = "1734--1744",
    year = "1970"
}

@article{Mandelstam:1970zy,
    author = "Mandelstam, S.",
    title = "{Relativistic quark model based on the veneziano representation. iii. baryon trajectories}",
    doi = "10.1103/PhysRevD.1.1745",
    journal = "Phys. Rev. D",
    volume = "1",
    pages = "1745--1753",
    year = "1970"
}

@article{Mandelstam:1969pkr,
    author = "Mandelstam, S.",
    title = "{Relativistic quark model based on the veneziano representation. i. meson trajectories}",
    doi = "10.1103/PhysRev.184.1625",
    journal = "Phys. Rev.",
    volume = "184",
    pages = "1625--1639",
    year = "1969"
}

@book{Becker:2006dvp,
    author = "Becker, K. and Becker, M. and Schwarz, J. H.",
    title = "{String theory and M-theory: A modern introduction}",
    doi = "10.1017/CBO9780511816086",
    isbn = "978-0-511-25486-4, 978-0-521-86069-7, 978-0-511-81608-6",
    publisher = "Cambridge University Press",
    month = "12",
    year = "2006"
}

@article{Coon:1974wh,
    author = "Coon, Darryl D. and Gunion, J. F. and Tran Thanh Van, J. and Blankenbecler, Richard",
    title = "{Phenomenology of High Momentum Transfer Elastic Processes}",
    reportNumber = "SLAC-PUB-1483",
    doi = "10.1103/PhysRevD.18.1451",
    journal = "Phys. Rev. D",
    volume = "18",
    pages = "1451",
    year = "1978"
}

@article{Brodsky:1973kr,
    author = "Brodsky, Stanley J. and Farrar, Glennys R.",
    title = "{Scaling Laws at Large Transverse Momentum}",
    reportNumber = "SLAC-PUB-1290",
    doi = "10.1103/PhysRevLett.31.1153",
    journal = "Phys. Rev. Lett.",
    volume = "31",
    pages = "1153--1156",
    year = "1973"
}

@article{Matveev:1973ra,
    author = "Matveev, V. A. and Muradian, R. M. and Tavkhelidze, A. N.",
    title = "{Automodellism in the large - angle elastic scattering and structure of hadrons}",
    doi = "10.1007/BF02728133",
    journal = "Lett. Nuovo Cim.",
    volume = "7",
    pages = "719--723",
    year = "1973"
}

@article{Gross:1973id,
    author = "Gross, David J. and Wilczek, Frank",
    editor = "Taylor, J. C.",
    title = "{Ultraviolet Behavior of Nonabelian Gauge Theories}",
    doi = "10.1103/PhysRevLett.30.1343",
    journal = "Phys. Rev. Lett.",
    volume = "30",
    pages = "1343--1346",
    year = "1973"
}

@article{Politzer:1973fx,
    author = "Politzer, H. David",
    editor = "Taylor, J. C.",
    title = "{Reliable Perturbative Results for Strong Interactions?}",
    doi = "10.1103/PhysRevLett.30.1346",
    journal = "Phys. Rev. Lett.",
    volume = "30",
    pages = "1346--1349",
    year = "1973"
}

@article{Irving:1977ea,
    author = "Irving, A. C. and Worden, R. P.",
    title = "{Regge Phenomenology}",
    doi = "10.1016/0370-1573(77)90010-2",
    journal = "Phys. Rept.",
    volume = "34",
    pages = "117--231",
    year = "1977"
}

@article{Gordon:1973vc,
    author = "Gordon, H. A. and Lai, Kwan-Wu and Scarr, J. M.",
    title = "{Rho-meson production in the reaction pi+- p ---\ensuremath{>} rho n at 6 gev/c}",
    doi = "10.1103/PhysRevD.8.779",
    journal = "Phys. Rev. D",
    volume = "8",
    pages = "779--799",
    year = "1973"
}

@article{tHooft:1973alw,
    author = "'t Hooft, Gerard",
    editor = "Taylor, J. C.",
    title = "{A Planar Diagram Theory for Strong Interactions}",
    reportNumber = "CERN-TH-1786",
    doi = "10.1016/0550-3213(74)90154-0",
    journal = "Nucl. Phys. B",
    volume = "72",
    pages = "461",
    year = "1974"
}

@article{McGuigan:1992bi,
    author = "McGuigan, Michael and Thorn, Charles B.",
    title = "{Quark - anti-quark Regge trajectories in large N(c) QCD}",
    eprint = "hep-ph/9205211",
    archivePrefix = "arXiv",
    reportNumber = "UFIFT-HEP-92-12",
    doi = "10.1103/PhysRevLett.69.1312",
    journal = "Phys. Rev. Lett.",
    volume = "69",
    pages = "1312--1315",
    year = "1992"
}

@article{Kruczenski:2004me,
    author = "Kruczenski, Martin and Pando Zayas, Leopoldo A. and Sonnenschein, Jacob and Vaman, Diana",
    title = "{Regge trajectories for mesons in the holographic dual of large-N(c) QCD}",
    eprint = "hep-th/0410035",
    archivePrefix = "arXiv",
    reportNumber = "MCTP-04-56",
    doi = "10.1088/1126-6708/2005/06/046",
    journal = "JHEP",
    volume = "06",
    pages = "046",
    year = "2005"
}

@article{RuizdeElvira:2010cs,
    author = "Ruiz de Elvira, J. and Pelaez, J. R. and Pennington, M. R. and Wilson, D. J.",
    title = "{Chiral Perturbation Theory, the ${1/N_c}$ expansion and Regge behaviour determine the structure of the lightest scalar meson}",
    eprint = "1009.6204",
    archivePrefix = "arXiv",
    primaryClass = "hep-ph",
    reportNumber = "DCPT-10-98, IPPP-10-49, JLAB-THY-10-1256",
    doi = "10.1103/PhysRevD.84.096006",
    journal = "Phys. Rev. D",
    volume = "84",
    pages = "096006",
    year = "2011"
}

@article{Dawid:2019vhl,
    author = "Dawid, Sebastian M. and Szczepaniak, Adam P.",
    title = "{The Coulomb flux tube revisited}",
    eprint = "1908.08874",
    archivePrefix = "arXiv",
    primaryClass = "hep-lat",
    reportNumber = "JLAB-THY-19-3018",
    doi = "10.1103/PhysRevD.100.074508",
    journal = "Phys. Rev. D",
    volume = "100",
    pages = "074508",
    year = "2019"
}

@article{Dawid:2024uqz,
    author = "Dawid, Sebastian M. and Smith, Wyatt A. and Rodas, Arkaitz and Perry, Robert J. and Fern\'andez-Ram\'\i{}rez, C\'esar and Swanson, Eric S. and Szczepaniak, Adam P.",
    title = "{Coulomb confinement in the Hamiltonian limit}",
    eprint = "2408.09007",
    archivePrefix = "arXiv",
    primaryClass = "hep-lat",
    doi = "10.1103/PhysRevD.110.094509",
    journal = "Phys. Rev. D",
    volume = "110",
    number = "9",
    pages = "094509",
    year = "2024"
}

@article{Greensite:2014bua,
    author = "Greensite, Jeff and Szczepaniak, Adam P.",
    title = "{Coulomb string tension, asymptotic string tension, and the gluon chain}",
    eprint = "1410.3525",
    archivePrefix = "arXiv",
    primaryClass = "hep-lat",
    reportNumber = "JLAB-THY-14-1925",
    doi = "10.1103/PhysRevD.91.034503",
    journal = "Phys. Rev. D",
    volume = "91",
    number = "3",
    pages = "034503",
    year = "2015"
}

@article{Greensite:2001nx,
    author = "Greensite, Jeff and Thorn, Charles B.",
    title = "{Gluon chain model of the confining force}",
    eprint = "hep-ph/0112326",
    archivePrefix = "arXiv",
    reportNumber = "LBNL-49311, UFIFT-HEP-01-26",
    doi = "10.1088/1126-6708/2002/02/014",
    journal = "JHEP",
    volume = "02",
    pages = "014",
    year = "2002"
}

@article{Bissey:2006bz,
    author = "Bissey, F. and Cao, F-G. and Kitson, A. R. and Signal, A. I. and Leinweber, D. B. and Lasscock, B. G. and Williams, A. G.",
    title = "{Gluon flux-tube distribution and linear confinement in baryons}",
    eprint = "hep-lat/0606016",
    archivePrefix = "arXiv",
    reportNumber = "ADP-06-04-T635",
    doi = "10.1103/PhysRevD.76.114512",
    journal = "Phys. Rev. D",
    volume = "76",
    pages = "114512",
    year = "2007"
}

@article{Bissey:2009gw,
    author = "Bissey, F. and Signal, A. I. and Leinweber, D. B.",
    title = "{Comparison of gluon flux-tube distributions for quark-diquark and quark-antiquark hadrons}",
    eprint = "0910.0958",
    archivePrefix = "arXiv",
    primaryClass = "hep-lat",
    reportNumber = "ADP-09-12-T690",
    doi = "10.1103/PhysRevD.80.114506",
    journal = "Phys. Rev. D",
    volume = "80",
    pages = "114506",
    year = "2009"
}

@article{Brodsky:1993fv,
    author = "Brodsky, Stanley J. and Tang, Wai-Keung and Thorn, Charles B.",
    title = "{The reggeon trajectory in exclusive and inclusive large momentum transfer reactions}",
    reportNumber = "SLAC-PUB-6227, UFIFT-HEP-93-21",
    doi = "10.1016/0370-2693(93)91807-Y",
    journal = "Phys. Lett. B",
    volume = "318",
    pages = "203--211",
    year = "1993"
}

@article{Collins:1983fg,
    author = "Collins, P. D. B. and Kearney, P. J.",
    title = "{Regge Theory and {QCD} in Large Angle Scattering}",
    reportNumber = "DTP-83/18",
    doi = "10.1007/BF01575793",
    journal = "Z. Phys. C",
    volume = "22",
    pages = "277",
    year = "1984"
}

@article{Kennett:1986wf,
    author = "Kennett, R. G. and Barnes, A. V. and Fox, G. C. and Walker, R. L. and Dahl, O. I. and Kenney, R. W. and Ogawa, A. and Pripstein, M.",
    title = "{The Production of Neutral Pions From 200-{GeV} $\pi^- p$ Collisions in the High X Region}",
    reportNumber = "CALT-68-1367",
    doi = "10.1016/0550-3213(87)90055-1",
    journal = "Nucl. Phys. B",
    volume = "284",
    pages = "653--673",
    year = "1987"
}

@article{Foldy:1963zz,
    author = "Foldy, Leslie L. and Peierls, Ronald F.",
    title = "{Isotopic Spin of Exchanged Systems}",
    doi = "10.1103/PhysRev.130.1585",
    journal = "Phys. Rev.",
    volume = "130",
    pages = "1585--1589",
    year = "1963"
}

@article{Feynman:1969wa,
    author = "Feynman, R. P.",
    title = "{The behavior of hadron collisions at extreme energies}",
    journal = "Conf. Proc. C",
    volume = "690905",
    pages = "237--258",
    year = "1969"
}

@article{Bibrzycki:2021rwh,
    author = "Bibrzycki, L. and Fernandez-Ramirez, C. and Mathieu, V. and Mikhasenko, M. and Albaladejo, M. and Blin, A. N. Hiller and Pilloni, A. and Szczepaniak, A. P.",
    collaboration = "JPAC",
    title = "{$\pi^-p\to\eta^{(\prime)}\, \pi^- p$ in the double-Regge region~
}",
    eprint = "2104.10646",
    archivePrefix = "arXiv",
    primaryClass = "hep-ph",
    reportNumber = "JLAB-THY-21-3354",
    doi = "10.1140/epjc/s10052-021-09594-8",
    journal = "Eur. Phys. J. C",
    volume = "81",
    pages = "647",
    year = "2021",
    note = "[Erratum: Eur.Phys.J.C 81, 915 (2021)]"
}

@article{Albaladejo:2020tzt,
    author = "Albaladejo, M. and Hiller Blin, A. N. and Pilloni, A. and Winney, D. and Fern\'andez-Ram\'\i{}rez, C. and Mathieu, V. and Szczepaniak, A.",
    collaboration = "JPAC",
    title = "{XYZ spectroscopy at electron-hadron facilities: Exclusive processes}",
    eprint = "2008.01001",
    archivePrefix = "arXiv",
    primaryClass = "hep-ph",
    reportNumber = "JLAB-THY-20-3231",
    doi = "10.1103/PhysRevD.102.114010",
    journal = "Phys. Rev. D",
    volume = "102",
    pages = "114010",
    year = "2020"
}

@article{Mathieu:2020zpm,
    author = "Mathieu, V. and Pilloni, A. and Albaladejo, M. and Bibrzycki, \L{}. and Celentano, A. and Fern\'andez-Ram\'\i{}rez, C. and Szczepaniak, A. P.",
    collaboration = "JPAC",
    title = "{Exclusive tensor meson photoproduction}",
    eprint = "2005.01617",
    archivePrefix = "arXiv",
    primaryClass = "hep-ph",
    reportNumber = "JLAB-THY-20-3188",
    doi = "10.1103/PhysRevD.102.014003",
    journal = "Phys. Rev. D",
    volume = "102",
    number = "1",
    pages = "014003",
    year = "2020"
}

@article{JPAC:2018zjz,
    author = "Silva-Castro, J. A. and Fernandez-Ramirez, C. and Albaladejo, M. and Danilkin, I. V. and Jackura, A. and Mathieu, V. and Nys, J. and Pilloni, A. and Szczepaniak, A. P. and Fox, G.",
    collaboration = "JPAC",
    title = "{Regge phenomenology of the $N^*$ and $\Delta^*$ poles}",
    eprint = "1809.01954",
    archivePrefix = "arXiv",
    primaryClass = "hep-ph",
    reportNumber = "JLAB-THY-18-2797",
    doi = "10.1103/PhysRevD.99.034003",
    journal = "Phys. Rev. D",
    volume = "99",
    number = "3",
    pages = "034003",
    year = "2019"
}

@article{Mathieu:2018xyc,
    author = "Mathieu, V. and Nys, J. and Fern\'andez-Ram\'\i{}rez, C. and Jackura, A. and Pilloni, A. and Sherrill, N. and Szczepaniak, A. P. and Fox, G.",
    collaboration = "JPAC",
    title = "{Vector Meson Photoproduction with a Linearly Polarized Beam}",
    eprint = "1802.09403",
    archivePrefix = "arXiv",
    primaryClass = "hep-ph",
    reportNumber = "JLAB-THY-18-2650",
    doi = "10.1103/PhysRevD.97.094003",
    journal = "Phys. Rev. D",
    volume = "97",
    number = "9",
    pages = "094003",
    year = "2018"
}

@article{JPAC:2016lnm,
    author = "Nys, J. and Mathieu, V. and Fern\'andez-Ram\'\i{}rez, C. and Hiller Blin, A. N. and Jackura, A. and Mikhasenko, M. and Pilloni, A. and Szczepaniak, A. P. and Fox, G. and Ryckebusch, J.",
    collaboration = "JPAC",
    title = "{Finite-energy sum rules in eta photoproduction off a nucleon}",
    eprint = "1611.04658",
    archivePrefix = "arXiv",
    primaryClass = "hep-ph",
    reportNumber = "JLAB-THY-16-2384",
    doi = "10.1103/PhysRevD.95.034014",
    journal = "Phys. Rev. D",
    volume = "95",
    number = "3",
    pages = "034014",
    year = "2017"
}

@article{JointPhysicsAnalysisCenter:2024kck,
    author = "Montana, Gloria and others",
    collaboration = "Joint Physics Analysis Center",
    title = "{Revisiting gauge invariance and Reggeization of pion exchange}",
    eprint = "2407.19577",
    archivePrefix = "arXiv",
    primaryClass = "hep-ph",
    reportNumber = "JLAB-THY-24-4129",
    doi = "10.1103/PhysRevD.110.114012",
    journal = "Phys. Rev. D",
    volume = "110",
    number = "11",
    pages = "114012",
    year = "2024"
}

@article{JointPhysicsAnalysisCenter:2024pat,
    author = "Winney, D. and others",
    collaboration = "Joint Physics Analysis Center",
    title = "{XYZ spectroscopy at electron-hadron facilities. III. Semi-inclusive processes with vector exchange}",
    eprint = "2404.05326",
    archivePrefix = "arXiv",
    primaryClass = "hep-ph",
    reportNumber = "JLAB-THY-24-4009",
    doi = "10.1103/PhysRevD.109.114035",
    journal = "Phys. Rev. D",
    volume = "109",
    number = "11",
    pages = "114035",
    year = "2024"
}

@article{Winney:2022tky,
    author = "Winney, D. and Pilloni, A. and Mathieu, V. and Hiller Blin, A. N. and Albaladejo, M. and Smith, W. A. and Szczepaniak, A.",
    collaboration = "Joint Physics Analysis Center",
    title = "{XYZ spectroscopy at electron-hadron facilities. II. Semi-inclusive processes with pion exchange}",
    eprint = "2209.05882",
    archivePrefix = "arXiv",
    primaryClass = "hep-ph",
    reportNumber = "JLAB-THY-22-3719",
    doi = "10.1103/PhysRevD.106.094009",
    journal = "Phys. Rev. D",
    volume = "106",
    number = "9",
    pages = "094009",
    year = "2022"
}

@article{Pelaez:2024uav,
    author = "Pel\'aez, J. R. and Rab\'an, P. and de Elvira, J. Ruiz",
    title = "{Global parametrizations of \ensuremath{\pi}\ensuremath{\pi} scattering with dispersive constraints: Beyond the S0 wave}",
    eprint = "2412.15327",
    archivePrefix = "arXiv",
    primaryClass = "hep-ph",
    reportNumber = "IPARCOS-UCM-24-062",
    doi = "10.1103/PhysRevD.111.074003",
    journal = "Phys. Rev. D",
    volume = "111",
    number = "7",
    pages = "074003",
    year = "2025"
}

@article{Pelaez:2019eqa,
    author = "Pelaez, J. R. and Rodas, A. and Ruiz De Elvira, J.",
    title = "{Global parameterization of $\pi \pi $ scattering up to 2 ${\mathrm {\,GeV}}$}",
    eprint = "1907.13162",
    archivePrefix = "arXiv",
    primaryClass = "hep-ph",
    doi = "10.1140/epjc/s10052-019-7509-6",
    journal = "Eur. Phys. J. C",
    volume = "79",
    number = "12",
    pages = "1008",
    year = "2019",
    note = "[Erratum: Eur.Phys.J.C 84, 1220 (2024)]"
}

@article{Caprini:2011ky,
    author = "Caprini, I. and Colangelo, G. and Leutwyler, H.",
    title = "{Regge analysis of the pi pi scattering amplitude}",
    eprint = "1111.7160",
    archivePrefix = "arXiv",
    primaryClass = "hep-ph",
    doi = "10.1140/epjc/s10052-012-1860-1",
    journal = "Eur. Phys. J. C",
    volume = "72",
    pages = "1860",
    year = "2012"
}

@article{Roy:1971tc,
    author = "Roy, S. M.",
    title = "{Exact integral equation for pion pion scattering involving only physical region partial waves}",
    doi = "10.1016/0370-2693(71)90724-6",
    journal = "Phys. Lett. B",
    volume = "36",
    pages = "353--356",
    year = "1971"
}

@article{Garcia-Martin:2011iqs,
    author = "Garcia-Martin, R. and Kaminski, R. and Pelaez, J. R. and Ruiz de Elvira, J. and Yndurain, F. J.",
    title = "{The Pion-pion scattering amplitude. IV: Improved analysis with once subtracted Roy-like equations up to 1100 MeV}",
    eprint = "1102.2183",
    archivePrefix = "arXiv",
    primaryClass = "hep-ph",
    doi = "10.1103/PhysRevD.83.074004",
    journal = "Phys. Rev. D",
    volume = "83",
    pages = "074004",
    year = "2011"
}

@article{Pelaez:2020gnd,
    author = "Pel\'aez, Jos\'e R. and Rodas, Arkaitz",
    title = "{Dispersive \ensuremath{\pi}K\textrightarrow{}\ensuremath{\pi}K and \ensuremath{\pi}\ensuremath{\pi}\textrightarrow{}KK amplitudes from scattering data, threshold parameters, and the lightest strange resonance \ensuremath{\kappa} or K0\ensuremath{*}(700)}",
    eprint = "2010.11222",
    archivePrefix = "arXiv",
    primaryClass = "hep-ph",
    reportNumber = "JLAB-THY-20-3276",
    doi = "10.1016/j.physrep.2022.03.004",
    journal = "Phys. Rept.",
    volume = "969",
    pages = "1--126",
    year = "2022"
}

@article{Ditsche:2012fv,
    author = "Ditsche, C. and Hoferichter, M. and Kubis, B. and Meissner, U. -G.",
    title = "{Roy-Steiner equations for pion-nucleon scattering}",
    eprint = "1203.4758",
    archivePrefix = "arXiv",
    primaryClass = "hep-ph",
    doi = "10.1007/JHEP06(2012)043",
    journal = "JHEP",
    volume = "06",
    pages = "043",
    year = "2012"
}

@article{Hoferichter:2015hva,
    author = "Hoferichter, Martin and Ruiz de Elvira, Jacobo and Kubis, Bastian and Mei\ss{}ner, Ulf-G.",
    title = "{Roy\textendash{}Steiner-equation analysis of pion\textendash{}nucleon scattering}",
    eprint = "1510.06039",
    archivePrefix = "arXiv",
    primaryClass = "hep-ph",
    reportNumber = "INT-PUB-15-050",
    doi = "10.1016/j.physrep.2016.02.002",
    journal = "Phys. Rept.",
    volume = "625",
    pages = "1--88",
    year = "2016"
}

@article{Londergan:2013dza,
    author = "Londergan, J. T. and Nebreda, J. and Pelaez, J. R. and Szczepaniak, A.",
    title = "{Identification of non-ordinary mesons from the dispersive connection between their poles and their Regge trajectories: The $f_0(500)$ resonance}",
    eprint = "1311.7552",
    archivePrefix = "arXiv",
    primaryClass = "hep-ph",
    reportNumber = "JLAB-THY-14-1837",
    doi = "10.1016/j.physletb.2013.12.061",
    journal = "Phys. Lett. B",
    volume = "729",
    pages = "9--14",
    year = "2014"
}

@article{Carrasco:2015fva,
    author = "Carrasco, J. A. and Nebreda, J. and Pelaez, J. R. and Szczepaniak, A. P.",
    title = "{Dispersive calculation of complex Regge trajectories for the lightest $f_2$ resonances and the K*(892)}",
    eprint = "1504.03248",
    archivePrefix = "arXiv",
    primaryClass = "hep-ph",
    reportNumber = "JLAB-THY-15-2060",
    doi = "10.1016/j.physletb.2015.08.019",
    journal = "Phys. Lett. B",
    volume = "749",
    pages = "399--406",
    year = "2015"
}

@article{Fernandez-Ramirez:2015fbq,
    author = "Fernandez-Ramirez, Cesar and Danilkin, Igor V. and Mathieu, Vincent and Szczepaniak, Adam P.",
    title = "{Understanding the Nature of $\Lambda$(1405) through Regge Physics}",
    eprint = "1512.03136",
    archivePrefix = "arXiv",
    primaryClass = "hep-ph",
    reportNumber = "JLAB-THY-15-2182",
    doi = "10.1103/PhysRevD.93.074015",
    journal = "Phys. Rev. D",
    volume = "93",
    number = "7",
    pages = "074015",
    year = "2016"
}

@article{Pelaez:2017sit,
    author = "Pelaez, J. R. and Rodas, A.",
    title = "{The non-ordinary Regge behavior of the $K^*_0(800)$ or $\kappa $ -meson versus the ordinary $K^*_0(1430)$}",
    eprint = "1703.07661",
    archivePrefix = "arXiv",
    primaryClass = "hep-ph",
    doi = "10.1140/epjc/s10052-017-4994-3",
    journal = "Eur. Phys. J. C",
    volume = "77",
    number = "6",
    pages = "431",
    year = "2017"
}

@article{GlueX:2020idb,
    author = "Adhikari, S. and others",
    collaboration = "GlueX",
    title = "{The GLUEX beamline and detector}",
    eprint = "2005.14272",
    archivePrefix = "arXiv",
    primaryClass = "physics.ins-det",
    reportNumber = "JLAB-PHY-20-3195",
    doi = "10.1016/j.nima.2020.164807",
    journal = "Nucl. Instrum. Meth. A",
    volume = "987",
    pages = "164807",
    year = "2021"
}

@article{COMPASS:2007rjf,
    author = "Abbon, P. and others",
    collaboration = "COMPASS",
    title = "{The COMPASS experiment at CERN}",
    eprint = "hep-ex/0703049",
    archivePrefix = "arXiv",
    reportNumber = "CERN-PH-EP-2007-001",
    doi = "10.1016/j.nima.2007.03.026",
    journal = "Nucl. Instrum. Meth. A",
    volume = "577",
    pages = "455--518",
    year = "2007"
}

@article{Hara:1971kj,
    author = "Hara, Y.",
    title = "{Crossing relations for helicity amplitudes}",
    reportNumber = "TUEP-70-20",
    doi = "10.1143/PTP.45.584",
    journal = "Prog. Theor. Phys.",
    volume = "45",
    pages = "584--595",
    year = "1971"
}

@article{GlueX:2023fcq,
    author = "Adhikari, S. and others",
    collaboration = "GlueX",
    title = "{Measurement of spin-density matrix elements in \ensuremath{\rho}(770) production with a linearly polarized photon beam at E\ensuremath{\gamma}=8.2\textendash{}8.8~GeV}",
    eprint = "2305.09047",
    archivePrefix = "arXiv",
    primaryClass = "nucl-ex",
    reportNumber = "JLAB-PHY-23-3838",
    doi = "10.1103/PhysRevC.108.055204",
    journal = "Phys. Rev. C",
    volume = "108",
    number = "5",
    pages = "055204",
    year = "2023"
}

@article{Schilling:1969um,
    author = "Schilling, K. and Seyboth, P. and Wolf, Guenter E.",
    title = "{On the Analysis of Vector Meson Production by Polarized Photons}",
    reportNumber = "SLAC-PUB-0683",
    doi = "10.1016/0550-3213(70)90070-2",
    journal = "Nucl. Phys. B",
    volume = "15",
    pages = "397--412",
    year = "1970",
    note = "[Erratum: Nucl.Phys.B 18, 332 (1970)]"
}

@article{Mathieu:2017jjs,
    author = "Mathieu, V. and Nys, J. and Fern\'andez-Ram\'\i{}rez, C. and Jackura, A. and Mikhasenko, M. and Pilloni, A. and Szczepaniak, A. P. and Fox, G.",
    title = "{On the $\eta$ and $\eta'$ Photoproduction Beam Asymmetry at High Energies}",
    eprint = "1704.07684",
    archivePrefix = "arXiv",
    primaryClass = "hep-ph",
    reportNumber = "JLAB-THY-17-2450",
    doi = "10.1016/j.physletb.2017.09.081",
    journal = "Phys. Lett. B",
    volume = "774",
    pages = "362--367",
    year = "2017"
}

@article{GlueX:2019adl,
    author = "Adhikari, S. and others",
    collaboration = "GlueX",
    title = "{Beam Asymmetry $\mathbf{\Sigma}$ for the Photoproduction of $\mathbf{\eta}$ and $\mathbf{\eta^{\prime}}$ Mesons at $\mathbf{E_{\gamma}=8.8}$GeV}",
    eprint = "1908.05563",
    archivePrefix = "arXiv",
    primaryClass = "nucl-ex",
    reportNumber = "GlueX-Doc 4093",
    doi = "10.1103/PhysRevC.100.052201",
    journal = "Phys. Rev. C",
    volume = "100",
    number = "5",
    pages = "052201",
    year = "2019"
}

@article{GlueX:2025kma,
    author = "Afzal, F. and others",
    collaboration = "GlueX",
    title = "{First Measurement of $a^0_2(1320)$ Polarized Photoproduction Cross Section}",
    eprint = "2501.03091",
    archivePrefix = "arXiv",
    primaryClass = "nucl-ex",
    month = "1",
    year = "2025"
}

@article{CLAS:2020ngl,
    author = "Carver, M. and others",
    collaboration = "CLAS",
    title = "{Photoproduction of the $f_2(1270)$ meson using the CLAS detector}",
    eprint = "2010.16006",
    archivePrefix = "arXiv",
    primaryClass = "nucl-ex",
    reportNumber = "JLAB-PHY-20-3279",
    doi = "10.1103/PhysRevLett.126.082002",
    journal = "Phys. Rev. Lett.",
    volume = "126",
    number = "8",
    pages = "082002",
    year = "2021"
}

@article{CLAS:2020rdz,
    author = "Celentano, A. and others",
    collaboration = "CLAS",
    title = "{First measurement of direct photoproduction of the $a_2(1320)^0$ meson on the proton}",
    eprint = "2004.05359",
    archivePrefix = "arXiv",
    primaryClass = "nucl-ex",
    reportNumber = "JLAB-PHY-20-3182",
    doi = "10.1103/PhysRevC.102.032201",
    journal = "Phys. Rev. C",
    volume = "102",
    number = "3",
    pages = "032201",
    year = "2020"
}

@article{CLAS:2003umf,
    author = "Mecking, B. A. and others",
    collaboration = "CLAS",
    title = "{The CEBAF Large Acceptance Spectrometer (CLAS)}",
    reportNumber = "JLAB-PHY-03-01",
    doi = "10.1016/S0168-9002(03)01001-5",
    journal = "Nucl. Instrum. Meth. A",
    volume = "503",
    pages = "513--553",
    year = "2003"
}

@article{JointPhysicsAnalysisCenter:2017del,
    author = "Nys, J. and Mathieu, V. and Fern\'andez-Ram\'\i{}rez, C. and Jackura, A. and Mikhasenko, M. and Pilloni, A. and Sherrill, N. and Ryckebusch, J. and Szczepaniak, A. P. and Fox, G.",
    collaboration = "Joint Physics Analysis Center",
    title = "{Features of $\pi \Delta$ Photoproduction at High Energies}",
    eprint = "1710.09394",
    archivePrefix = "arXiv",
    primaryClass = "hep-ph",
    reportNumber = "JLAB-THY-17-2580",
    doi = "10.1016/j.physletb.2018.01.075",
    journal = "Phys. Lett. B",
    volume = "779",
    pages = "77--81",
    year = "2018"
}

@article{Szczepaniak:2001qz,
    author = "Szczepaniak, Adam P. and Swat, Maciej",
    title = "{Role of photoproduction in exotic meson searches}",
    eprint = "hep-ph/0105329",
    archivePrefix = "arXiv",
    reportNumber = "IUNTC-01-01",
    doi = "10.1016/S0370-2693(01)00905-4",
    journal = "Phys. Lett. B",
    volume = "516",
    pages = "72--76",
    year = "2001"
}

@article{Shimada:1978sx,
    author = "Shimada, T. and Martin, Alan D. and Irving, A. C.",
    title = "{DOUBLE REGGE EXCHANGE PHENOMENOLOGY}",
    reportNumber = "Print-78-0676 (DURHAM)",
    doi = "10.1016/0550-3213(78)90209-2",
    journal = "Nucl. Phys. B",
    volume = "142",
    pages = "344--364",
    year = "1978"
}

@article{COMPASS:2014vkj,
    author = "Adolph, C. and others",
    collaboration = "COMPASS",
    title = "{Odd and even partial waves of $\eta\pi^-$ and $\eta'\pi^-$ in $\pi^-p\to\eta^{(\prime)}\pi^-p$ at $191\,\textrm{GeV}/c$}",
    eprint = "1408.4286",
    archivePrefix = "arXiv",
    primaryClass = "hep-ex",
    reportNumber = "CERN-PH-EP-2014-204",
    doi = "10.1016/j.physletb.2014.11.058",
    journal = "Phys. Lett. B",
    volume = "740",
    pages = "303--311",
    year = "2015",
    note = "[Erratum: Phys.Lett.B 811, 135913 (2020)]"
}

@article{Brambilla:2019esw,
    author = "Brambilla, Nora and Eidelman, Simon and Hanhart, Christoph and Nefediev, Alexey and Shen, Cheng-Ping and Thomas, Christopher E. and Vairo, Antonio and Yuan, Chang-Zheng",
    title = "{The $XYZ$ states: experimental and theoretical status and perspectives}",
    eprint = "1907.07583",
    archivePrefix = "arXiv",
    primaryClass = "hep-ex",
    reportNumber = "TUM-EFT 125/19",
    doi = "10.1016/j.physrep.2020.05.001",
    journal = "Phys. Rept.",
    volume = "873",
    pages = "1--154",
    year = "2020"
}

@article{Esposito:2016noz,
    author = "Esposito, A. and Pilloni, A. and Polosa, A. D.",
    title = "{Multiquark Resonances}",
    eprint = "1611.07920",
    archivePrefix = "arXiv",
    primaryClass = "hep-ph",
    reportNumber = "JLAB-THY-16-2301",
    doi = "10.1016/j.physrep.2016.11.002",
    journal = "Phys. Rept.",
    volume = "668",
    pages = "1--97",
    year = "2017"
}

@article{Guo:2017jvc,
    author = "Guo, Feng-Kun and Hanhart, Christoph and Mei\ss{}ner, Ulf-G. and Wang, Qian and Zhao, Qiang and Zou, Bing-Song",
    title = "{Hadronic molecules}",
    eprint = "1705.00141",
    archivePrefix = "arXiv",
    primaryClass = "hep-ph",
    doi = "10.1103/RevModPhys.90.015004",
    journal = "Rev. Mod. Phys.",
    volume = "90",
    number = "1",
    pages = "015004",
    year = "2018",
    note = "[Erratum: Rev.Mod.Phys. 94, 029901 (2022)]"
}

@article{Albrow:2010yb,
    author = "Albrow, M. G. and Coughlin, T. D. and Forshaw, J. R.",
    title = "{Central Exclusive Particle Production at High Energy Hadron Colliders}",
    eprint = "1006.1289",
    archivePrefix = "arXiv",
    primaryClass = "hep-ph",
    reportNumber = "FERMILAB-PUB-10-193-E",
    doi = "10.1016/j.ppnp.2010.06.001",
    journal = "Prog. Part. Nucl. Phys.",
    volume = "65",
    pages = "149--184",
    year = "2010"
}

@article{Huang:2008nr,
    author = "Huang, F. and Sibirtsev, A. and Krewald, S. and Hanhart, C. and Haidenbauer, J. and Meissner, Ulf-G.",
    title = "{Pion-nucleon charge-exchange amplitudes above 2-GeV}",
    eprint = "0810.2680",
    archivePrefix = "arXiv",
    primaryClass = "hep-ph",
    reportNumber = "FZJ-IKP-TH-2008-17",
    doi = "10.1140/epja/i2008-10728-9",
    journal = "Eur. Phys. J. A",
    volume = "40",
    pages = "77--87",
    year = "2009"
}

@article{Becher:2001hv,
    author = "Becher, Thomas and Leutwyler, H.",
    title = "{Low energy analysis of pi N ---\ensuremath{>} pi N}",
    eprint = "hep-ph/0103263",
    archivePrefix = "arXiv",
    reportNumber = "CLNS-01-1727, BUTP-01-11",
    doi = "10.1088/1126-6708/2001/06/017",
    journal = "JHEP",
    volume = "06",
    pages = "017",
    year = "2001"
}

@article{Ananthanarayan:2000ht,
    author = "Ananthanarayan, B. and Colangelo, G. and Gasser, J. and Leutwyler, H.",
    title = "{Roy equation analysis of pi pi scattering}",
    eprint = "hep-ph/0005297",
    archivePrefix = "arXiv",
    reportNumber = "IISC-CTS-12-99, ZU-TH-10-00, BUTP-99-33",
    doi = "10.1016/S0370-1573(01)00009-6",
    journal = "Phys. Rept.",
    volume = "353",
    pages = "207--279",
    year = "2001"
}

@article{Pelaez:2003ky,
    author = "Pelaez, J. R. and Yndurain, F. J.",
    title = "{Regge analysis of pion pion (and pion kaon) scattering for energy s**1/2 \ensuremath{>} 1.4-GeV}",
    eprint = "hep-ph/0312187",
    archivePrefix = "arXiv",
    reportNumber = "FTUAM-03-19",
    doi = "10.1103/PhysRevD.69.114001",
    journal = "Phys. Rev. D",
    volume = "69",
    pages = "114001",
    year = "2004"
}

@article{Buettiker:2003pp,
    author = "Buettiker, Paul and Descotes-Genon, S. and Moussallam, B.",
    title = "{A new analysis of pi K scattering from Roy and Steiner type equations}",
    eprint = "hep-ph/0310283",
    archivePrefix = "arXiv",
    reportNumber = "HISKP-TH-03-18, IPNO-DR-03-08, LPT-ORSAY-03-76",
    doi = "10.1140/epjc/s2004-01591-1",
    journal = "Eur. Phys. J. C",
    volume = "33",
    pages = "409--432",
    year = "2004"
}

@article{Steiner:1971ms,
    author = "Steiner, F.",
    title = "{Partial wave crossing relations for meson-baryon scattering}",
    doi = "10.1002/prop.19710190302",
    journal = "Fortsch. Phys.",
    volume = "19",
    pages = "115--159",
    year = "1971"
}

@article{Hite:1973pm,
    author = "Hite, G. E. and Steiner, F.",
    title = "{New dispersion relations and their application to partial-wave amplitudes}",
    doi = "10.1007/BF02722827",
    journal = "Nuovo Cim. A",
    volume = "18",
    pages = "237--270",
    year = "1973"
}

@article{Fiore:2000fp,
    author = "Fiore, R. and Jenkovszky, Laszlo L. and Magas, V. and Paccanoni, F. and Papa, A.",
    title = "{Analytic model of Regge trajectories}",
    eprint = "hep-ph/0011035",
    archivePrefix = "arXiv",
    reportNumber = "DFPD-00-TH-50, UNICAL-TH-00-8",
    doi = "10.1007/s100500170133",
    journal = "Eur. Phys. J. A",
    volume = "10",
    pages = "217--221",
    year = "2001"
}

@article{Fiore:2004xb,
    author = "Fiore, R. and Jenkovszky, Laszlo L. and Paccanoni, F. and Prokudin, A.",
    title = "{Baryonic regge trajectories with analyticity constraints}",
    eprint = "hep-ph/0404021",
    archivePrefix = "arXiv",
    reportNumber = "DFCAL-TH-04-2",
    doi = "10.1103/PhysRevD.70.054003",
    journal = "Phys. Rev. D",
    volume = "70",
    pages = "054003",
    year = "2004"
}

@article{Szanyi:2023ano,
    author = "Szanyi, Istv\'an and Bir\'o, Tam\'as and Jenkovszky, L\'aszl\'o and Libov, Vladyslav",
    title = "{Nonlinear Regge trajectories and saturation of the Hagedorn spectrum}",
    eprint = "2302.00838",
    archivePrefix = "arXiv",
    primaryClass = "hep-ph",
    doi = "10.1103/PhysRevC.107.024904",
    journal = "Phys. Rev. C",
    volume = "107",
    number = "2",
    pages = "024904",
    year = "2023"
}

@article{Brisudova:1998wq,
    author = "Brisudova, Martina M. and Burakovsky, L. and Goldman, J. Terrance",
    title = "{Effect of color screening on heavy quarkonia Regge trajectories}",
    eprint = "hep-ph/9810296",
    archivePrefix = "arXiv",
    reportNumber = "LA-UR-98-4357",
    doi = "10.1016/S0370-2693(99)00732-7",
    journal = "Phys. Lett. B",
    volume = "460",
    pages = "1--7",
    year = "1999"
}

@article{Kholodkov:1991hx,
    author = "Kholodkov, A. V. and Paccanoni, F. and Stepanov, S. S. and Tutik, R. S.",
    title = "{Regge trajectories for the dipole field interaction and the quark confinement}",
    reportNumber = "DFPD-91-TH-34",
    doi = "10.1088/0954-3899/18/6/002",
    journal = "J. Phys. G",
    volume = "18",
    pages = "985--992",
    year = "1992"
}

@article{Bali:2005fu,
    author = "Bali, Gunnar S. and Neff, Hartmut and Duessel, Thomas and Lippert, Thomas and Schilling, Klaus",
    collaboration = "SESAM",
    title = "{Observation of string breaking in QCD}",
    eprint = "hep-lat/0505012",
    archivePrefix = "arXiv",
    doi = "10.1103/PhysRevD.71.114513",
    journal = "Phys. Rev. D",
    volume = "71",
    pages = "114513",
    year = "2005"
}

@article{MacDowell:1959zza,
    author = "MacDowell, S. W.",
    title = "{Analytic Properties of Partial Amplitudes in Meson-Nucleon Scattering}",
    doi = "10.1103/PhysRev.116.774",
    journal = "Phys. Rev.",
    volume = "116",
    pages = "774--778",
    year = "1959"
}

@article{Loring:2001kx,
    author = "Loring, Ulrich and Metsch, Bernard C. and Petry, Herbert R.",
    title = "{The Light baryon spectrum in a relativistic quark model with instanton induced quark forces: The Nonstrange baryon spectrum and ground states}",
    eprint = "hep-ph/0103289",
    archivePrefix = "arXiv",
    reportNumber = "TK-01-06",
    doi = "10.1007/s100500170105",
    journal = "Eur. Phys. J. A",
    volume = "10",
    pages = "395--446",
    year = "2001"
}

@article{Capstick:1986ter,
    author = "Capstick, Simon and Isgur, Nathan",
    title = "{Baryons in a relativized quark model with chromodynamics}",
    doi = "10.1103/physrevd.34.2809",
    journal = "Phys. Rev. D",
    volume = "34",
    number = "9",
    pages = "2809--2835",
    year = "1986"
}

@article{Meissner:2020khl,
    author = "Mei\ss{}ner, Ulf-G.",
    title = "{Two-pole structures in QCD: Facts, not fantasy!}",
    eprint = "2005.06909",
    archivePrefix = "arXiv",
    primaryClass = "hep-ph",
    doi = "10.3390/sym12060981",
    journal = "Symmetry",
    volume = "12",
    number = "6",
    pages = "981",
    year = "2020"
}

@article{Mai:2020ltx,
    author = "Mai, Maxim",
    title = "{Review of the ${\Lambda }$(1405) A curious case of a strangeness resonance}",
    eprint = "2010.00056",
    archivePrefix = "arXiv",
    primaryClass = "nucl-th",
    doi = "10.1140/epjs/s11734-021-00144-7",
    journal = "Eur. Phys. J. ST",
    volume = "230",
    number = "6",
    pages = "1593--1607",
    year = "2021"
}

@article{Field:1974fg,
    author = "Field, R. D. and Fox, G. C.",
    title = "{Triple Regge and Finite Mass Sum Rule Analysis of the Inclusive Reaction $p$ + $p \to p$ + x}",
    reportNumber = "CALT-68-434",
    doi = "10.1016/0550-3213(74)90495-7",
    journal = "Nucl. Phys. B",
    volume = "80",
    pages = "367--402",
    year = "1974"
}

@article{Ganguli:1980mp,
    author = "Ganguli, S. N. and Roy, D. P.",
    title = "{Regge phenomenology and inclusive reactions}",
    doi = "10.1016/0370-1573(80)90067-8",
    journal = "Phys. Rept.",
    volume = "67",
    pages = "201--395",
    year = "1980"
}

@article{Kim:2022zgz,
    author = "Kim, Hee-Jin and Clymton, Samson and Kim, Hyun-Chul",
    title = "{Triple Regge exchange and transverse single-spin asymmetries of the very forward neutral pion production in polarized p+p collisions}",
    eprint = "2206.02184",
    archivePrefix = "arXiv",
    primaryClass = "hep-ph",
    reportNumber = "INHA-NTG-05/2022",
    doi = "10.1103/PhysRevD.106.054001",
    journal = "Phys. Rev. D",
    volume = "106",
    number = "5",
    pages = "054001",
    year = "2022"
}

@article{Adler:1964um,
    author = "Adler, Stephen L.",
    title = "{Consistency conditions on the strong interactions implied by a partially conserved axial vector current}",
    doi = "10.1103/PhysRev.137.B1022",
    journal = "Phys. Rev.",
    volume = "137",
    pages = "B1022--B1033",
    year = "1965"
}

@article{Fischer:2014xha,
    author = "Fischer, Christian S. and Kubrak, Stanislav and Williams, Richard",
    title = "{Mass spectra and Regge trajectories of light mesons in the Bethe-Salpeter approach}",
    eprint = "1406.4370",
    archivePrefix = "arXiv",
    primaryClass = "hep-ph",
    doi = "10.1140/epja/i2014-14126-6",
    journal = "Eur. Phys. J. A",
    volume = "50",
    pages = "126",
    year = "2014"
}

@article{Nedelko:2016gdk,
    author = "Nedelko, Sergei N. and Voronin, Vladimir E.",
    title = "{Regge spectra of excited mesons, harmonic confinement and QCD vacuum structure}",
    eprint = "1603.01447",
    archivePrefix = "arXiv",
    primaryClass = "hep-ph",
    doi = "10.1103/PhysRevD.93.094010",
    journal = "Phys. Rev. D",
    volume = "93",
    number = "9",
    pages = "094010",
    year = "2016"
}

@article{Sonnenschein:2018fph,
    author = "Sonnenschein, Jacob and Weissman, Dorin",
    title = "{Excited mesons, baryons, glueballs and tetraquarks: Predictions of the Holography Inspired Stringy Hadron model}",
    eprint = "1812.01619",
    archivePrefix = "arXiv",
    primaryClass = "hep-ph",
    doi = "10.1140/epjc/s10052-019-6828-y",
    journal = "Eur. Phys. J. C",
    volume = "79",
    number = "4",
    pages = "326",
    year = "2019"
}

@article{Meyer:2004jc,
    author = "Meyer, Harvey B. and Teper, Michael J.",
    title = "{Glueball Regge trajectories and the pomeron: A Lattice study}",
    eprint = "hep-ph/0409183",
    archivePrefix = "arXiv",
    doi = "10.1016/j.physletb.2004.11.036",
    journal = "Phys. Lett. B",
    volume = "605",
    pages = "344--354",
    year = "2005"
}

@article{Cao:2023ntr,
    author = "Cao, Xiong-Hui and Li, Qu-Zhi and Guo, Zhi-Hui and Zheng, Han-Qing",
    title = "{Roy equation analyses of \ensuremath{\pi}\ensuremath{\pi} scatterings at unphysical pion masses}",
    eprint = "2303.02596",
    archivePrefix = "arXiv",
    primaryClass = "hep-ph",
    doi = "10.1103/PhysRevD.108.034009",
    journal = "Phys. Rev. D",
    volume = "108",
    number = "3",
    pages = "034009",
    year = "2023"
}

@article{Rossi:1977cy,
    author = "Rossi, G. C. and Veneziano, G.",
    title = "{A Possible Description of Baryon Dynamics in Dual and Gauge Theories}",
    reportNumber = "CERN-TH-2287",
    doi = "10.1016/0550-3213(77)90178-X",
    journal = "Nucl. Phys. B",
    volume = "123",
    pages = "507--545",
    year = "1977"
}

@article{Veneziano:1976wm,
    author = "Veneziano, G.",
    title = "{Some Aspects of a Unified Approach to Gauge, Dual and Gribov Theories}",
    reportNumber = "CERN-TH-2200",
    doi = "10.1016/0550-3213(76)90412-0",
    journal = "Nucl. Phys. B",
    volume = "117",
    pages = "519--545",
    year = "1976"
}

@article{tHooft:1974pnl,
    author = "'t Hooft, Gerard",
    title = "{A Two-Dimensional Model for Mesons}",
    reportNumber = "CERN-TH-1820",
    doi = "10.1016/0550-3213(74)90088-1",
    journal = "Nucl. Phys. B",
    volume = "75",
    pages = "461--470",
    year = "1974"
}

@article{Wheeler:1937zz,
    author = "Wheeler, John A.",
    title = "{On the Mathematical Description of Light Nuclei by the Method of Resonating Group Structure}",
    doi = "10.1103/PhysRev.52.1107",
    journal = "Phys. Rev.",
    volume = "52",
    pages = "1107--1122",
    year = "1937"
}

@article{Eden:1971fm,
    author = "Eden, R. J.",
    title = "{Theorems on high energy collisions of elementary particles}",
    doi = "10.1103/RevModPhys.43.15",
    journal = "Rev. Mod. Phys.",
    volume = "43",
    pages = "15--35",
    year = "1971"
}

@article{Roy:1972xa,
    author = "Roy, S. M.",
    title = "{High energy theorems for strong interactions and their comparison with experimental data}",
    doi = "10.1016/0370-1573(72)90005-1",
    journal = "Phys. Rept.",
    volume = "5",
    pages = "125--196",
    year = "1972"
}

@article{Rodas:2023nec,
    author = "Rodas, Arkaitz and Dudek, Jozef J. and Edwards, Robert G.",
    collaboration = "Hadron Spectrum",
    title = "{Determination of crossing-symmetric {\ensuremath{\pi}}{\ensuremath{\pi}} scattering amplitudes and the quark mass evolution of the {\ensuremath{\sigma}} constrained by lattice QCD}",
    eprint = "2304.03762",
    archivePrefix = "arXiv",
    primaryClass = "hep-lat",
    reportNumber = "JLAB-THY-23-3791",
    doi = "10.1103/PhysRevD.109.034513",
    journal = "Phys. Rev. D",
    volume = "109",
    number = "3",
    pages = "034513",
    year = "2024"
}

@article{Cao:2024zuy,
    author = "Cao, Xiong-Hui and Guo, Feng-Kun and Guo, Zhi-Hui and Li, Qu-Zhi",
    title = "{Rigorous Roy-Steiner equation analysis of {\ensuremath{\pi}}K scattering at unphysical quark masses}",
    eprint = "2412.03374",
    archivePrefix = "arXiv",
    primaryClass = "hep-ph",
    doi = "10.1103/38b9-m57d",
    journal = "Phys. Rev. D",
    volume = "112",
    number = "3",
    pages = "L031503",
    year = "2025"
}

@article{Cao:2025hqm,
    author = "Cao, Xiong-Hui and Guo, Feng-Kun and Guo, Zhi-Hui and Li, Qu-Zhi",
    title = "{Revisiting Roy-Steiner-equation analysis of pion-kaon scattering from lattice QCD data}",
    eprint = "2506.10619",
    archivePrefix = "arXiv",
    primaryClass = "hep-ph",
    doi = "10.1103/hn8j-95vn",
    journal = "Phys. Rev. D",
    volume = "112",
    number = "3",
    pages = "034042",
    year = "2025"
}

\end{document}